\documentclass[fleqn,usenatbib]{mnras}

\usepackage{amsmath}

\usepackage{newtxtext,newtxmath}

\usepackage[T1]{fontenc}

\DeclareRobustCommand{\VAN}[3]{#2}
\let\VANthebibliography\thebibliography
\def\thebibliography{\DeclareRobustCommand{\VAN}[3]{##3}\VANthebibliography}

\usepackage{graphicx}	
\usepackage{amsmath}	
\newcommand\textlcsc[1]{\textsc{\MakeLowercase{#1}}}

\title[What drives bar rotation?]{What drives bar rotation? The effect of internal properties and galaxy interactions on bar pattern speeds}

\author[A. Merrow et al.]{
Alex Merrow,$^{1}$\thanks{E-mail: a.j.cooke@2022.ljmu.ac.uk}
Francesca Fragkoudi,$^{2,3}$
Robert J. J. Grand,$^{1}$
Marie Martig,$^{1}$
\\
$^{1}$Astrophysics Research Institute, Liverpool John Moores University, 146 Brownlow Hill, Liverpool, L3 5RF, UK\\
$^{2}$Department of Physics, Durham University, South Road, Durham DH1 3LE, UK\\
$^{3}$Institute of Computational Cosmology, Department of Physics, Durham University, South Road, Durham DH1 3LE, UK
}

\date{Accepted XXX. Received YYY; in original form ZZZ}

\pubyear{2025}

\begin{document}
\label{firstpage}
\pagerange{\pageref{firstpage}--\pageref{lastpage}}
\maketitle

\begin{abstract}
%
One of the main properties of galactic bars is their rotation (or pattern) speed, which is driven by both internal galactic properties, as well as external interactions. To assess the influence of these internal and external drivers on bar rotation in a cosmological setting, we use the Auriga suite of cosmological hydrodynamical zoom-in simulations. We calculate the bar pattern speed and the bar rotation rate -- the ratio of corotation radius to bar length -- at the time of bar formation and at $z=0$, and compare these to bar age, bar strength, baryon dominance, galaxy stellar mass, and the history of external galaxy interactions. We find that galaxies which are more baryon dominated at $z=0$ -- and which lie above the observed stellar mass-halo mass abundance matching relation -- host faster bars, while more dark matter dominated galaxies host slower bars. Baryon-dominated galaxies also form their bars earlier and their rotation rates stay constant or even decrease over time; this leads to older bars being faster than their younger counterparts --  in contrast to the expectation of bar slow-down from dynamical friction imparted by the dark matter halo. We also find a trend in stellar mass, with `faster' bars being hosted in more massive galaxies, which could be driven by the underlying higher baryon-dominance of more massive galaxies. Furthermore, we find that external interactions, such as mergers and flybys, correlate with lower bar rotation rates, particularly for strong interactions that occur around bar formation time. This correlation is relatively weak, leaving internal baryon-dominance as the main driver of fast bar rotation rates.

\end{abstract}




\section{Introduction}
\label{intro}

Galactic bars are prevalent structures present in approximately $70\%$ of disc galaxies at redshift zero \citep[$z=0$; ][]{2000Es,2007Ma,2007Me,2008Sh,2015Bu,2018Er} and have been detected out to cosmic noon, with their fraction decreasing to below 10\% by $z\sim4$ \citep{2008Sh,2024LC,2025HC,2025Ge,2025Am,2025Gu,2026LC}.  These numbers imply that galactic bars have been present in a significant number of disc galaxies for much of their evolution: indeed, observations have found some bars to be ancient structures as old as 13.5 Gyr  \citep{2015Ga,2019Bo,2020Gr,2025DSF}. Additionally, simulations of disc galaxies in \citet{1987Sp}, \citet{2007Ma}, \citet{2012Kr}, \citet{2013At} and \citet{2025Fr} confirm that bars are generally long-lived structures, which evolve significantly across cosmic time.

Bars mediate the exchange of angular momentum and energy of galactic material and, given their prevalence and longevity in disc galaxies, they are significant drivers of secular evolution. This angular momentum exchange is most efficient at resonances -- set by the pattern speed of the bar, $\Omega_\mathrm{p}$ -- where the orbits of stars and other matter can by significantly altered by becoming trapped in or pushed away from the resonance \citep[e.g.][]{1977Co,2000De}. In the disc, these resonances have been seen to have strong effects on the presence and properties of spiral arms \citep{1976Sa,1984Sc,2003Bi}, the migration and mixing of stellar populations across the disc \citep{2000De,2010Mi,2011Br}, and the driving of gas towards the centre of the galaxy, which can in turn form structures such as nuclear discs and rings \citep{1977Ma,1980Si,1999Sa,2005Sh,2014Co}. Meanwhile, the stellar and dark matter halos absorb angular momentum from the bar, which can also spin these components up and change the overall mass profile of the halo \citep{1980Se,1987Ta,1992He,2002We,2003At,2013At}. Therefore, in order to understand the evolution of disc galaxies, it is essential to understand the evolution of their bars and their pattern speeds in particular.

Angular momentum exchange with the rest of the galaxy has an effect on the bar pattern speed in turn. Studies tend to find that, as they evolve, bars lengthen and slow while trapping more stars on elongated bar-supporting orbits and transferring their angular momentum outwards \citep{1984Tr2,2006MV,2007El}. Bars can also slow down (without significantly lengthening) due to dynamical friction imparted by the dark matter halo \citep{1984Tr2,1985We,1991Li,2000De2,2003At,2003ON,2006Se,2019Ka}.
To explore the effect of dynamical friction on the slow down of bars, we can use
the ratio of their corotation radius (the radius at which circular orbits within the disc have angular velocity matching the rotation of the bar) to their semi-major axis length \citep[e.g.][]{1998De,2000De2}. This ratio is sometimes referred to as the `rotation rate' \citep[e.g.][]{2019Cu,2023Bu} and denoted by $\mathscr{R}$. $\mathscr{R}$ increases when the growth of corotation radius -- due to the decrease of the pattern speed -- outpaces the lengthening of the bar.
Bars with $\mathscr{R}>1.4$ are considered slow and bars with $1<\mathscr{R}<1.4$ are considered fast \citep{2000De2}. Bars with $\mathscr{R}<1$ are called ultra-fast, since consideration of the main bar-supporting orbits theoretically places the ends of any bar within the corotation radius \citep{1980Co}. In practice, observations of ultra-fast bars are not unusual \citep{2009Bu,2015Ag,2019Gu} but may be a result of errors such as artificial lengthening of the bar caused by alignment of spiral arms \citep{2020Hi,2021Cu}.

There are a number of both internal and external factors that can affect the slow down (or lack thereof) of the bar, leading to departures from the picture of inevitable slow-down in both $\Omega_\mathrm{p}$ and $\mathscr{R}$. For example, bars in highly baryon-dominated discs will tend to experience less of an impact from dynamical friction \citep{1998De,2000De2,2003At}. Additionally, \citet{2023Li} show that if the dark matter halo has a high spin, then slow down from dynamical friction can be halted for an extended period of up to $5\,\mathrm{Gyr}$, due to the wake of the bar in the dark matter following the stellar bar very closely and so imparting very little torque. \citet{2003At}, \citet{2009Kl} and \citet{2025Zh} also find that a higher vertical velocity dispersion in the disc can reduce the efficiency of angular momentum exchange and delay bar slow down.

Simulations of barred galaxies in \citet{2007Be}, \citet{2008RD}, \citet{2010VV}, \citet{2014At} and \citet{2023Be} find that bars in galaxies containing higher fractions of gas slow down less compared to those with lower gas fractions: the former either maintain a constant pattern speed or speed up. The speed-up occurs as the gas is directed towards the centre of the galaxy by the bar, but the exact mechanism by which the bar slow-down is delayed/halted is unclear. \citet{2005Bo} and \citet{2023Be} claim that the inflowing gas produces a positive torque on the bar directly, whereas \citet{2007Be} and \citet{2008RD} find that it is instead the build up of mass in the centre of the galaxy due to the inflow which both weakens the bar and speeds it up.

Another source of angular momentum for existing bars is interactions and mergers with external galaxies. Simulations find that interactions can affect bars in a variety of ways: interactions can slow-down bars \citep[as in][]{1993Su,1996At}, speed-up bars \citep[see][]{1990Ge,2014Lo,2024Se}, or even destroy them in extreme cases \citep{2015Ka,2022Ca,2022Bi} -- although this last scenario is rare \citep{2018Za}. While the effects of individual interactions are determined in the above papers, there is no comprehensive understanding of the net effects of multiple galaxy interactions and mergers on a population of galactic bars in a cosmological setting.

Galaxy interactions and mergers can also trigger bar formation in unbarred discs through induced perturbations in the host galaxy \citep[e.g.][]{1987No,1990Ge,2014La,2014Lo,2017Mo}, including potentially in our own Galaxy \citep{2024Me}. \citet{1998Mi}, \citet{2017MV}, \citet{2017Ga} and \citet{2022Bi} find that tidally triggered bars tend to be slower than their secularly formed counterparts in simulations. However, \citet{2025Zh} argue that this trend is due to the internal properties of galaxy simulations which have been set up to require an interaction in order to form their bar: initial conditions for idealised simulations of tidally-induced bars are often tuned such that a bar does not form in isolation. Tuning in this way means that properties like the velocity dispersion of discs are higher than their secular bar-forming counterparts. \citet{2025Zh} conclude that this bias in disc initial conditions (mainly the vertical velocity dispersion in the disc, although the circular velocity in the inner disc differs slightly between models) results in slower bars in tidal interaction simulations. However, \citet{2018Pe} find no consistent difference in pattern speed between tidally induced bars and bars formed in isolation, and \citet{2019Pe} even find that tidally induced bars tend to be slightly faster than their secularly formed counterparts (in terms of rotation rate $\mathscr{R}$) in the Illustris cosmological simulation \citep{2014Vo}.

Observations in \citet{2008Ra}, \citet{2011Co}, \citet{2015Ag} and \citet{2019Gu} point towards most bars in the nearby Universe being fast, or close to fast. While more recent data suggest that there may instead be an approximately equal number of slow and fast bars \citep{2022GO,2023Ge}, these bars are not extremely slow, with bars slower than $\mathscr{R}=2.5$ being relatively uncommon.
In contrast, cosmological simulations show either a lack of bars, as \citet{2022Re} shows in the NewHorizon simulation \citep{2021Du}, or a prevalence of slow bars. \citet{2017Al}, \citet{2019Pe}, \citet{2021Ro} and \citet{2024Ha} show that the bars in the EAGLE \citet{2015Cr,2015Sc}, Illustris \citep{2014Vo} and IllustrisTNG -- both TNG100 \citep{2018Ma,2018Na,2018Ne,2018Pi,2018Sp} and TNG50 \citep{2019Ne,2019Pi} -- simulations have rotation rates much slower than those from observations (most have $\mathscr{R}>2.5$), and \citet{2025An} find no fast bars (i.e. with $\mathscr{R}<1.4$) in the FIRE-2 simulations \citep{2018Ho,2023We} despite lower average rotation rates. \citet{2022Fr} attribute the lack of fast bars in the IllustrisTNG simulations to the bar lengths being too short on average in comparison to observations, which is exacerbated for low numerical resolution.
However, \citet{2021Fr} show that the Auriga cosmological zoom-in simulations do produce fast bars, in keeping with observations. They argue that these faster bars are because Auriga galaxies have a higher stellar mass to dark matter mass ratio compared to other cosmological simulations, which minimises the slow-down impact from dynamical friction. In this paper,
we expand on previous work \citep{2021Fr}, by increasing the sample of barred galaxies for which we explore bar pattern speed to lower masses and also by exploring how the bar pattern speed evolution is affected by internal galactic properties and external perturbations, such as interactions and mergers.


The paper is structured as follows. We first outline the Auriga suite of simulations in Sec.~\ref{auriga} before establishing the measurements we extract in Sec.~\ref{method:main}. In Sec.~\ref{barresults}, we present the time evolution of bar pattern speeds in our sample, and subsequently examine how they have been impacted by baryon fraction in Sec.~\ref{internalresults} and from external interactions in Sec.~\ref{externalresults}. We discuss the implications and context of these results in Sec.~\ref{ratediscuss} and how they compare to observations in Sec.~\ref{obscomparison}. Lastly, we summarise and conclude our results in Sec.~\ref{conc}.

\section{Methodology}
\label{meth}

\subsection{The Auriga simulations}
\label{auriga}

Throughout this work, we use the Auriga suite of simulations \citep{2017Gr,2019Gr}, a set of 39 magneto-hydrodynamical cosmological zoom-in simulations of Milky~Way mass halos. The initial conditions for the simulations were generated by first identifying halos with $0.5< M_{200}/10^{12}\,\mathrm{M}_\odot<2$ from the $z=0$ snapshot of the dark matter-only run of the Lambda cold dark matter ($\Lambda$CDM) EAGLE cosmological simulations \citep{2015Cr,2015Sc}. The most isolated quartile of these halos was randomly sampled \citep[see][for details]{2017Gr}. Once identified, the resolution of the dark matter particles of the Lagrangian region of each halo at $z=127$ is refined to higher resolution, whereas the particle resolution outside of this region is progressively worsened to increase computational efficiency while retaining the large-scale gravitational tidal field from the Cosmic Web. Dark matter particles are then split into dark matter-baryon pairs according to cosmological parameters taken from \citet{2014Pl}: $h=0.677$, $\Omega_{\mathrm{b}}=0.048$, $\Omega_{\mathrm{m}}=0.307$, and $\Omega_{\mathrm{\Lambda}}=0.693$. Typical mass resolutions in the highest resolution region are $5\times10^4\,\mathrm{M}_\odot$ for baryons and $3\times10^5\,\mathrm{M}_\odot$ for dark matter. The dark matter and stars are given a softening length of $500h^{-1}\,\mathrm{cpc}$ up to being fixed at $369\,\mathrm{pc}$ from $z=1$ onward. The softening length of a gas cell varies from these values as a minimum up to a maximum of $1.85\,\mathrm{kpc}$, scaling with the size of the gas cell. The simulations are run with the moving-mesh magneto-hydrodynamic code \textlcsc{AREPO} \citep{2010Sp,2016Pa}, which includes a TreePM gravity solver and the \textlcsc{Auriga} sub-grid galaxy formation physics model.
The subgrid physics include: uniform UV background radiation until $z=6$ \citep{2009FG}; a subgrid model of the interstellar medium with two phases \citep{2003Sp}; star particles forming stochastically out of gas denser than $0.1$ atoms $\rm cm^{-3}$ representing simple stellar populations; primordial and metal line cooling; stellar evolution with feedback from type Ia and type II supernovae and AGB stars; black hole seeding with subsequent accretion and AGN feedback; a model of magnetic fields \citep{2014Pa,2017Pa}.
For full details of the \textlcsc{Auriga} code, see \citet{2013Vo}, \citet{2014Ma} and \citet{2017Gr}. For further information on the details of the Auriga simulations as a whole, see \citet{2017Gr} and references therein. 

The initial conditions of our simulations are the same as those of the \texttt{Original}/4 and \texttt{LowMassMws}/4 suites described in \citet{2024Gr}\footnote{These initial conditions and simulations are publicly available and accessible via \href{https://wwwmpa.mpa-garching.mpg.de/auriga/data.html}{https://wwwmpa.mpa-garching.mpg.de/auriga/data.html}.}. However, our analysis requires a much finer snapshot cadence than these simulations provide in order to accurately measure the patterns speeds of bars. Therefore, we re-simulated each halo that produces a barred galaxy at $z=0$ \citep[according to][]{2025Fr} with additional 5 Myr-cadence snapshots of the stellar particles only. These higher cadence outputs, which we call ``snipshots'', allow us to compute pattern speed directly from the change in the phase angle of the bar without using any assumptions such as those required for the commonly used Tremaine-Weinberg method \citep{1984Tr}. In addition, we approximately double the number of full snapshots (which contain dark matter particles, gas cells, black holes, and tracer particles as well as star particles) such that their cadence becomes $\sim60\,\mathrm{Myr}$ at late times\footnote{Au18 is an exception since each of its 3000 snapshots contain information for all simulated matter. Therefore, all types of matter are output at $\sim5\,\mathrm{Myr}$ intervals rather than just the star particles.}. These snapshots are accompanied by a merger tree, which tracks the paths and merger histories of each dark matter halo in the simulation.

\subsection{Galaxy and interaction properties}
\label{method:main}

\subsubsection{Disc and bar parameters}
\label{method:internal}

\begin{figure}
	\includegraphics[width=\columnwidth]{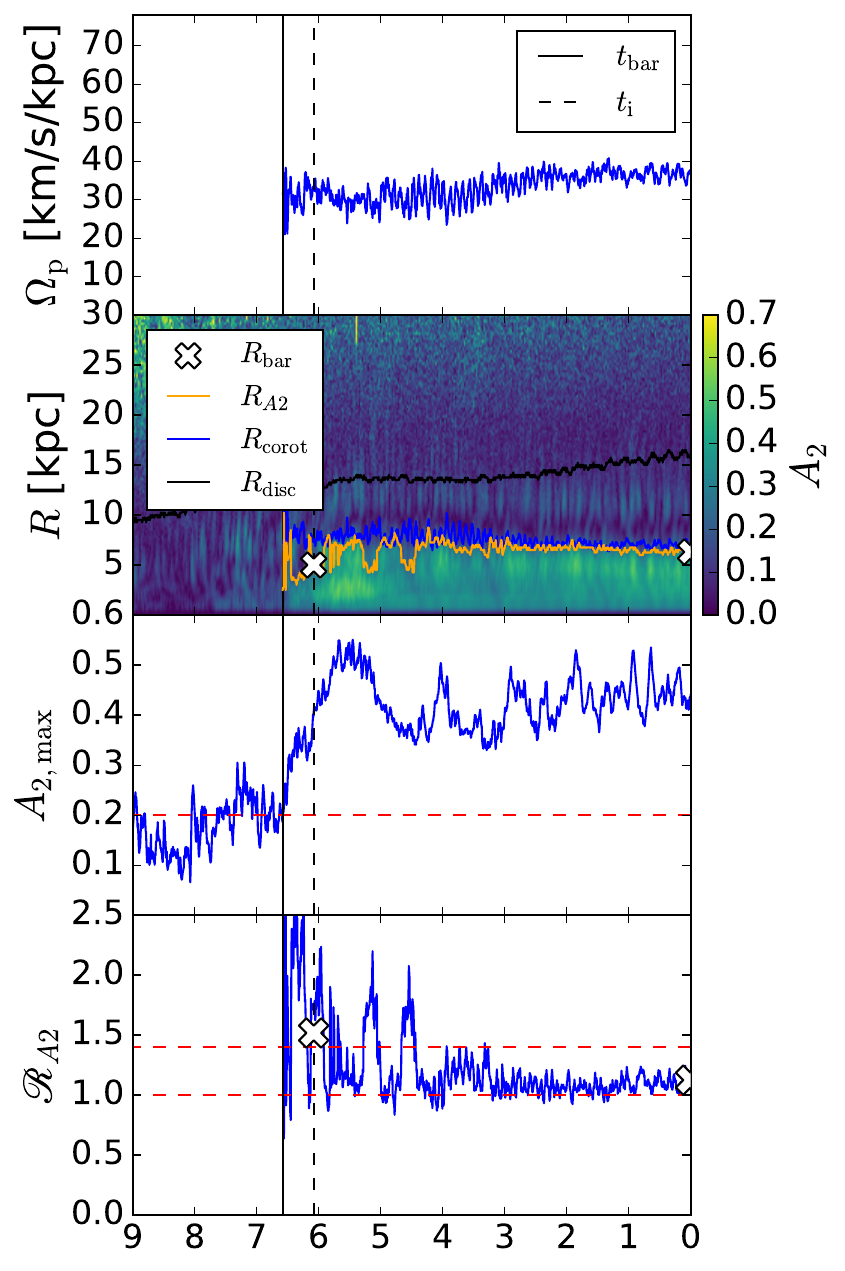}
    \caption{Summary of the main bar parameters used for halo 22 of the Auriga simulations plotted against lookback time. The top panel shows the pattern speed of the bar, from the bar formation time onwards. The second panel shows the bar length calculated from the $A_2$ Fourier mode (orange) and the corotation radius (blue) from the bar formation time onwards, alongside the disc extent (black). The colour scale shows the radial profile of the $A_2$ Fourier mode with lighter, yellower colours indicating higher values. The white crosses give the values of bar length from ellipse fitting. The third panel shows the maximum of the $A_2$ Fourier mode at each time, which we also define as our bar strength; the red dashed line indicates $A_{2\mathrm{,max}}=0.2$, our cut-off for bar formation. The bottom panel shows the rotation rate as calculated from the bar length with $A_2$ and corotation rate (blue line), with the red dashed lines indicating rotation rates of $1$ and $1.4$. The white crosses give the values of rotation rate using bar lengths from ellipse fitting, as used in our results. In all panels, the vertical solid black line shows the time of bar formation $t_\mathrm{bar}$, and the vertical black dashed line shows $0.5\,\mathrm{Gyr}$ later, when we take our measurements at $t_\mathrm{i}$.}
    \label{fig:24}
\end{figure}

Before analysing each snipshot, we first centre and rotate the galaxy. Beginning with the interpolated coordinates from the merger tree at the snapshot times, we centre the galaxy by taking the centre of mass of the stars within a sphere of radius $30\,\mathrm{kpc}$, repeating for spheres of radius $10\,\mathrm{kpc}$ and $4\,\mathrm{kpc}$ to converge on the true centre. We then align the mean angular momentum vector of the stars within the effective (half-mass) radius of the galaxy with the z-axis, such that the region containing most of the mass, and therefore most of the bar, lies within the x-y plane.

We also find the angle between the bar and the x-axis for each snipshot, using the angle of the second Fourier mode $A_2$ of the stellar mass profile in the inner $4\,\mathrm{kpc}$, excluding the inner $0.5\,\mathrm{kpc}$ and restricting to within $1\,\mathrm{kpc}$ of the disc plane. This angle is calculated as

\begin{equation}
    \label{FourierAngle}
    \Theta_2 = \frac{1}{2}\arctan{\frac{\sum_k m_k \sin{2 \theta_k}}{\sum_k m_k \cos{2 \theta_k}}}
\end{equation}

with an associated magnitude for the mode given by

\begin{equation}
    \label{Fourier}
    A_2 = \sqrt{\left( \sum_k m_k \cos{2 \theta_k} \right) ^2 + \left( \sum_k m_k \sin{2 \theta_k} \right) ^2}
\end{equation}

where $m_k$ and $\theta_k$ are the mass and angle from the positive x-axis respectively for each particle in the chosen volume. The magnitude is normalised to the $A_0$ Fourier mode, the total mass in the same volume. The pattern speed, $\Omega_\mathrm{p}$, is then the rate of change of the bar angle between adjacent snipshots. We show an example of the pattern speed evolution as calculated by this method for Au22 in the top panel of Fig.~\ref{fig:24}, converted from $\mathrm{rad}\,\mathrm{Gyr}^{-1}$ to $\mathrm{km}\,\mathrm{s}^{-1}\,\mathrm{kpc}^{-1}$ to match the typical units used in observations. We can see that in this case, the pattern speed is nearly constant, except for a small increase from $4-2\,\mathrm{Gyr}$ lookback time.

To find the time of bar formation, we first use the maximum of the $A_2$ mode magnitude to measure bar strength. We calculate $A_2$ of stars in $250\,\mathrm{pc}$-wide radial annuli within $1\,\mathrm{kpc}$ of the disc plane. The 2D histogram in the second panel of Fig.~\ref{fig:24} shows the dependence of the normalised $A_2$ Fourier mode on both galaxy radius and time. Consistent higher values in the inner regions indicate the presence of a bar and this feature can be seen in the lower portion of the panel from $\sim6\,\mathrm{Gyr}$ lookback time onwards. We therefore use, for any given time, the maximum value for any bin in the inner $6\,\mathrm{kpc}$ as our measure of bar strength, $A_{2,\mathrm{max}}$. We show bar strength in the third panel of Fig.~\ref{fig:24}, with the red dashed line indicating $A_{2,\mathrm{max}}=0.2$. In Au22, as is common for many of our halos, the bar strength exhibits some variation at early times due to transient features, before quickly increasing past $0.2$ and remaining above this line for the remainder of the simulation. We set the bar formation time as the time of the snipshot for which $A_{2,\mathrm{max}}$ last surpasses this value of $0.2$\footnote{We note that this is a different value of $A_{2,\mathrm{max}}$ to the $0.25$ used in \citet{2025Fr} for the same suite of simulations. We make this change due to some of the halos in our rerun suite now with $0.2<A_{2,\mathrm{max}}<0.25$ but visible bars in face-on projection images.}. The vertical solid line shown across all panels of Fig.~\ref{fig:24} indicates the bar formation time calculated in this way, $t_\mathrm{bar}$. $0.5\,\mathrm{Gyr}$ after $t_\mathrm{bar}$ is the time at which we take initial measurements of bar properties, $t_\mathrm{i}$, to give values just after the bar forms, but while allowing the bar to somewhat settle from any chaotic perturbations involved in its formation. If a halo has $A_{2,\mathrm{max}}<0.2$ at $z=0$, we say that the halo is unbarred.

Although, following the definition above, the bar formation time is known to within half a snipshot, the definition itself is somewhat arbitrary due to the selection of $A_{2\mathrm{,max}}$ cut-off. Therefore, we examine the sensitivity of the bar formation time by varying the limit from $A_{2\mathrm{,max}}$ by $10\%$ to see how the change affects the resultant bar formation time. We find that the bar formation time is changed by less than $2.5\%$ for the oldest 10 bars, with only 5 others (Au12, Au20, Au28, AuL2, and AuL10) changing by more than $0.5\,\mathrm{Gyr}$. Each of these 5 bars is younger than $3\,\mathrm{Gyr}$ and have bar strengths at $z=0$ weaker than $0.31$. We conclude that our definition of bar formation time is stable for older, stronger bars, but is sensitive to the choice of limits for weak and young bars.

Note that we make 3 exceptions to our definition of bar formation time: Au17 which forms its bar extremely early in its evolution, Au23 which periodically produces snipshots with $A_{2,\mathrm{max}}<0.2$ due to a double barred system, and AuL2 which has a bar that is destroyed by $z=0$. In the latter two cases, we choose a time when the bar is clearly present in the face-on stellar distribution, but before any subsequent dips in $A_{2,\mathrm{max}}$ below $0.2$, and only use the snipshots before this time to calculate the bar formation time.
For Au17, we delay our bar formation time until a bar has become visually present in the face-on stellar distribution, since prominent, consistent spirals which reach to the inner $6\,\mathrm{kpc}$ lead to a consistently high `bar strength' before a genuine bar has formed.


In the following, we primarily use the rotation rate $\mathscr{R}=\frac{R_\mathrm{corot}}{R_\mathrm{bar}}$ as our measure of how fast or slow the bar is, since this gives a greater insight into the disc dynamics than pattern speed (see Sec.~\ref{intro}). $\mathscr{R}$ requires the bar length and corotation radius, which we calculate at two times for each halo: $0.5\,\mathrm{Gyr}$ after the time of bar formation, $t_\mathrm{i}$, and at the end of the simulation, at $z=0$. We find the bar length from ellipse fittings to the contours of the face-on stellar surface density profile -- see Appendix~\ref{lengthapp} for further details. For the corotation radius, we take the pattern speed $\Omega_\mathrm{p}$ and the circular rotation velocity curve (as calculated from the total mass inside spherical bins centred on the origin) at each snapshot, and interpolate the rotation curve across each snipshot. We then interpolate the resulting circular rotation curve with respect to radius to match the pattern speed and find the corotation radius.

The second panel of Fig.~\ref{fig:24} shows illustrative versions of these lengths at each snipshot. The blue line is the corotation radius as calculated above, and the white crosses are the bar lengths from ellipse fitting which we use throughout the rest of the paper. The black line gives the radius of the disc, here defined as the first radial bin in the disc plane for which the surface density drops below $1\,\mathrm{M}_\odot\,\mathrm{pc}^{-2}$. We also show a second version of bar length in this figure, which we call $R_{A2}$, in the orange line. This takes the first radial bin to drop below $0.7\,A_{2,\mathrm{max}}$ in each snipshot. This calculation is more easily automated, so helps to show the continuous evolution of parameters for this figure, although we only use the ellipse fitting derived $R_\mathrm{bar}$ for the results that follow, since $R_{A2}$ is more susceptible to being artificially extended by galaxy interactions. Observations also tend to use ellipse fitting to find bar lengths in external galaxies, so taking a similar method improves comparison with observations. We note that, despite bar lengths from ellipses tending to be longer than those from Fourier decomposition \citep{2024Gh}, our results are qualitatively broadly similar when using $R_{A2}$ instead of $R_\mathrm{bar}$. The bottom panel shows that the resulting rotation rate, $\mathscr{R}_{A2}$, is approximately constant, after initially decreasing for the first few $\mathrm{Gyr}$ signifying an early speed up. The white crosses in the bottom panel show $\mathscr{R}$ calculated from $R_\mathrm{bar}$ instead -- in this case the two versions of bar length are very similar, and so the white crosses match the blue line at both times.

In addition to these parameters, in what follows we also use the virial mass, the stellar mass within $20\,\mathrm{kpc}$, and the baryon dominance within one effective radius. The last of these is calculated as the ratio of the circular velocity at the effective radius induced by the enclosed stellar mass to the circular velocity at the effective radius from the total enclosed mass. Many of these parameters vary apparently randomly on short timescales. For most parameters, we account for their dispersion by taking the mean value across $200\,\mathrm{Myr}$ (40 snipshots), with the standard deviation from the same interval giving a basis for the error bars shown below -- these error bars are propagated through as uncertainties for derived quantities. Since we do not calculate bar length for every snipshot, we treat these error bars slightly differently and describe this in Appendix \ref{lengthapp}.

\subsubsection{Galaxy interaction parameters}
\label{method:external}

\begin{figure}
	\includegraphics[width=\columnwidth]{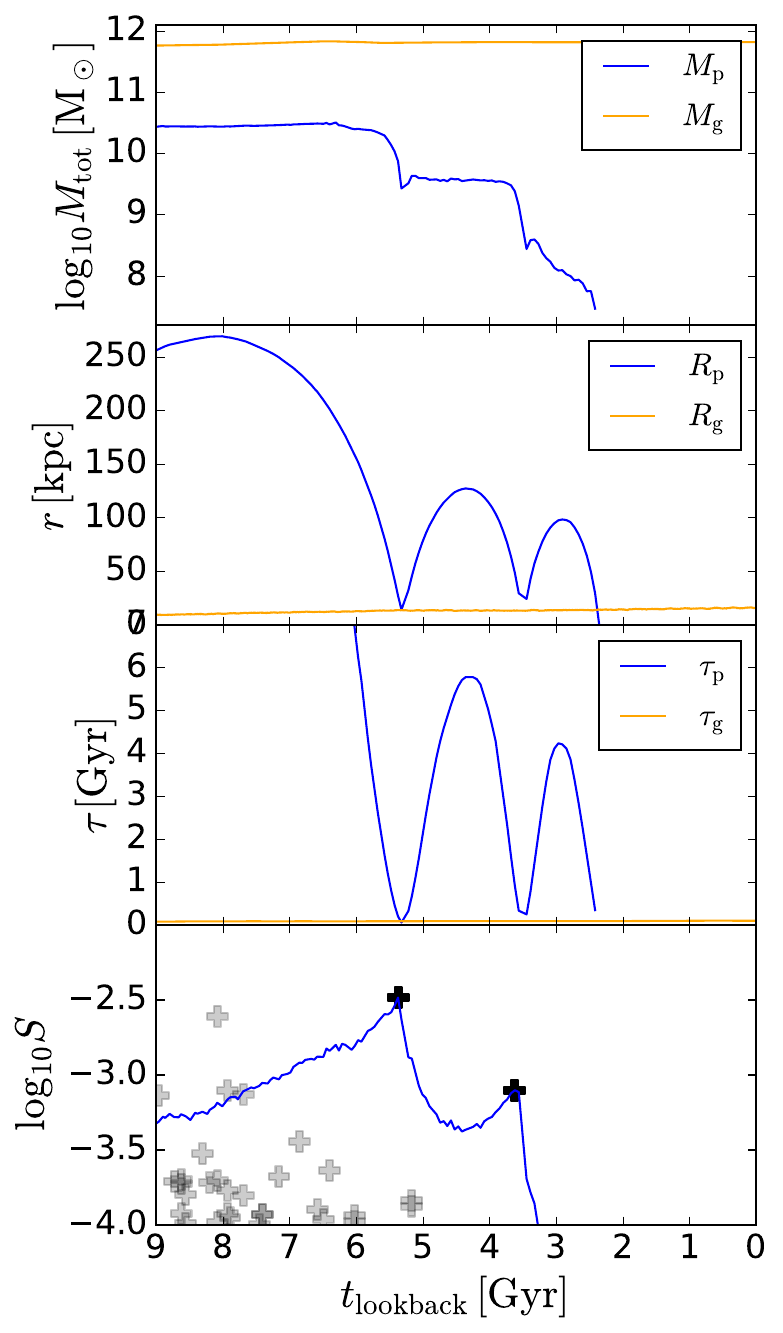}
    \caption{An illustrative example of the Elmegreen parameter calculation for one interacting halo in Au22. Shown against lookback time, the panels show from top to bottom: the halo mass for the perturber (blue) and main halo (orange); the distance between the main halo and the perturber (blue) and the extent of the disc in the main halo (orange); the time to complete $1\,\mathrm{rad}$ of orbit around the centre of the main halo for the perturber (blue) and stars at the edge of the disc (orange); the resulting Elmegreen parameter (blue line) with its local maxima (black crosses). Other interactions in Au22 are shown by the grey crosses.}
    \label{fig:intex}
\end{figure}

To parametrise the accretion and interaction history of each Auriga halo, we use the masses, positions, and velocities from their merger trees. Using the merger tree, we calculate, for each snapshot output time, the Elmegreen parameter $S$ \citep{1991El} for each halo within $50\,\mathrm{kpc}$ of the main halo\footnote{The Elmegreen parameter is strictly intended for use at pericentre passage, but the quantities used to calculate it are meaningful at any time}. This parameter gives a dimensionless measure of the tidal forces imparted during the course of an interaction. The Elmegreen parameter $S$ is given by

\begin{equation}
    \label{Elme}
    S = \left( \frac{G m_\star M_\mathrm{p} R_\mathrm{g}}{R_\mathrm{p}\,^3} \tau_\mathrm{p} \right) \left( \frac{m_\star R_\mathrm{g}}{\tau_\mathrm{g}} \right)^{-1}
    = \frac{M_\mathrm{p}}{M_\mathrm{g}} \left( \frac{R_\mathrm{p}}{R_\mathrm{g}} \right)^{-3} \frac{\tau_\mathrm{p}}{\tau_\mathrm{g}}
\end{equation}

where $M_\mathrm{p}$ is the total mass of the perturbing halo, $M_\mathrm{g}$ is the total mass of the central halo, $R_\mathrm{p}$ is the distance between the centres of the two halos, $R_\mathrm{g}$ is the extent of the disc in the main halo, $\tau_\mathrm{p}$ is the time taken for the perturber to orbit $1\,\mathrm{rad}$ around the central halo (from its angular velocity at that time), and $\tau_\mathrm{g} = \sqrt{\frac{R_\mathrm{g}^3}{G M_\mathrm{g}}}$ is the time taken for a star (of mass $m_\star$) on a circular orbit at the edge of the disc to orbit $1\,\mathrm{rad}$.

The first expression for $S$ in equation \ref{Elme} is arrived at directly from the physical motivation for the parameter: it is the ratio of the tidal impulse imparted to a star in the outer disc by the perturber (left bracket) and the momentum of such a star in its assumed circular orbit (right bracket). This, in effect, gives the relative impact of the tidal force on stars in the outer disc over the course of the interaction. While the first expression of equation \ref{Elme} is already dimensionless, we can remove constants from the expression by substituting for $\tau_\mathrm{g}$, thereby arriving at the second expression for $S$ in equation \ref{Elme}. For further details, see \citet{1991El}.

In Fig.~\ref{fig:intex}, we illustrate $S$ and its component parameters for one example interaction from Au22. The top panel shows the mass terms, with $M_\mathrm{p}$ in blue and $M_\mathrm{g}$ in orange. The perturber is less massive than the main halo, and over the course of their extended interaction the main halo accretes matter from the perturber. The accretion concludes with a merger at $\sim2.5\,\mathrm{Gyr}$ lookback time, with $M_\mathrm{p}$ dropping significantly before being fully accreted into the main halo. The Elmegreen parameter is larger for higher mass ratios, since these have a larger impact on the main galaxy; as accretion progresses, this contributes a decreasing factor to $S$. The second panel shows the distance terms, with $R_\mathrm{p}$ in blue and $R_\mathrm{g}$ in orange. The disc extent $R_\mathrm{g}$ is calculated according to the definition given in Sec.~\ref{method:internal}. We can see the disc growing over time, while the perturber experiences 3 pericentres in a decaying orbit. Since closer interactions produce stronger forces, the Elmegreen parameter is larger when the perturber is closer to the centre of the main halo relative to the disc extent, so is increased around pericentres. The third panel shows the time terms, with $\tau_\mathrm{p}$ in blue and $\tau_\mathrm{g}$ in orange near the lower axis of the panel. The total force imparted by the perturber is larger when it acts over a longer time (greater $\tau_\mathrm{p}$) and this is included in the Elmegreen parameter, normalised to $\tau_\mathrm{g}$. As such, increases in $S$ from the short separation at pericentres are somewhat reduced by the shorter times involved as the perturber speeds up near pericentre.

The final panel of Fig.~\ref{fig:intex} shows $S$ at each snapshot for this example interaction, compiling the behaviours of the individual quantities described above into one value, shown as the blue line. From the evolution of $S$, we take the local maxima (defined as any snapshot whose value of $S$ is larger than any other snapshot within $\sim1\,\mathrm{Gyr}$) to be the strength and time of the interaction. For this example, our approach yields 2 interactions for the perturber from its first 2 pericentres -- by the time of the merger, the perturber's mass is too low to produce a significant tidal effect. We show these maxima of the blue line as black crosses. We also show the maxima for other interacting halos with Au22 as grey crosses. We can see there are a few other small interactions after the bar forms at $\sim6\,\mathrm{Gyr}$, but the example highlighted here with the blue line and black crosses dominates the external forces on the main halo.


We employ this method of local maxima rather than simply calculating $S$ at pericentre for two reasons. Firstly, for a merging galaxy there is no well defined pericentre for its final accretion as it disappears from the merger tree, whereas the maximum $S$ can be reliably found in all cases. Secondly, during close orbits and mergers, the halos of the two galaxies can mix, most often artificially decreasing the recorded mass of the interacting galaxy for single snapshots \citep[see Sec~7.4 of][for an examination of this effect]{2021Sp} and producing an artificially lower value of $S$. By taking the local maximum of $S$, we avoid an extremely low value if this effect is significant for the snapshot at pericentre, taking the next best snapshot instead.

\section{Results}
\label{allresults}

In this section, we use the methods described in Sec.~\ref{meth} to analyse the evolution of the pattern speeds of bars. We begin with a description of the evolutions of the bars over their lifetimes in Sec.~\ref{barresults}, followed by a discussion of how this is affected by the changing composition of the host galaxies in Sec.~\ref{internalresults} and by interactions with external galaxies in Sec.~\ref{externalresults}.

\subsection{Evolution of bar properties}
\label{barresults}

\begin{figure*}
	\includegraphics[width=2\columnwidth]{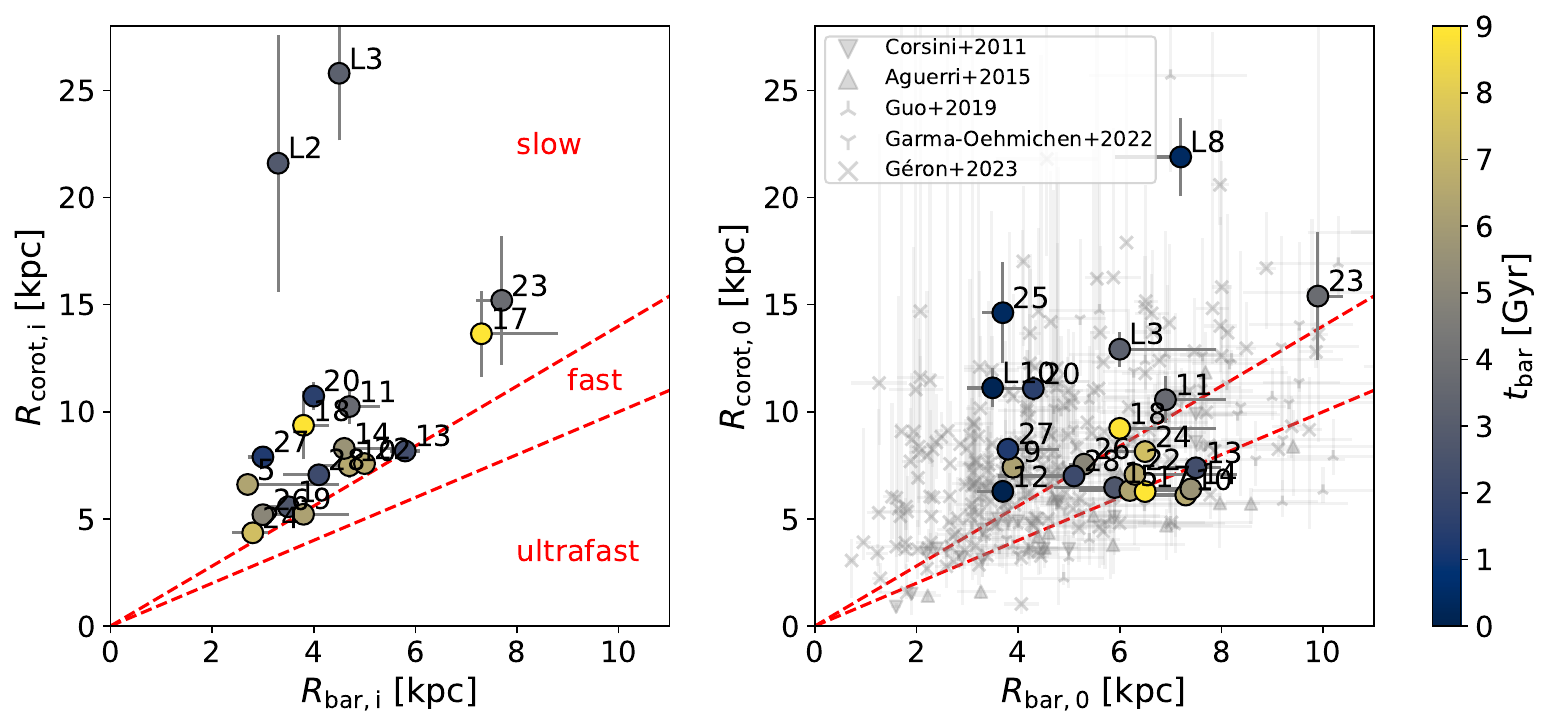}
    \caption{Bar length and corotation radius for each halo in our barred sample, with uncertainties, at $t_\mathrm{i}$ (left panel) and $z=0$ (right panel). The colour scale indicates the bar formation time in each case. In the right panel, we also plot in grey with uncertainties the observational bar lengths and corotation radii from \citet{2011Co} (downwards triangles), \citet{2015Ag} \citep[with redshifts from][]{2002Pa,2015GB} (upwards triangles), \citet{2019Gu} (upwards tri-points), \citet{2022GO} (downwards tri-points), and \citet{2023Ge} (crosses). The upper and lower dashed red lines correspond to ratios of $1.4$ and $1$ respectively, with the red labels in the left panel indicating the resulting slow, fast, and ultrafast regions.}
    \label{fig:lengths}
\end{figure*}

Fig.~\ref{fig:lengths} shows the bar length and corotation radius for each of the 22 bar-forming simulations from our sample, at $z=0$  in the right panel and at the initial time $t_\mathrm{i}$ ($0.5\,\mathrm{Gyr}$ after the bar forms) in the left panel, alongside $z\sim0$ measurements from the local Universe taken from \citet{2011Co}, \citet{2015Ag}, \citet{2019Gu}, \citet{2022GO} and \citet{2023Ge}. Of the 22 simulations, 4 halos form their bars less than $0.5\,\mathrm{Gyr}$ before the end of the simulation, so we only show these values for $z=0$. There is also one case, AuL2, of a bar which is weakened beyond our bar formation threshold before the simulation ends; for this halo we only show initial values of bar properties. We show only one time for these 5 halos throughout the rest of the paper.

We see that the bars in our sample have lengths in the range of $3.5-8\,\mathrm{kpc}$, with the exception of Au23 whose bar is length is almost $10\,\mathrm{kpc}$ by $z=0$\footnote{Au23 also has an unusual bar since it is the outer bar of a double-barred system.}. These lengths are consistent with measurements of bar lengths from the local Universe in \citet{2011Ga}, which finds a median of $4.5\,\mathrm{kpc}$, a standard deviation of $1.9\,\mathrm{kpc}$ and some values reaching up to $13\,\mathrm{kpc}$. The corotation radii found in our sample tend to be larger, ranging from $\sim5\,\mathrm{kpc}$ up to $\sim24\,\mathrm{kpc}$ in a handful of cases with very slowly rotating bars. By comparing with the observations in the right panel, we can see that our sample at $z=0$ generally overlaps well with the deprojected properties of real bars, albeit with a scarcity of short bars with similarly small corotation radii. It appears that the bars from Auriga have properties compatible with those of the bars in similar mass galaxies in the local Universe, at least in terms of the radii shown here, but may not cover the full range observed.

Using the red dashed lines to indicate $\frac{R_\mathrm{corot}}{R_\mathrm{bar}}=\mathscr{R}=1,1.4$, we see that, at initial times, the majority of the bars in our sample are slow (points lying above both lines). However, most of the bars at $z=0$ are fast or very close to fast (between the two lines) and 4 are ultra-fast (Au10, Au13, Au14, and Au17 lie below both lines). Since the fast bar region of Fig.~\ref{fig:lengths} is predominantly populated by $z=0$ bars, the bars appear to have sped up on average. For the largest cluster of points, presumably representing the more typical population of barred galaxies, the points mostly move from left to right; this suggests that the primary driver of the bar speed up is an increasing bar length, with corotation radius remaining almost constant.

\begin{figure}
	\includegraphics[width=\columnwidth]{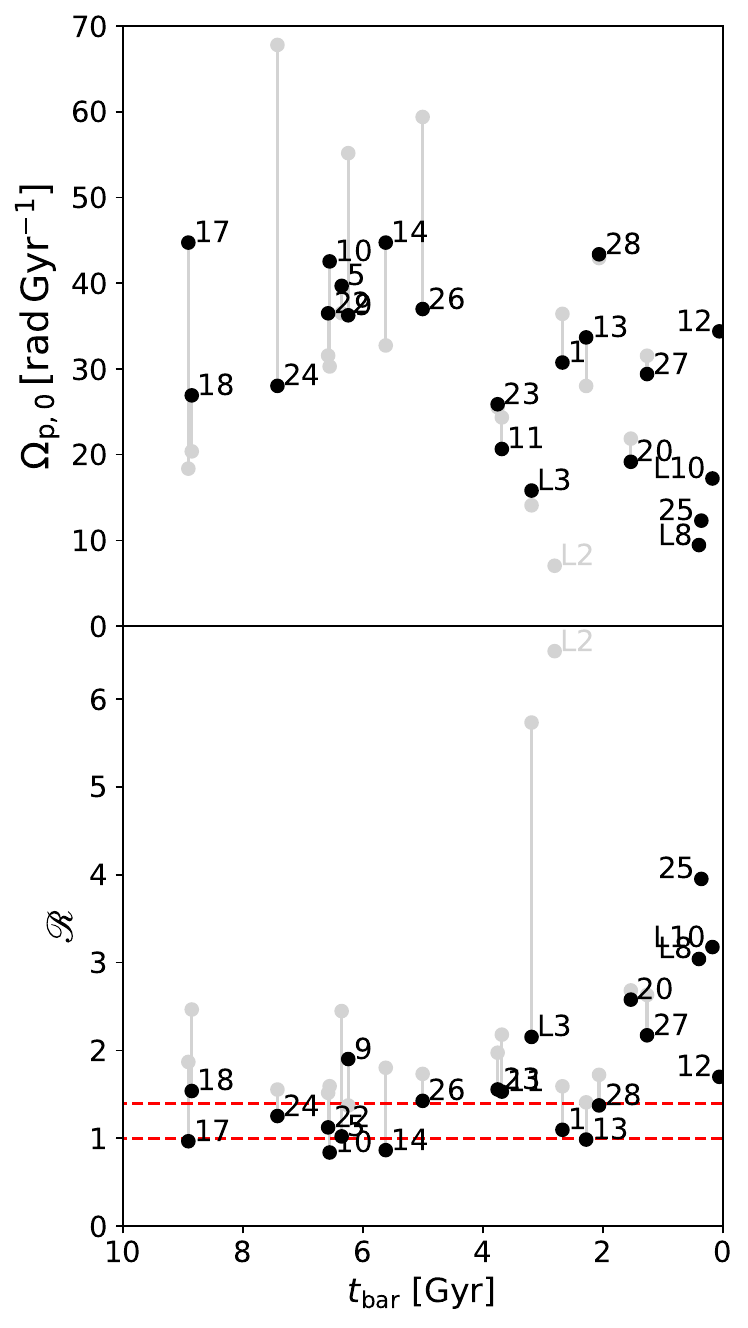}
    \caption{Top: the relation between bar formation time and pattern speed at $z=0$, shown as the black points, and at $0.5\,\mathrm{Gyr}$ after bar formation, shown as the connected light grey points. Bottom: the same, but for $\mathscr{R}$ instead of pattern speed. The red dashed lines indicate $\mathscr{R}=1,1.4$.}
    \label{fig:time}
\end{figure}

The evolution of the bar pattern speed and rotation rate in Auriga is shown in Fig.~\ref{fig:time}, which shows pattern speed versus bar formation time in the top panel and $\mathscr{R}$ versus bar formation time in the bottom panel. Grey points show the initial $\mathscr{R}_\mathrm{i}$ and black points the final $\mathscr{R}_0$. From the top panel, we see that while a significant number of older bars have sped up or slowed down, there is no reliable net change in speed across the population. In contrast, it is clear in the bottom panel that the majority of the bars have a higher value of $\mathscr{R}$ at $t_\mathrm{i}$, so are initially slower in this measure. At $z=0$, we can also see the trend that older bars are faster on average in both panels. This trend is most obvious in $\mathscr{R}$ for the bars younger than $2\,\mathrm{Gyr}$, which are almost all -- except for Au12 -- slower than any of the older bars. The same trend is also apparent, but with a more gradual slope for bars older than $\gtrsim 2$ Gyr. The speed-up in the Auriga bars does not appear to be a constant, gradual process, since the relation between change in $\mathscr{R}$ and bar age is relatively weak (see panel (f) in Fig.~\ref{fig:app_wide}). Additionally, many of the younger bars are also slow at $t_i$, which indicates an effect on rotation rate before the bar can age.

\begin{figure}
    \includegraphics[width=\columnwidth]{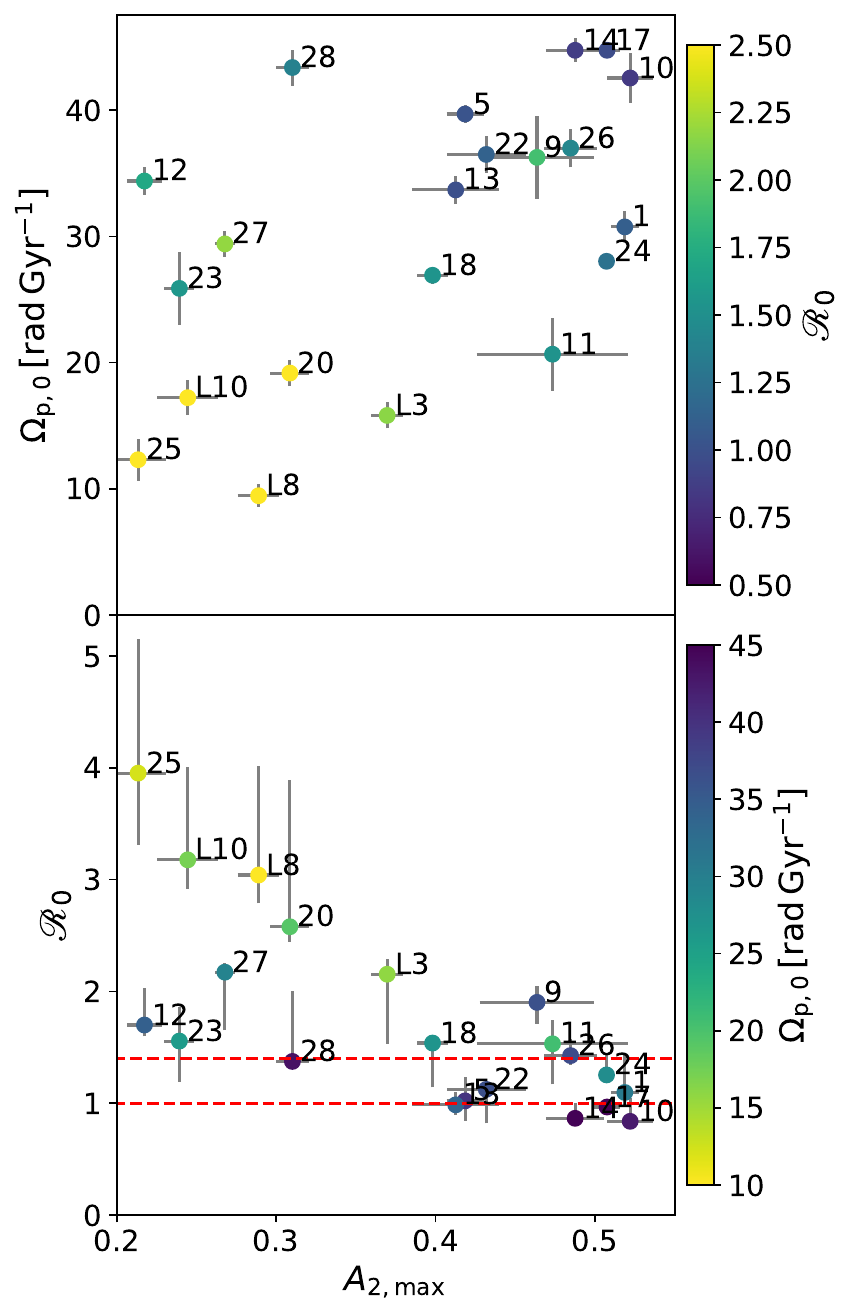}
    \caption{The dependence at $z=0$ of bar pattern speed and rotation rate on bar strength. The top panel shows the pattern speed at $z=0$ $\Omega_{\mathrm{p,}0}$, coloured by the rotation rate at $z=0$ $\mathscr{R}_0$, while the bottom panel shows $\mathscr{R}_0$ on the axis and $\Omega_{\mathrm{p,}0}$ in the colour scale. Both have error bars shown. The bottom panel shows red dashed lines indicating $\mathscr{R}=1,1.4$.}
    \label{fig:strength}
\end{figure}

In Fig.~\ref{fig:strength}, we show the bar pattern speed in the top panel and the normalised rotation rate $\mathscr{R}$ in the bottom panel. In the bottom panel, we see that stronger bars tend to be faster when using $\mathscr{R}$. This mirrors the trend seen in Fig.~\ref{fig:time}, since age and strength are positively correlated in our sample (see panel (e) in Fig.~\ref{fig:app_wide}), but appears stronger here across the whole sample than the trend between $\mathscr{R}_0$ and $t_\mathrm{bar}$. As seen in the top panel, this relation with bar strength holds when using pattern speed instead of rotation rate, albeit with significantly greater scatter. Stronger bars still tend to be faster, but there is a wide spread especially for the weaker bars, indicating that the clearer trend with $\mathscr{R}$ relies on both pattern speed and bar length.

Put together, we find a typical picture of bar evolution in the Auriga simulations: the bar forms somewhat slow, or very slow if it forms late, but as it evolves over time it becomes stronger and longer, and lowers its rotation rate to become faster. This is perhaps rather surprising, given that we tend to expect bars to slow with time \citep{1985We,1991Li,2000De2,2003At}. In the following sections, we explore the causes of this trend, examining how the internal and external properties affect the rotation rate.

\subsection{The effect of changes in the host galaxy}
\label{internalresults}

Bars do not exist in static systems, instead belonging to host galaxies that have a wide variety of evolving properties. This section examines how the wider galaxy affects the rotation rate, focusing on the mass and radial distributions of the stellar and dark matter components of the galaxies. Sec.~\ref{globalresults} explores this at the whole galaxy and halo scales, while Sec.~\ref{innerresults} restricts this to the inner few $\mathrm{kpc}$.

\subsubsection{Global galaxy composition}
\label{globalresults}

\begin{figure*}
    \includegraphics[width=2\columnwidth]{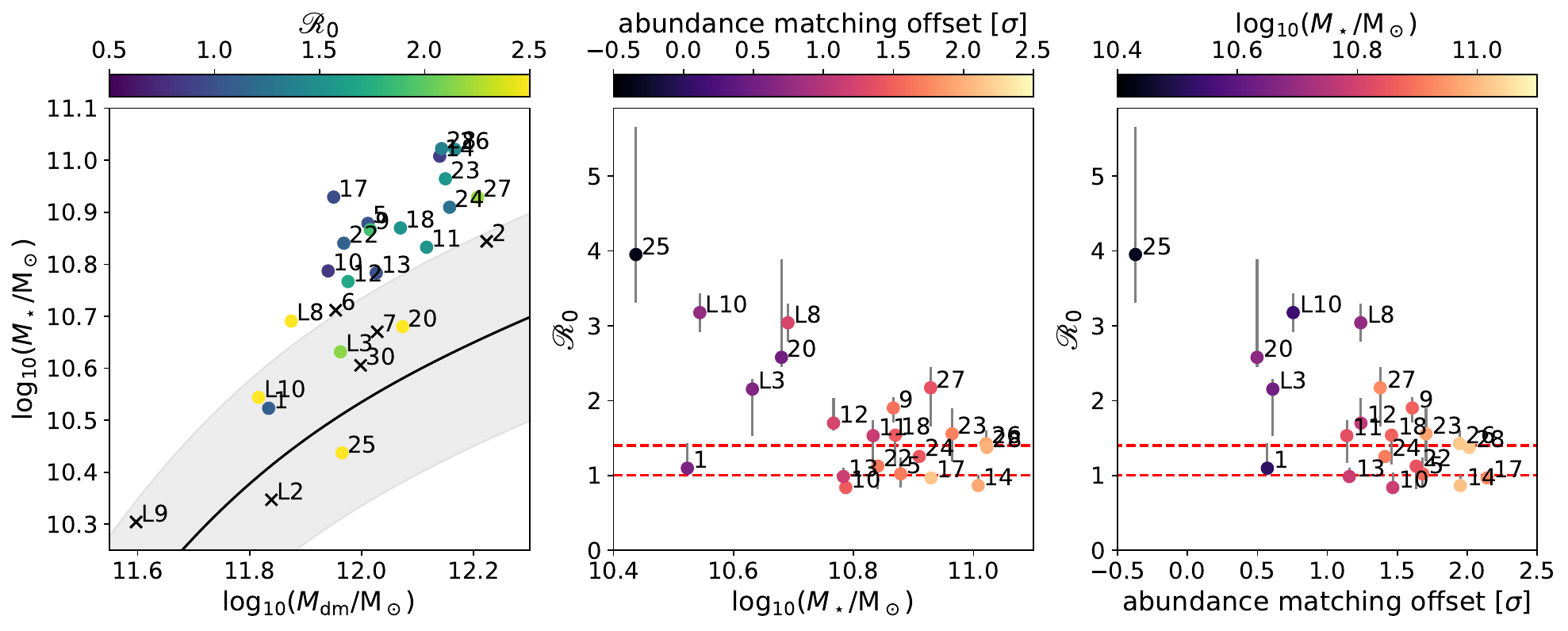}
    \caption{Left: Final dark matter halo mass compared with final stellar mass for each halo in our sample. The colour scale shows the final value of $\mathscr{R}$ for each bar. Halos which are unbarred at $z=0$ are marked as black crosses. The black line shows the abundance matching relation from \citet{2013Mo}, with the grey shaded region showing the $1\sigma$ dispersion at each halo mass. Middle: final $\mathscr{R}$ plotted against final stellar mass, for those halos which are barred at $z=0$. The colour scale shows the vertical offset of the halo from the abundance matching relation in the left panel, given in units of the $\sigma$ dispersion at the relevant dark matter mass. Right: final $\mathscr{R}$ plotted against the vertical offset from the abundance matching relation, for those halos which are barred at $z=0$. The colour scale shows the final stellar masses of the halos. In the two rightmost panels, error bars are shown for $\mathscr{R}_0$ and the red dashed lines, as before, indicate $\mathscr{R}=1,1.4$.}
    \label{fig:abundance}
\end{figure*}

The left panel of Fig.~\ref{fig:abundance} shows the stellar mass halo mass relation, comparing the masses of the halos in our sample to the semi-empirical abundance matching relation from \citet{2013Mo}. We first note that our sample sees a steeper slope than the observational relation, with the barred galaxies in particular lying significantly above the relation at higher masses, in keeping with previous results from the Auriga simulations \citep[see][for more details]{2025Fr}. We can also see from the colour scale that the halos in the upper right portion of the figure, with higher stellar masses and a greater offset from the abundance matching relation, tend to host faster bars.

We show $\mathscr{R}$ as a function of stellar mass in the middle panel of Fig.~\ref{fig:abundance}, where we can see that the lower mass galaxies in our sample tend to host slower bars, with just one fast bar for galaxies with stellar masses below $6\times10^{10}\,\mathrm{M}_\odot$. This relation may link to the trend in the left panel of more massive galaxies lying further above the abundance matching relation. The right panel explores this, showing the vertical offset from the abundance matching relation (the difference in $\mathrm{log}_{10} (M_{\star}/M_{\odot})$ between a galaxy's stellar mass and the stellar mass predicted by the abundance matching relation at the same halo mass), normalised to the observed $1\sigma$ dispersion of the relation. It confirms that, at $z=0$, faster bars are indeed hosted by halos lying at higher abundance matching offsets. Although there is a large scatter in $\mathscr{R}_0$, this trend appears across the full range of the sample, suggesting it may be a stronger trend compared to the relation between $\mathscr{R}_0$ and stellar mass, which only appears to affect rotation rates in galaxies below $6\times10^{10}\,\mathrm{M}_\odot$. Indeed, we find Spearman correlation coefficients with $\mathscr{R}_0$ of -0.57 for the abundance matching offset and -0.45 for the stellar mass. As such, the rotation rate appears to be more directly dependent on the relative masses of the galaxy and its dark matter halo than the stellar mass alone.

The trend with stellar mass acts as a prediction of the Auriga model that can be tested with observations. In particular, this provides a test of $\Lambda$CDM cosmology, since the cause of slower bars in less massive spiral galaxies appears to be lower fractions of baryons. Since baryon fraction is not meaningful in cosmology models without dark matter \citep[e.g. in a Modified Newtonian Dynamics framework;][]{1983Mi}, we would not expect to see a dependence of rotation rate on galaxy mass in such models. Indeed, bars in galaxies with masses $10^9-10^{11}\,\mathrm{M}_\odot$ in \citet{2023Na} under modified Newtonian gravity (MOND) have $\mathscr{R}\sim1.6$ for each of 3 simulations. While a direct comparison with Auriga is unavailable due to a lack of cosmological simulations with MOND and a comparatively small number of isolated examples, this indicates that simulations using MOND produce relatively fast bars at all masses. We compare our relation with stellar mass to those found in observations in Sec.~\ref{obscomparison}.

\subsubsection{Baryon dominance in the inner regions}
\label{innerresults}

Bars are structures of the inner few $\mathrm{kpc}$ of the stellar disc, and as such their formation and evolution is especially closely tied to the properties of the host galaxy in its inner region, in particular the relative properties of the stars and the dark matter. We quantify the relative contributions of stars and dark matter to the mass of the inner galaxy using the ratio of the circular velocity due to the stellar component within $1$ effective radius of the centre of the halo, to the circular velocity from the total mass within the same region. We refer to this as the baryon dominance. The effective radius for the simulated galaxies in our sample is in the region of $2-10\,\mathrm{kpc}$, matching the range of bar lengths as shown in Fig.~\ref{fig:lengths}. This definition of baryon dominance therefore describes the prevalence of stars compared to dark matter in the inner bar-forming regions.

\begin{figure*}
	\includegraphics[width=2\columnwidth]{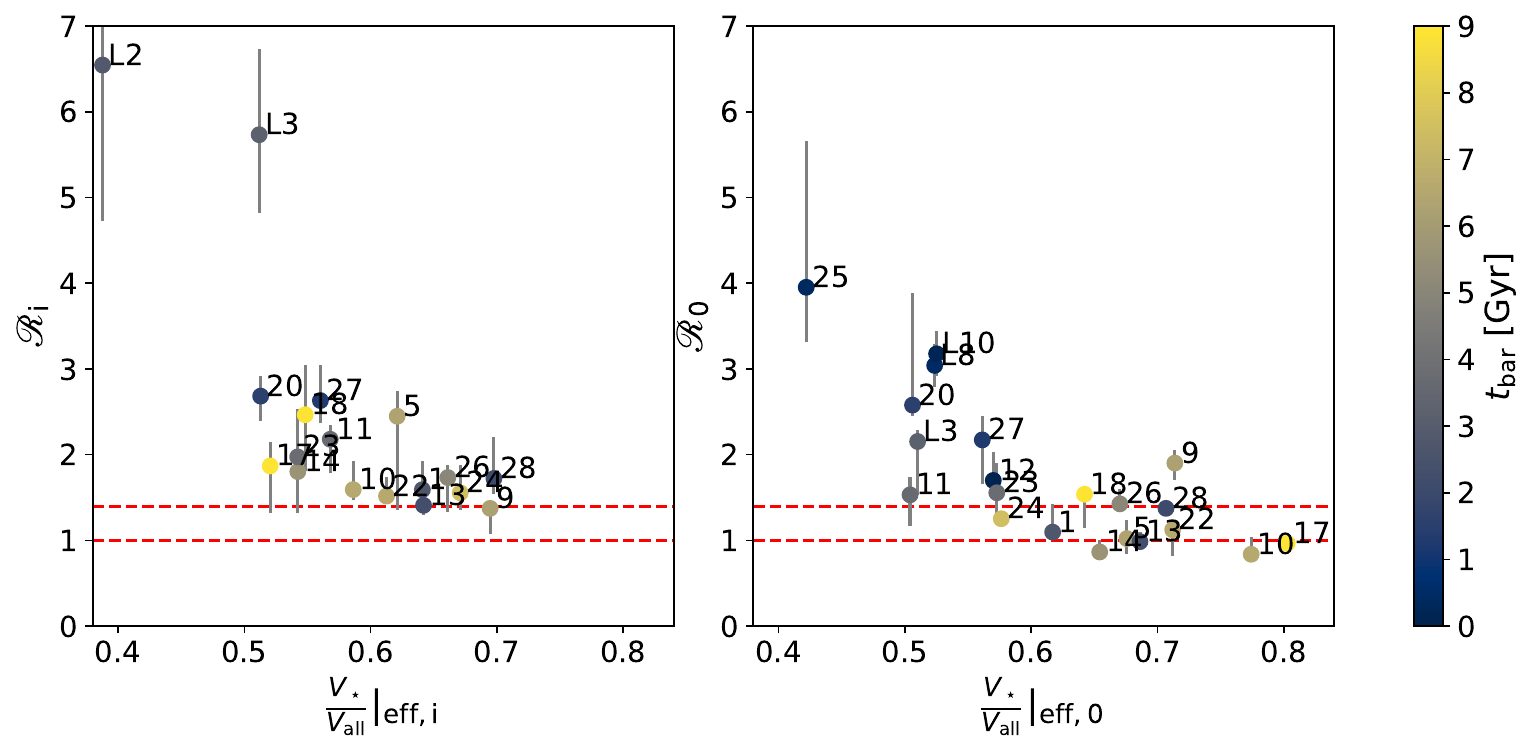}
    \caption{The relation between baryon dominance (quantified as the ratio of circular velocities for stars and for all matter, evaluated at the effective radius) and $\mathscr{R}$. Both axes are shown for $z=0$ on the right and for $0.5\,\mathrm{Gyr}$ after bar formation on the left, coloured by the bar formation time. Error bars are shown for $\mathscr{R}$ in both panels. The red dashed lines indicate $\mathscr{R}=1,1.4$ in both panels.}
    \label{fig:baryondom}
\end{figure*}

Fig.~\ref{fig:baryondom} shows the dependence of $\mathscr{R}$ at initial (left panel) and final (right panel) times on the baryon dominance of the host galaxy at the same time. We clearly see, at both bar formation and the end of the simulation, that rotation rate is strongly dependent on the baryon dominance of its host galaxy at the corresponding times. Focusing on the left panel, we see that all but one bar are slow initially, although the faster bars at $t_\mathrm{i}$ reside in the more baryon dominated hosts, with a gradual increase in $\mathscr{R}$ as baryon dominance decreases. The colour scale shows that this trend is largely independent of the bar formation time. In the right panel, at $z=0$, we see the same trend of faster bars tending to be hosted by more baryon dominated galaxies, but now with a high proportion of fast bars for $\frac{V_\star}{V_\mathrm{all}}|_{\mathrm{eff,}0}>0.6$. The change is largely due to the older bars, whose galaxies become more baryon dominated since their bars formed and whose bars correspondingly decrease their rotation rates to occupy the fast regime. For example, Au17, the oldest bar in the sample, begins as a slow bar in the fourth least baryon dominated halo in the left panel of Fig.~\ref{fig:baryondom}, but by $z=0$ is the most baryon dominated galaxy in the right panel, and hosts an ultra-fast bar.

\begin{figure}
	\includegraphics[width=\columnwidth]{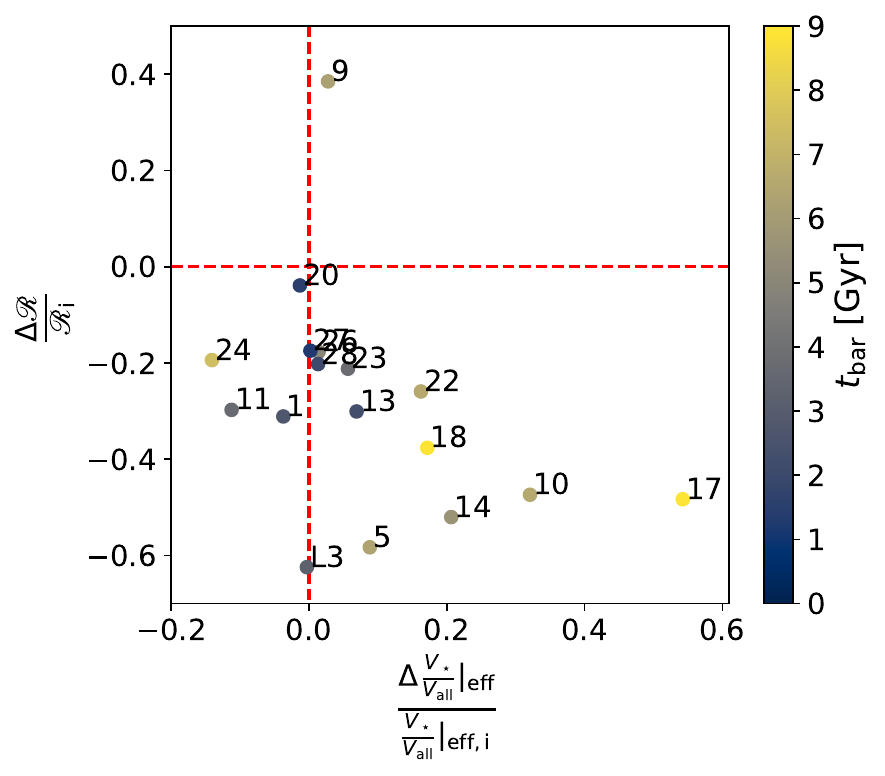}
    \caption{The fractional change in $\mathscr{R}$ from bar formation to $z=0$ shown against the fractional change in baryon dominance (quantified as in Fig.~\ref{fig:baryondom}) over the same time period. The colour scale shows the lookback time to bar formation. The red dashed lines indicate no change in $\mathscr{R}$ (horizontal) and baryon dominance (vertical). Bars below the horiztonal line have sped up and halos to the right of the vertical line have increased their baryon dominance.}
    \label{fig:baryonchange}
\end{figure}

In Fig.~\ref{fig:baryonchange} we show how each halo changes from $t_\mathrm{i}$ to $z=0$, with fractional changes in $\mathscr{R}$ and baryon dominance. There appears to be a trend present, with significant scatter: galaxies that have experienced the largest increase in baryon dominance generally host bars that have sped up more. Additionally, the single bar that has instead slowed down experienced just a $5\%$ increase in baryon dominance over its $6\,\mathrm{Gyr}$ lifetime, far less than many of the other halos. The colour scale indicates some weak correlation with the bar formation time -- here the time over which the change is measured -- with the oldest bars seeing some of the largest changes, speeding up while their host galaxies become more baryon dominated.

\subsection{The effect of galaxy interactions}
\label{externalresults}

Within a cosmological setting, galaxies evolve not just by secular processes, but also by interactions with other galaxies. These interactions can induce torques, and thus change the angular momentum of discs and their internal structures, such as bars, and therefore may affect their properties. We parametrise here the strength of such interactions by the Elmegreen parameter as outlined in Sec.~\ref{method:external}, and focus on the strongest interactions in two time intervals to describe their potential impact on the rotation rates of bars. We are firstly interested in the time around bar formation, since this affects the conditions under which the bar first appears and has the potential to set its behaviour early in its evolution. Following this, we investigate the continued effect of interactions on the rotation rate during the lifetime of the bar, which may alter the bar as it evolves up to the end of the simulation.

\begin{figure}
\includegraphics[width=\columnwidth]{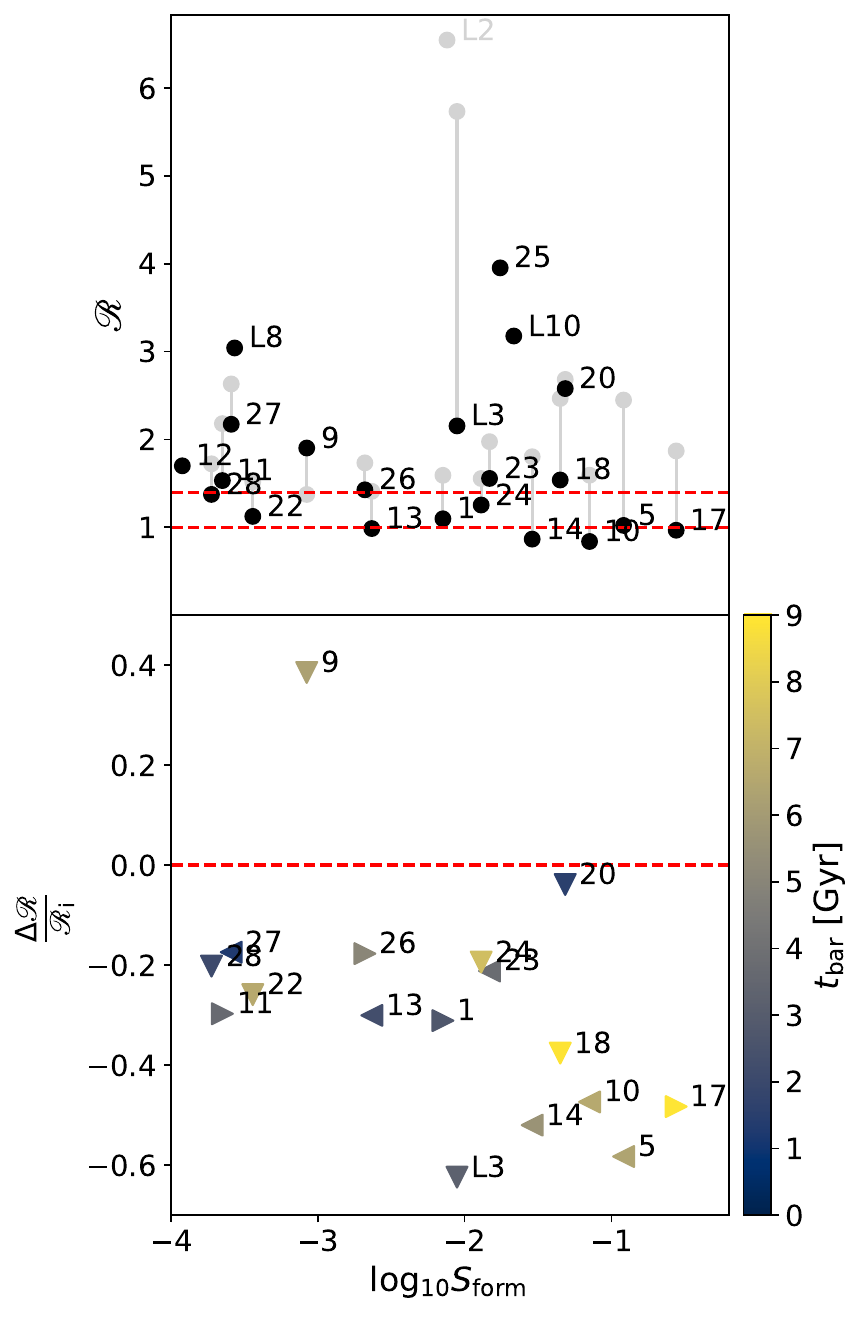}
    \caption{The relation between $\mathscr{R}$ and the maximum Elmegreen parameter ($S$) for any interaction from $1\,\mathrm{Gyr}$ before to $0.5\,\mathrm{Gyr}$ after bar formation. The top panel shows $\mathscr{R}$ at $z=0$ (black points) and at $0.5\,\mathrm{Gyr}$ after bar formation (grey points connected by the grey lines). The red dashed lines indicate $\mathscr{R}=1,1.4$. The bottom panel shows the fractional change in $\mathscr{R}$ between these two times. The colour scale shows the bar formation time for each halo, and the left/down/right pointing triangles indicate retrograde/polar/prograde oriented interactions respectively. The red dashed line in the lower panel indicates no change in $\mathscr{R}$; bars below this line have sped up.}
    \label{fig:formation}
\end{figure}

Fig.~\ref{fig:formation} shows the dependence of rotation rate on the logarithm of the maximum interaction strength around the time of bar formation (the $1.5\,\mathrm{Gyr}$ immediately preceding our $t_\mathrm{i}$). We denote this as $\log_{10}S_\mathrm{form}$. The top panel of Fig.~\ref{fig:formation} shows the initial $\mathscr{R}_\mathrm{i}$ and final $\mathscr{R}_0$ in grey and black points, respectively. Despite a wide spread, we see a weak trend in $\mathscr{R}_0$ where the fastest bars by $z=0$ formed alongside some of the strongest interactions. We note that the bars in Au20, Au25, and AuL10, while contradicting this trend, are each younger than $2\,\mathrm{Gyr}$ -- this may be the reason for their high rotation rates given the trend in Fig.~\ref{fig:formation}. There is, however, no discernable trend in their initial rotation rates. This is unexpected, since interactions around the bar formation time would intuitively be more likely to affect bar properties around the same time. Instead, we see little to no effect at the initial time itself, with a weak trend only arising at later times. We note that there is a bias present: stronger interactions are more likely to occur at earlier times in the simulation. This results in earlier forming bars being more likely to form amid strong interactions, thereby having larger values of $\log_{10}S_\mathrm{form}$, and, as we saw in Sec.~\ref{barresults}, earlier forming bars are also faster on average. The interdependence of $\log_{10}S_\mathrm{form}$ and $t_\mathrm{bar}$ can be seen in the thick-bordered panel (c) in Fig.~\ref{fig:app_wide}.

Our finding that older bars experience stronger interactions at their time of formation relative to younger bars contrasts with results from the TNG50 simulation \citep{2019Ne,2019Pi} presented in \citet{2025Fr2}. These authors argue that secularly formed bars are older, because they find: i) that older bars form with higher baryon dominance and; ii) a weak anti-correlation between baryon dominance and interaction strength around the bar formation time (note that they use different formulations of these quantities than those used here). Our results are consistent with the former (see panel (d) in Fig.~\ref{fig:app_wide}), but not with the latter: we do not see any relation between baryon dominance and $S_\mathrm{form}$ (see panel (b) in Fig.~\ref{fig:app_wide}), which accounts for the differing results between simulations.

The bottom panel of Fig.~\ref{fig:formation} shows the fractional change in rotation rate across the lifetime of the bar, $\frac{\Delta\mathscr{R}}{\mathscr{R}_\mathrm{i}}$. We see a somewhat stronger trend with $\log_{10}S_\mathrm{form}$ when looking at change rather than instantaneous $\mathscr{R}$, although the scatter is still significant: bars forming alongside stronger interactions tend to have sped up more by $z=0$ than those which were formed in relatively isolated galaxies. This is not entirely explained by the aforementioned bias from bar formation times, as can be seen by the colour scale (e.g. Au9 and Au-22 which have not sped up by more than $25\%$ and have not had strong interactions, but are relatively old bars). In this panel, we also show the orientation of the interaction relative to the spin of the disc by the shape of the points: right-pointing triangles indicate prograde interactions where the orbit of the perturber is aligned with the rotation of the disc to within $60^\circ$; left-pointing triangles indicate retrograde interactions where the orbit of the perturber is aligned opposite to the rotation of the disc to within $60^\circ$; upwards-pointing triangles indicate polar interactions where the orbit of the perturber is intermediate to these two cases, i.e. approaching at or close to perpendicular to the disc plane. There seems to be little coherent effect from the orientation of the interactions, despite expectations that prograde fly-by interactions should have greater effects on the galaxy \citep{1972To,2010DO,2014La,2018Lo}.


There is also a strong relation between $\log_{10}S_\mathrm{form}$ and the fractional change in baryon dominance (see panel (a) in Fig.~\ref{fig:app_wide}) which suggests some interdependency between these parameters since both show a similar trend with $\frac{\Delta\mathscr{R}}{\mathscr{R}_\mathrm{i}}$. It is possible that a strong interaction around the time of bar formation produces an inflow of baryonic material, increasing the baryon dominance of the inner regions of the galaxy over a timescale of a few $\mathrm{Gyr}$. It is unclear which relation is the genuine causation -- whether increase in baryon dominance and maximum interaction strength around bar formation are mechanically linked, or they simply both produce a similar speeding up of the rotation rate.

\begin{figure}
    \includegraphics[width=\columnwidth]{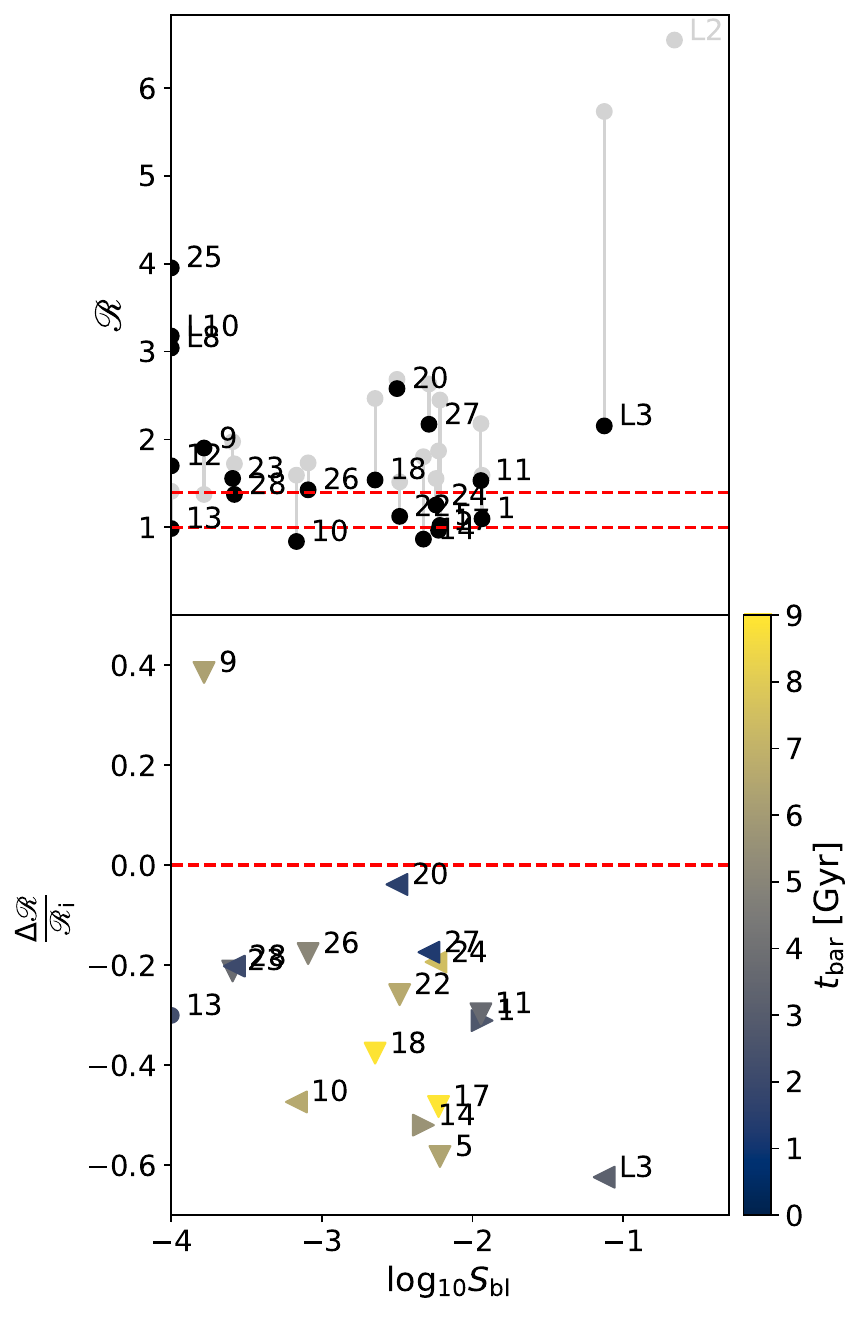}
    \caption{The relation between $\mathscr{R}$ and the maximum Elmegreen parameter ($S$) for any interaction following $0.5\,\mathrm{Gyr}$ after bar formation. The top panel shows $\mathscr{R}$ at $z=0$ (black points) and at $t_\mathrm{i}$ (grey points connected by the grey lines). The red dashed lines indicate $\mathscr{R}=1,1.4$. The bottom panel shows the fractional change in $\mathscr{R}$ between these two times. The colour scale shows the bar formation time for each halo, and the left/down/right pointing triangles indicate retrograde/polar/prograde oriented interactions, respectively. The red dashed line in the lower panel indicates no change in $\mathscr{R}$; bars below this line have sped up.}
    \label{fig:interactions}
\end{figure}

Fig.~\ref{fig:interactions} shows the highest Elmegreen parameter for any interaction since $0.5\,\mathrm{Gyr}$ after bar formation, covering the lifetime of the bar up to the end of the simulation. We denote this maximum as $\log_{10}S_\mathrm{bl}$, using `bl' to denote the bar lifetime rather than `form' which denoted the bar formation time. From the top panel we see little correlation between $\mathscr{R}$ and $\log_{10}S_\mathrm{bl}$ at either of these times. However, in the bottom panel we see that the bars that have experienced stronger galaxy interactions speed up more. This correlation is quite weak, but does notably account for the significant decreases in rotation rates of AuL3, which is unusual in Fig.~\ref{fig:baryonchange}, having not increased in baryon dominance, but here can be seen to have experienced the strongest interactions since bar formation of the entire sample with surviving bars at $z=0$. The pattern speed in AuL3 increases steadily following its interaction until the end of the simulation, dramatically decreasing its large corotation radius and thus decreasing $\mathscr{R}$.

Together with the results shown in Sec.~\ref{barresults} and Sec.~\ref{internalresults}, interactions with external galaxies seem to have only a secondary effect on the rotation rates of their bars, with weaker correlations than those seen with bar strength and with baryon dominance.

\section{Discussion}
\label{discuss}

We first discuss the physical processes that may lie behind the low, decreasing bar rotation rates in Auriga, given the correlations with baryon dominance and interaction strength presented above, and how these compare to the literature. We then discuss the extent to which the trends with bar rotation rate above are supported by observational studies, in particular those found with both bar strength and galaxy stellar mass.

\subsection{Decreasing rotation rates}
\label{ratediscuss}

The prevalent mode of evolution for the bars in the Auriga suite appears to be lengthening without significant slowing of pattern speed, resulting in bars which speed up in rotation rate through their lifetimes (see Fig.~\ref{fig:time}). While an increasing bar length is expected for bars evolving secularly, the expectation from theory and tailored simulations is that their pattern speeds, and therefore rotation rates, will slow down, primarily due to dynamical friction from the dark matter halo \citep{1985We,1991Li,2000De2,2003At}. In some cases, the increase in corotation radius from a slowing pattern speed is not sufficient to outpace the increase in bar length, resulting in the bar maintaining a fast rotation rate, i.e., low $\mathscr{R}$ \citep[e.g.][]{2003ON,2006MV}. Additionally, \citet{2006Se} find that the presence of a gravitational perturbation which increases the pattern speed briefly -- such as an interaction with an external galaxy --
can halt the slow down from dynamical friction for a few $\mathrm{Gyr}$. However, they also show that this state of `metastability' is sensitive to further perturbations, which are frequent in a cosmological setting. Therefore, the metastable state is unlikely to last long in reality, allowing slow down from dynamical friction to resume.

In this study, we do not seem to find dynamical friction acting as a persistent, gradual mechanism, which would slow down bars on long timescales. In contrast, we find that dynamical friction is likely acting on much shorter timescales, since Fig.~\ref{fig:abundance} and Fig.~\ref{fig:baryondom} show that bar rotation rates are mainly dependent on the ratio of the stellar to dark matter (with more baryon dominated galaxies hosting faster bars) at formation time, as well as at subsequent times. This trend is established by $t_\mathrm{i}$ (just $0.5\,\mathrm{Gyr}$ after the bar forms), which is a much shorter timescale than the $5-10\,\mathrm{Gyr}$ found in \citet{1991Li}, \citet{2000De2}, \citet{2003At}, \citet{2003ON} and \citet{2006Se} for dynamical friction, potentially on the same timescale as bar formation. Instead, the correlation of rotation rate and baryon dominance which we see in our sample may suggest the fast-acting dynamical friction originally described in \citet{1985We} which acts over a few bar rotation times (roughly $0.1-1\,\mathrm{Gyr}$ here) for bars that form fast.

In addition to acting over a short timescale, dynamical friction is relatively weak in the baryon-dominated systems we explore since bars remain fast and even decrease in rotation rate as they evolve. 
As mentioned in Sec.~\ref{intro}, some studies suggest that there could be mechanisms arising from channelling gas towards the centre of the galaxy that can speed the bar up \citep{2005Bo,2007Be,2008RD,2010VV,2014At,2023Be}. Even if this effect is not enough to counter dynamical friction on its own, \citet{2006Se} finds that gas inflows may be able to reliably produce the perturbation in pattern speed required to trigger a metastable state.

Interactions with other galaxies are an unlikely main candidate for speeding up bars in Auriga, since the associated trends in Sec.~\ref{externalresults} are relatively weak, and there is no evidence for tidal forces during an interaction reliably increasing the pattern speed of a bar, instead showing a mixture of speeding and slowing the bar as in e.g. \citet{1990Ge}, \citet{2014Lo} and \citet{2024Se}. Panel (a) in Fig.~\ref{fig:app_wide} does, however, indicate a correlation between stronger interactions around the time of bar formation and greater changes in baryon dominance between $t_\mathrm{i}$ and $z=0$. This correlation may be linked to inflows of gas which could increase the bar rotation rate, or higher rates of gas accretion onto the galaxy which could add angular momentum to the disc and the bar. Both mechanisms require a lasting effect on the radial motion of gas in order to cause a speed up of the bar over its lifetime, consistently maintaining the influence of the interaction for at least $1\,\mathrm{Gyr}$ after its pericentre.

Further clues may come from comparison with other cosmological simulations in which the speeding up of bars seen here are not present. As mentioned in Sec.~\ref{intro}, most cosmological simulations which form bars tend to produce bars which are too slow, particularly in rotation rate \citep{2017Al,2019Pe,2021Ro,2024Ha,2025An}, potentially as a result of excessive slow down from dynamical friction \citep[significant slow down is seen in][]{2017Al,2024Ha} or a lack of lengthening through their evolution \citep{2022Fr}. These studies examine bar rotation rate in 5 different simulations, which we compare with Auriga in turn.

In Illustris \citep{2014Vo}, \citet{2019Pe} find a mean bar rotation rate of $5.6\pm3.8$. In EAGLE \citep{2015Cr,2015Sc}, \citet{2017Al} find rotation rates of $1.4-10$ for the strong bars in EAGLE Ref-L100N1504 (the $100\,\mathrm{cMpc}$ box run). Additionally, \citet{2021Ro} find mean rotation rates of $2.4^{+1.1}_{-0.8}$ in EAGLE Ref-L100N1504 and $1.9^{+1.7}_{-0.9}$ in EAGLE Ref-L050N0752 (the $50\,\mathrm{cMpc}$ box run). For these two simulations, \citet{2021Fr} find that the higher values of baryon dominance found in Auriga galaxies are the likely reason for its faster bars. Increased baryon dominance in the inner regions of galaxies can significantly reduce the effects of dynamical friction \citep{1998De} and allow the bar to lengthen without significantly decreasing pattern speed.

In the IllustrisTNG simulations, \citet{2021Ro} find mean rotation rates of $2.8^{+1.6}_{-1.0}$ in TNG100 \citep{2018Ma,2018Na,2018Ne,2018Pi,2018Sp} and $3.0^{+1.8}_{-1.1}$ in TNG50 \citep{2019Ne,2019Pi}. Also, \citet{2024Ha} find most bars have rotation rates of $1.4-5$ in TNG50. Using the stellar mass fractions for discs in TNG100 \citep[from][]{2020Ma} and for barred galaxies in TNG50 \citep[from][]{2022RG} we can see that both simulation runs also have lower baryon dominance than in Auriga, with TNG100 being comparable with the original Illustris and TNG50 being slightly more baryon dominated but still less so than Auriga. Therefore, following the conclusions of \citet{2021Fr}, this likely accounts for the most of the difference in bar rotation rates.

Since TNG50 shows less of a difference in baryon dominance with Auriga than the other simulations mentioned, it is worthwhile to consider other factors which may contribute to the wider difference in rotation rates. Firstly, the lower resolution of a full-box cosmological simulation may artificially slow down the bar due to stochasticity \citep{2023Ha}, although \citet{2021Fr} do investigate the effects of resolution on bar pattern speeds within Auriga and find that the effects of resolution alone are not enough to explain the differences between Auriga and EAGLE or Illustris.


Additionally, \citet{2023Li} find from tailored N-body simulations that a high value of dark matter halo spin can halt slowdown from dynamical friction. Milky Way-mass halos in TNG50 have spin parameters in the range $0.02-0.07$, skewing towards lower values for barred galaxies \citep{2022RG}, whereas all original Auriga halos have spin parameters above $0.08$ \citep{2017Gr} -- this is in the regime where \citet{2023Li} find spin to halt bar slowdown for up to $5\,\mathrm{Gyr}$. It is also notable that \citet{2024Ha} find that 2 of the 3 bars in TNG50 which have sped up significantly are also in the minority that experience major mergers after $z=1$, pointing towards galaxy interactions playing a key role in speeding up bars. We see a similar, albeit very weak, trend in Sec.~\ref{externalresults}.

Lastly, in the FIRE-2 simulations \citep{2018Ho,2023We}, \citet{2025An} find bar rotation rates of $1.6-6.5$. Examining the stellar masses and halos masses from \citet{2018Ho}, we see that FIRE-2 also has lower baryon dominance than Auriga, providing an explanation for its slower bars. We also note that the bars in FIRE-2 are very weak, with none stronger than $A_{2\mathrm{,max}}=0.21$ (only this strongest bar would be considered a barred galaxy in this work). \citet{2025An} find that this is likely a result of an initially high fraction of central gas triggering extreme stellar feedback processes. Since we find that rotation rate is negatively correlated with bar strength, we would expect such weak bars to also be slow in Auriga. Indeed, the strongest bar in FIRE-2 also has the lowest rotation rate.

Across these 5 other cosmological simulations, we see that the disc galaxy population in Auriga has a higher baryon dominance than each, in addition to significantly lower average bar rotation rates. This indicates that having a high baryon dominance is the most important factor in forming fast bars within cosmological simulations. Other ingredients, potentially including properties of merger histories, feedback physics and dark matter halo spin, may contribute to increasing the fraction of fast bars formed.

\subsection{Comparison with observations}
\label{obscomparison}

While there are significant challenges of observationally measuring bar pattern speeds \citep[e.g.][]{2003De,2019Zo,2020GO}, here we discuss some of the studies that have explored bar pattern speeds and compared these to different parameters. In particular, we compare our results of trends with bar strength and with stellar mass, since these are measured more easily in observations than the other quantities used in this paper.

We first address the correlation between bar strength and rotation rate. \citet{2008Ra} used models of the rotating bar potential, fitted based on the response features produced by numerical simulations of the disc, to find bar pattern speeds for 38 $z\sim0$ bright spiral galaxies from the Ohio State University Bright Spiral Galaxy Survey \citep{2002Es}. Although using a different measure of bar strength than that used in this paper -- the ratio of tangential and radial forces exerted by the bar potential in their best fit model -- they find that all but one of the slow ($\mathscr{R}>1.4$) bars in their sample have a bar strength less than $0.3$. This matches the tendency for slow bars to be weaker in Fig.~\ref{fig:strength}. It is, however, worth noting that their sample only contains 6 other galaxies stronger than this limit, meaning the majority of their fast bars are also weaker than $0.3$. \citet{2023Ge}, use the Tremaine-Weinberg method \citep{1984Tr} to find the pattern speeds of 255 bars from the MaNGA survey \citep{2015Bu2}; they also find that strong bars have lower rotation rates in agreement with Auriga. The bars in this work are classified by Galaxy Zoo \citep{2008Li,2022Wa} into strong and weak categories which have rotation rates of $1.53^{+0.74}_{-0.53}$ and $1.88^{+1.08}_{-0.75}$ respectively.

Other works find no such relation between rotation rate and bar strength in observations. \citet{2019Gu} also use the Tremaine-Weinberg method to find pattern speeds in a smaller sample (53) from the MaNGA survey and find no correlation with bar strength. \citet{2020Cu} adds 18 bars from the literature and 31 bars from the CALIFA survey \citep{2012Sa,2014Wa} to this sample, making 77 bars once additional quality cuts are imposed, and finds the same lack of correlation. It is unsurprising that these studies agree with each other, given the large overlap in galaxy sample. Another similarity which may affect their results, is that they find corotation radii by assuming that the rotation curve of the disc very quickly flattens to a constant circular velocity. \citet{2008Ra} and \citet{2023Ge}, the two studies which do see a correlation between bar strength and rotation rate, instead model the rotation curve for a (potentially) more accurate corotation radius. In summary, although not all observations support a correlation between bar strength and rotation rate, the relatively large sample and accurate method for measuring the corotation radius of \citet{2023Ge} is compelling evidence for a correlation.

The correlation between stellar mass and bar rotation rate is of particular interest since, as mentioned in Sec.~\ref{globalresults}, it provides a test of $\Lambda$CDM cosmology. The correlation could be caused by the underlying trend of higher stellar mass galaxies having higher baryon dominance. Due to the lack of dark matter, in a Modified Newtonian Dynamics framework \citep{1983Mi}, we would not expect to see a dependence of rotation rate on galaxy mass in such scenarios (since there is no trend or change in `baryon-dominance'). Indeed, the study of  \citet{2023Na} explored bars in a MOND framework, showing that galaxies with masses $10^9-10^{11}\,\mathrm{M}_\odot$ all have similar rotation rates of $\mathscr{R}\sim1.6$.

In observations, \citet{2023Ge} show a weak negative correlation between bar rotation rate and stellar mass in their Fig.~13. This figure shows that weak bars tend to have higher rotation rates in lower mass galaxies for $M_\star \lesssim 3\times10^{10}\,\mathrm{M}_\odot$. Above this mass, bars have rotation rates which are lower, but approximately constant with stellar mass. We see a similar shape in the rotation rate-stellar mass relation for Auriga in Fig.~\ref{fig:abundance}, but with a slightly higher transitional mass of $6\times10^{10}\,\mathrm{M}_\odot$: 5 of the 6 galaxies with stellar masses below $6\times10^{10}\,\mathrm{M}_\odot$ in our sample have slow bars, all of which have weaker bars than the median.

Two other works examine the relation between bar rotation rate and stellar mass: \citet{2020GO} which investigates 18 galaxies from the MaNGA and CALIFA surveys, and \citet{2022GO} which applies the Tremaine-Weinberg method to 97 galaxies from the MaNGA survey with masses and morphology similar to the Milky Way. Both studies find no correlation between rotation rate and stellar mass. We note, however, that both studies have a smaller sample size than \citet{2023Ge}, especially \citet{2020GO} which only finds a convergent pattern speed for 13 galaxies due to its focus on galaxies which present challenges to the Tremaine-Weinberg method. \citet{2022GO} also has a lower stellar mass limit of $2\times10^{10}\,\mathrm{M}_\odot$ -- this would likely hide a trend like in \citet{2023Ge} which only appears below $3\times10^{10}\,\mathrm{M}_\odot$, especially since the relation is presented as a correlation coefficient only.

\citet{2019Gu} additionally investigates the dark matter fraction within an effective radius, the inverse of baryon dominance. In their Fig.~11, there is a weak negative correlation between rotation rate and dark matter fraction (more slow bars have higher dark matter fractions, corresponding to lower baryon dominance) in the discs with the most reliable position angles. This is in agreement with our results in Fig.~\ref{fig:baryondom}, although the observations show a much weaker correlation.

The relationships between bar rotation rate and bar strength, stellar mass, and baryon dominance found in some observations are consistent with our findings. While not all observational studies agree with our work, those that do have either have larger sample sizes or fewer assumptions in calculating their rotation rates, which lends tentative support to the bar evolution in the Auriga simulations, the baryon physics involved, and the $\Lambda$CDM cosmology producing it. Future observations may clarify the picture currently presented, allowing the trends we have presented here to give a more robust test of bar evolution in Auriga. In particular, studies of bars in lower mass disc galaxies may provide a crucial test for the relations that we find.

\section{Conclusions}
\label{conc}

In this paper, we investigated the internal and external physical processes that set the bar pattern speed and drive its evolution in simulated Milky Way-mass galaxies. We used 27 high temporal resolution reruns from the Auriga suite of magneto-hydrodynamic cosmological zoom-in simulations to compare bar rotation rates, $\mathscr{R}$ (the ratio of corotation radius to bar length), with bar formation times and strengths, as well as with the baryon/dark matter composition and interaction histories of their host galaxies.

We summarise our conclusions as follows:

\begin{itemize}
\item All but one of the bars in the Auriga suite of simulations decrease in rotation rate from their formation to the present day, largely due to an increase in bar length and approximately constant corotation radius (e.g. Fig.~\ref{fig:lengths} and Fig.~\ref{fig:time}). This speed-up in $\mathscr{R}$ is unexpected when compared to the literature, and implies that any dynamical friction acts on short timescales, having little effect following $0.5\,\mathrm{Gyr}$ after the bar forms.
\item Younger (older) bars tend to have a slower (faster) rotation rate (ratio of corotation radius to bar length) by the end of the simulation (Fig.~\ref{fig:time}). Younger bars are also slower than older bars shortly after the bar formation time, suggesting that the mechanism responsible for the difference acts mostly within the first $0.5\,\mathrm{Gyr}$ of bar evolution.
\item Bars that are stronger at $z=0$ tend to be faster (Fig.~\ref{fig:strength}). This is a relation that can be tested in the nearby Universe and is confirmed in some, but not all, observational studies.
\item Galaxies with higher stellar masses tend to host faster bars (Fig.~\ref{fig:abundance}). This relation acts as a test of the $\Lambda$CDM model of cosmology, since it seems to be driven by the higher fractions of stellar to dark matter in high mass galaxies. This relation has so far been observed in weak bars at stellar masses below $3\times10^{10}\,\mathrm{M}_\odot$.
\item The ratio of stellar to dark matter (or baryon dominance) within galaxies appears to be the strongest factor in setting the rotation rate at both the start and end of the lifetime of their bars up to $z=0$. Regardless of epoch, bars in more baryon dominated environments are faster (Fig.~\ref{fig:abundance} and Fig.~\ref{fig:baryondom}); additionally, bars tend to speed up when their host galaxies increase in baryon dominance (Fig.~\ref{fig:baryonchange}).
\item Galaxy interactions at the time of bar formation have little immediate effect on pattern speed, however bars formed alongside stronger interactions tend to speed up more through their life and become slightly faster by $z=0$ (Fig.~\ref{fig:formation}). The mechanisms behind this are unclear, but may relate to the trends with baryon dominance -- possibly due to interactions driving in-situ baryons towards the galactic centre or by adding gas to the system directly. Regardless of mechanism, this contradicts the expectation from literature that tidally formed bars are slower on average.
\item There is a weak trend between stronger galaxy interactions after the bar has formed and a greater decrease in bar rotation rate (Fig.~\ref{fig:interactions}). This effect is, however, minimal when compared to the other correlations presented here.
\end{itemize}

The Auriga simulations give an insight into how fast bars in Milky Way-mass halos can arise in cosmological simulations. A high baryon dominance in the inner regions appears to be crucial, as do relatively early-forming bars. Comparison with bars in other cosmological simulations supports the need for high baryon dominance in order to produce fast bars in Milky~Way-mass discs. Having this baryon dominance likely counters dynamical friction in Auriga to stop these bars slowing down excessively by present day and even allowing the majority to speed up. Our study highlights the potential constraining power of bar rotation rates on the nature of dark matter and baryonic physics. 

\section*{Acknowledgements}

AM is supported by a joint STFC and LJMU FET doctoral studentship. FF is supported by a UKRI Future Leaders Fellowship (grant no. MR/X033740/1). RG is supported by an STFC Ernest Rutherford Fellowship (ST/W003643/1). The simulations presented in this paper were run on the Freya computer cluster at the Max Planck Institute for Astrophysics and the DiRAC@Durham facility managed by the Institute for Computational Cosmology on behalf of the STFC DiRAC HPC Facility (www.dirac.ac.uk). The analysis in this work used the latter. The equipment was funded by BEIS capital funding via STFC capital grants ST/K00042X/1, ST/P002293/1 and ST/R002371/1, Durham University and STFC operations grant ST/R000832/1. DiRAC is part of the National e-Infrastructure.

\section*{Data Availability}

The data presented in this article will be shared on reasonable request to the corresponding author.



\bibliographystyle{mnras}
\bibliography{example.bib} 

@ARTICLE{2013Mo,
       author = {{Moster}, Benjamin P. and {Naab}, Thorsten and {White}, Simon D.~M.},
        title = "{Galactic star formation and accretion histories from matching galaxies to dark matter haloes}",
      journal = {\mnras},
         year = 2013,
        month = feb,
       volume = {428},
       number = {4},
        pages = {3121-3138},
          doi = {10.1093/mnras/sts261},
archivePrefix = {arXiv},
       eprint = {1205.5807},
 primaryClass = {astro-ph.CO},
       adsurl = {https://ui.adsabs.harvard.edu/abs/2013MNRAS.428.3121M}
}

@ARTICLE{1983Mi,
       author = {{Milgrom}, M.},
        title = "{A modification of the Newtonian dynamics as a possible alternative to the hidden mass hypothesis.}",
      journal = {\apj},
         year = 1983,
        month = jul,
       volume = {270},
        pages = {365-370},
          doi = {10.1086/161130},
       adsurl = {https://ui.adsabs.harvard.edu/abs/1983ApJ...270..365M}
}

@ARTICLE{1991El,
       author = {{Elmegreen}, D.~M. and {Sundin}, M. and {Elmegreen}, B. and {Sundelius}, B.},
        title = "{Properties and simulations of interacting spiral galaxies with transient ``ocular'' shapes.}",
      journal = {\aap},
         year = 1991,
        month = apr,
       volume = {244},
        pages = {52},
       adsurl = {https://ui.adsabs.harvard.edu/abs/1991A&A...244...52E}
}

@ARTICLE{2021Sp,
       author = {{Springel}, Volker and {Pakmor}, R{\"u}diger and {Zier}, Oliver and {Reinecke}, Martin},
        title = "{Simulating cosmic structure formation with the GADGET-4 code}",
      journal = {\mnras},
         year = 2021,
        month = sep,
       volume = {506},
       number = {2},
        pages = {2871-2949},
          doi = {10.1093/mnras/stab1855},
archivePrefix = {arXiv},
       eprint = {2010.03567},
 primaryClass = {astro-ph.IM},
       adsurl = {https://ui.adsabs.harvard.edu/abs/2021MNRAS.506.2871S}
}

@ARTICLE{2017Gr,
       author = {{Grand}, Robert J.~J. and {G{\'o}mez}, Facundo A. and {Marinacci}, Federico and {Pakmor}, R{\"u}diger and {Springel}, Volker and {Campbell}, David J.~R. and {Frenk}, Carlos S. and {Jenkins}, Adrian and {White}, Simon D.~M.},
        title = "{The Auriga Project: the properties and formation mechanisms of disc galaxies across cosmic time}",
      journal = {\mnras},
         year = 2017,
        month = may,
       volume = {467},
       number = {1},
        pages = {179-207},
          doi = {10.1093/mnras/stx071},
archivePrefix = {arXiv},
       eprint = {1610.01159},
 primaryClass = {astro-ph.GA},
       adsurl = {https://ui.adsabs.harvard.edu/abs/2017MNRAS.467..179G}
}

@ARTICLE{2019Gr,
       author = {{Grand}, Robert J.~J. and {van de Voort}, Freeke and {Zjupa}, Jolanta and {Fragkoudi}, Francesca and {G{\'o}mez}, Facundo A. and {Kauffmann}, Guinevere and {Marinacci}, Federico and {Pakmor}, R{\"u}diger and {Springel}, Volker and {White}, Simon D.~M.},
        title = "{Gas accretion and galactic fountain flows in the Auriga cosmological simulations: angular momentum and metal redistribution}",
      journal = {\mnras},
         year = 2019,
        month = dec,
       volume = {490},
       number = {4},
        pages = {4786-4803},
          doi = {10.1093/mnras/stz2928},
archivePrefix = {arXiv},
       eprint = {1909.04038},
 primaryClass = {astro-ph.GA},
       adsurl = {https://ui.adsabs.harvard.edu/abs/2019MNRAS.490.4786G}
}

@ARTICLE{2015Sc,
       author = {{Schaye}, Joop and {Crain}, Robert A. and {Bower}, Richard G. and {Furlong}, Michelle and {Schaller}, Matthieu and {Theuns}, Tom and {Dalla Vecchia}, Claudio and {Frenk}, Carlos S. and {McCarthy}, I.~G. and {Helly}, John C. and {Jenkins}, Adrian and {Rosas-Guevara}, Y.~M. and {White}, Simon D.~M. and {Baes}, Maarten and {Booth}, C.~M. and {Camps}, Peter and {Navarro}, Julio F. and {Qu}, Yan and {Rahmati}, Alireza and {Sawala}, Till and {Thomas}, Peter A. and {Trayford}, James},
        title = "{The EAGLE project: simulating the evolution and assembly of galaxies and their environments}",
      journal = {\mnras},
         year = 2015,
        month = jan,
       volume = {446},
       number = {1},
        pages = {521-554},
          doi = {10.1093/mnras/stu2058},
archivePrefix = {arXiv},
       eprint = {1407.7040},
 primaryClass = {astro-ph.GA},
       adsurl = {https://ui.adsabs.harvard.edu/abs/2015MNRAS.446..521S}
}

@ARTICLE{2024Gr,
       author = {{Grand}, Robert J.~J. and {Fragkoudi}, Francesca and {G{\'o}mez}, Facundo A. and {Jenkins}, Adrian and {Marinacci}, Federico and {Pakmor}, R{\"u}diger and {Springel}, Volker},
        title = "{Overview and public data release of the augmented Auriga Project: cosmological simulations of dwarf and Milky Way-mass galaxies}",
      journal = {\mnras},
         year = 2024,
        month = aug,
       volume = {532},
       number = {2},
        pages = {1814-1831},
          doi = {10.1093/mnras/stae1598},
archivePrefix = {arXiv},
       eprint = {2401.08750},
 primaryClass = {astro-ph.GA},
       adsurl = {https://ui.adsabs.harvard.edu/abs/2024MNRAS.532.1814G}
}

@ARTICLE{2010Sp,
       author = {{Springel}, Volker},
        title = "{E pur si muove: Galilean-invariant cosmological hydrodynamical simulations on a moving mesh}",
      journal = {\mnras},
         year = 2010,
        month = jan,
       volume = {401},
       number = {2},
        pages = {791-851},
          doi = {10.1111/j.1365-2966.2009.15715.x},
archivePrefix = {arXiv},
       eprint = {0901.4107},
 primaryClass = {astro-ph.CO},
       adsurl = {https://ui.adsabs.harvard.edu/abs/2010MNRAS.401..791S}
}

@ARTICLE{2016Pa,
       author = {{Pakmor}, R{\"u}diger and {Springel}, Volker and {Bauer}, Andreas and {Mocz}, Philip and {Munoz}, Diego J. and {Ohlmann}, Sebastian T. and {Schaal}, Kevin and {Zhu}, Chenchong},
        title = "{Improving the convergence properties of the moving-mesh code AREPO}",
      journal = {\mnras},
         year = 2016,
        month = jan,
       volume = {455},
       number = {1},
        pages = {1134-1143},
          doi = {10.1093/mnras/stv2380},
archivePrefix = {arXiv},
       eprint = {1503.00562},
 primaryClass = {astro-ph.GA},
       adsurl = {https://ui.adsabs.harvard.edu/abs/2016MNRAS.455.1134P}
}

@ARTICLE{2014Pl,
       author = {{Planck Collaboration} and {Ade}, P.~A.~R. and {Aghanim}, N. and {Armitage-Caplan}, C. and {Arnaud}, M. and {Ashdown}, M. and {Atrio-Barandela}, F. and {Aumont}, J. and {Baccigalupi}, C. and {Banday}, A.~J. and {Barreiro}, R.~B. and {Bartlett}, J.~G. and {Battaner}, E. and {Benabed}, K. and {Beno{\^\i}t}, A. and {Benoit-L{\'e}vy}, A. and {Bernard}, J. -P. and {Bersanelli}, M. and {Bielewicz}, P. and {Bobin}, J. and {Bock}, J.~J. and {Bonaldi}, A. and {Bond}, J.~R. and {Borrill}, J. and {Bouchet}, F.~R. and {Bridges}, M. and {Bucher}, M. and {Burigana}, C. and {Butler}, R.~C. and {Calabrese}, E. and {Cappellini}, B. and {Cardoso}, J. -F. and {Catalano}, A. and {Challinor}, A. and {Chamballu}, A. and {Chary}, R. -R. and {Chen}, X. and {Chiang}, H.~C. and {Chiang}, L. -Y. and {Christensen}, P.~R. and {Church}, S. and {Clements}, D.~L. and {Colombi}, S. and {Colombo}, L.~P.~L. and {Couchot}, F. and {Coulais}, A. and {Crill}, B.~P. and {Curto}, A. and {Cuttaia}, F. and {Danese}, L. and {Davies}, R.~D. and {Davis}, R.~J. and {de Bernardis}, P. and {de Rosa}, A. and {de Zotti}, G. and {Delabrouille}, J. and {Delouis}, J. -M. and {D{\'e}sert}, F. -X. and {Dickinson}, C. and {Diego}, J.~M. and {Dolag}, K. and {Dole}, H. and {Donzelli}, S. and {Dor{\'e}}, O. and {Douspis}, M. and {Dunkley}, J. and {Dupac}, X. and {Efstathiou}, G. and {Elsner}, F. and {En{\ss}lin}, T.~A. and {Eriksen}, H.~K. and {Finelli}, F. and {Forni}, O. and {Frailis}, M. and {Fraisse}, A.~A. and {Franceschi}, E. and {Gaier}, T.~C. and {Galeotta}, S. and {Galli}, S. and {Ganga}, K. and {Giard}, M. and {Giardino}, G. and {Giraud-H{\'e}raud}, Y. and {Gjerl{\o}w}, E. and {Gonz{\'a}lez-Nuevo}, J. and {G{\'o}rski}, K.~M. and {Gratton}, S. and {Gregorio}, A. and {Gruppuso}, A. and {Gudmundsson}, J.~E. and {Haissinski}, J. and {Hamann}, J. and {Hansen}, F.~K. and {Hanson}, D. and {Harrison}, D. and {Henrot-Versill{\'e}}, S. and {Hern{\'a}ndez-Monteagudo}, C. and {Herranz}, D. and {Hildebrandt}, S.~R. and {Hivon}, E. and {Hobson}, M. and {Holmes}, W.~A. and {Hornstrup}, A. and {Hou}, Z. and {Hovest}, W. and {Huffenberger}, K.~M. and {Jaffe}, A.~H. and {Jaffe}, T.~R. and {Jewell}, J. and {Jones}, W.~C. and {Juvela}, M. and {Keih{\"a}nen}, E. and {Keskitalo}, R. and {Kisner}, T.~S. and {Kneissl}, R. and {Knoche}, J. and {Knox}, L. and {Kunz}, M. and {Kurki-Suonio}, H. and {Lagache}, G. and {L{\"a}hteenm{\"a}ki}, A. and {Lamarre}, J. -M. and {Lasenby}, A. and {Lattanzi}, M. and {Laureijs}, R.~J. and {Lawrence}, C.~R. and {Leach}, S. and {Leahy}, J.~P. and {Leonardi}, R. and {Le{\'o}n-Tavares}, J. and {Lesgourgues}, J. and {Lewis}, A. and {Liguori}, M. and {Lilje}, P.~B. and {Linden-V{\o}rnle}, M. and {L{\'o}pez-Caniego}, M. and {Lubin}, P.~M. and {Mac{\'\i}as-P{\'e}rez}, J.~F. and {Maffei}, B. and {Maino}, D. and {Mandolesi}, N. and {Maris}, M. and {Marshall}, D.~J. and {Martin}, P.~G. and {Mart{\'\i}nez-Gonz{\'a}lez}, E. and {Masi}, S. and {Massardi}, M. and {Matarrese}, S. and {Matthai}, F. and {Mazzotta}, P. and {Meinhold}, P.~R. and {Melchiorri}, A. and {Melin}, J. -B. and {Mendes}, L. and {Menegoni}, E. and {Mennella}, A. and {Migliaccio}, M. and {Millea}, M. and {Mitra}, S. and {Miville-Desch{\^e}nes}, M. -A. and {Moneti}, A. and {Montier}, L. and {Morgante}, G. and {Mortlock}, D. and {Moss}, A. and {Munshi}, D. and {Murphy}, J.~A. and {Naselsky}, P. and {Nati}, F. and {Natoli}, P. and {Netterfield}, C.~B. and {N{\o}rgaard-Nielsen}, H.~U. and {Noviello}, F. and {Novikov}, D. and {Novikov}, I. and {O'Dwyer}, I.~J. and {Osborne}, S. and {Oxborrow}, C.~A. and {Paci}, F. and {Pagano}, L. and {Pajot}, F. and {Paladini}, R. and {Paoletti}, D. and {Partridge}, B. and {Pasian}, F. and {Patanchon}, G. and {Pearson}, D. and {Pearson}, T.~J. and {Peiris}, H.~V. and {Perdereau}, O. and {Perotto}, L. and {Perrotta}, F. and {Pettorino}, V. and {Piacentini}, F. and {Piat}, M. and {Pierpaoli}, E. and {Pietrobon}, D. and {Plaszczynski}, S. and {Platania}, P. and {Pointecouteau}, E.},
        title = "{Planck 2013 results. XVI. Cosmological parameters}",
      journal = {\aap},
         year = 2014,
        month = nov,
       volume = {571},
          eid = {A16},
        pages = {A16},
          doi = {10.1051/0004-6361/201321591},
archivePrefix = {arXiv},
       eprint = {1303.5076},
 primaryClass = {astro-ph.CO},
       adsurl = {https://ui.adsabs.harvard.edu/abs/2014A&A...571A..16P}
}

@ARTICLE{1984Tr,
       author = {{Tremaine}, S. and {Weinberg}, M.~D.},
        title = "{A kinematic method for measuring the pattern speed of barred galaxies.}",
      journal = {\apjl},
         year = 1984,
        month = jul,
       volume = {282},
        pages = {L5-L7},
          doi = {10.1086/184292},
       adsurl = {https://ui.adsabs.harvard.edu/abs/1984ApJ...282L...5T}
}

@ARTICLE{2013Vo,
   author = {{Vogelsberger}, M. and {Genel}, S. and {Sijacki}, D. and {Torrey}, P. and 
	{Springel}, V. and {Hernquist}, L.},
    title = "{A model for cosmological simulations of galaxy formation physics}",
  journal = {\mnras},
archivePrefix = "arXiv",
   eprint = {1305.2913},
     year = 2013,
    month = dec,
   volume = 436,
    pages = {3031-3067},
      doi = {10.1093/mnras/stt1789},
   adsurl = {http://adsabs.harvard.edu/abs/2013MNRAS.436.3031V}
}

@ARTICLE{2014Ma,
   author = {{Marinacci}, F. and {Pakmor}, R. and {Springel}, V.},
    title = "{The formation of disc galaxies in high-resolution moving-mesh cosmological simulations}",
  journal = {\mnras},
archivePrefix = "arXiv",
   eprint = {1305.5360},
     year = 2014,
    month = jan,
   volume = 437,
    pages = {1750-1775},
      doi = {10.1093/mnras/stt2003},
   adsurl = {http://adsabs.harvard.edu/abs/2014MNRAS.437.1750M}
}

@ARTICLE{2009FG,
   author = {{Faucher-Gigu{\`e}re}, C.-A. and {Lidz}, A. and {Zaldarriaga}, M. and 
	{Hernquist}, L.},
    title = "{A New Calculation of the Ionizing Background Spectrum and the Effects of He II Reionization}",
  journal = {\apj},
archivePrefix = "arXiv",
   eprint = {0901.4554},
 primaryClass = "astro-ph.CO",
     year = 2009,
    month = oct,
   volume = 703,
    pages = {1416-1443},
      doi = {10.1088/0004-637X/703/2/1416},
   adsurl = {http://adsabs.harvard.edu/abs/2009ApJ...703.1416F}
}

@ARTICLE{2003Sp,
   author = {{Springel}, V. and {Hernquist}, L.},
    title = "{Cosmological smoothed particle hydrodynamics simulations: a hybrid multiphase model for star formation}",
  journal = {\mnras},
   eprint = {astro-ph/0206393},
     year = 2003,
    month = feb,
   volume = 339,
    pages = {289-311},
      doi = {10.1046/j.1365-8711.2003.06206.x},
   adsurl = {http://adsabs.harvard.edu/abs/2003MNRAS.339..289S}
}

@ARTICLE{2014Pa,
   author = {{Pakmor}, R. and {Marinacci}, F. and {Springel}, V.},
    title = "{Magnetic Fields in Cosmological Simulations of Disk Galaxies}",
  journal = {\apjl},
archivePrefix = "arXiv",
   eprint = {1312.2620},
     year = 2014,
    month = mar,
   volume = 783,
      eid = {L20},
    pages = {L20},
      doi = {10.1088/2041-8205/783/1/L20},
   adsurl = {http://adsabs.harvard.edu/abs/2014ApJ...783L..20P}
}

@ARTICLE{2017Pa,
   author = {{Pakmor}, R. and {G{\'o}mez}, F.~A. and {Grand}, R.~J.~J. and 
	{Marinacci}, F. and {Simpson}, C.~M. and {Springel}, V. and 
	{Campbell}, D.~J.~R. and {Frenk}, C.~S. and {Guillet}, T. and 
	{Pfrommer}, C. and {White}, S.~D.~M.},
    title = "{Magnetic field formation in the Milky Way like disc galaxies of the Auriga project}",
  journal = {\mnras},
     year = 2017,
    month = aug,
   volume = 469,
    pages = {3185-3199},
      doi = {10.1093/mnras/stx1074},
   adsurl = {http://adsabs.harvard.edu/abs/2017MNRAS.469.3185P}
}

@ARTICLE{2021Fr,
       author = {{Fragkoudi}, F. and {Grand}, R.~J.~J. and {Pakmor}, R. and {Springel}, V. and {White}, S.~D.~M. and {Marinacci}, F. and {Gomez}, F.~A. and {Navarro}, J.~F.},
        title = "{Revisiting the tension between fast bars and the {\ensuremath{\Lambda}}CDM paradigm}",
      journal = {\aap},
         year = 2021,
        month = jun,
       volume = {650},
          eid = {L16},
        pages = {L16},
          doi = {10.1051/0004-6361/202140320},
archivePrefix = {arXiv},
       eprint = {2011.13942},
 primaryClass = {astro-ph.GA},
       adsurl = {https://ui.adsabs.harvard.edu/abs/2021A&A...650L..16F}
}

@ARTICLE{2025Fr,
       author = {{Fragkoudi}, Francesca and {Grand}, Robert J.~J. and {Pakmor}, R{\"u}diger and {G{\'o}mez}, Facundo and {Marinacci}, Federico and {Springel}, Volker},
        title = "{Bar formation and evolution in the cosmological context: inputs from the Auriga simulations}",
      journal = {\mnras},
         year = 2025,
        month = apr,
       volume = {538},
       number = {3},
        pages = {1587-1608},
          doi = {10.1093/mnras/staf389},
archivePrefix = {arXiv},
       eprint = {2406.09453},
 primaryClass = {astro-ph.GA},
       adsurl = {https://ui.adsabs.harvard.edu/abs/2025MNRAS.538.1587F}
}

@ARTICLE{2014La,
       author = {{Lang}, Meagan and {Holley-Bockelmann}, Kelly and {Sinha}, Manodeep},
        title = "{Bar Formation from Galaxy Flybys}",
      journal = {\apjl},
         year = 2014,
        month = aug,
       volume = {790},
       number = {2},
          eid = {L33},
        pages = {L33},
          doi = {10.1088/2041-8205/790/2/L33},
archivePrefix = {arXiv},
       eprint = {1405.5832},
 primaryClass = {astro-ph.GA},
       adsurl = {https://ui.adsabs.harvard.edu/abs/2014ApJ...790L..33L}
}

@ARTICLE{2018Lo,
       author = {{{\L}okas}, Ewa L.},
        title = "{Formation of Tidally Induced Bars in Galactic Flybys: Prograde versus Retrograde Encounters}",
      journal = {\apj},
         year = 2018,
        month = apr,
       volume = {857},
       number = {1},
          eid = {6},
        pages = {6},
          doi = {10.3847/1538-4357/aab4ff},
archivePrefix = {arXiv},
       eprint = {1803.09465},
 primaryClass = {astro-ph.GA},
       adsurl = {https://ui.adsabs.harvard.edu/abs/2018ApJ...857....6L}
}

@ARTICLE{1972To,
       author = {{Toomre}, Alar and {Toomre}, Juri},
        title = "{Galactic Bridges and Tails}",
      journal = {\apj},
         year = 1972,
        month = dec,
       volume = {178},
        pages = {623-666},
          doi = {10.1086/151823},
       adsurl = {https://ui.adsabs.harvard.edu/abs/1972ApJ...178..623T}
}

@ARTICLE{2010DO,
       author = {{D'Onghia}, Elena and {Vogelsberger}, Mark and {Faucher-Giguere}, Claude-Andre and {Hernquist}, Lars},
        title = "{Quasi-resonant Theory of Tidal Interactions}",
      journal = {\apj},
         year = 2010,
        month = dec,
       volume = {725},
       number = {1},
        pages = {353-368},
          doi = {10.1088/0004-637X/725/1/353},
archivePrefix = {arXiv},
       eprint = {1009.3927},
 primaryClass = {astro-ph.CO},
       adsurl = {https://ui.adsabs.harvard.edu/abs/2010ApJ...725..353D}
}

@ARTICLE{2025Am,
       author = {{Amvrosiadis}, A. and {Lange}, S. and {Nightingale}, J.~W. and {He}, Q. and {Frenk}, C.~S. and {Oman}, K.~A. and {Smail}, I. and {Swinbank}, A.~M. and {Fragkoudi}, F. and {Gadotti}, D.~A. and {Cole}, S. and {Borsato}, E. and {Robertson}, A. and {Massey}, R. and {Cao}, X. and {Li}, R.},
        title = "{The onset of bar formation in a massive galaxy at z \raisebox{-0.5ex}\textasciitilde 3.8}",
      journal = {\mnras},
         year = 2025,
        month = feb,
       volume = {537},
       number = {2},
        pages = {1163-1181},
          doi = {10.1093/mnras/staf048},
archivePrefix = {arXiv},
       eprint = {2404.01918},
 primaryClass = {astro-ph.GA},
       adsurl = {https://ui.adsabs.harvard.edu/abs/2025MNRAS.537.1163A}
}

@ARTICLE{2025Gu,
       author = {{Guo}, Yuchen and {Jogee}, Shardha and {Wise}, Eden and {Pritchett}, Keith and {McGrath}, Elizabeth J. and {Finkelstein}, Steven L. and {Iyer}, Kartheik G. and {Arrabal Haro}, Pablo and {Bagley}, Micaela B. and {Dickinson}, Mark and {Kartaltepe}, Jeyhan S. and {Koekemoer}, Anton M. and {Papovich}, Casey and {Pirzkal}, Nor and {Yung}, L.~Y. Aaron and {Backhaus}, Bren E. and {Bell}, Eric F. and {Bhatawdekar}, Rachana and {Cheng}, Yingjie and {Costantin}, Luca and {de la Vega}, Alexander and {Giavalisco}, Mauro and {Hathi}, Nimish P. and {Holwerda}, Benne W. and {Kurczynski}, Peter and {Lucas}, Ray A. and {Mobasher}, Bahram and {P{\'e}rez-Gonz{\'a}lez}, Pablo G. and {Pacucci}, Fabio},
        title = "{The Abundance and Properties of Barred Galaxies out to z {\ensuremath{\sim}} 4 Using JWST CEERS Data}",
      journal = {\apj},
         year = 2025,
        month = jun,
       volume = {985},
       number = {2},
          eid = {181},
        pages = {181},
          doi = {10.3847/1538-4357/adc8a7},
archivePrefix = {arXiv},
       eprint = {2409.06100},
 primaryClass = {astro-ph.GA},
       adsurl = {https://ui.adsabs.harvard.edu/abs/2025ApJ...985..181G}
}

@ARTICLE{2025Ge,
       author = {{G{\'e}ron}, Tobias and {Smethurst}, R.~J. and {Dickinson}, Hugh and {Fortson}, L.~F. and {Garland}, Izzy L. and {Kruk}, Sandor and {Lintott}, Chris and {Makechemu}, Jason Shingirai and {Mantha}, Kameswara Bharadwaj and {Masters}, Karen L. and {O'Ryan}, David and {Roberts}, Hayley and {Simmons}, B.~D. and {Walmsley}, Mike and {Calabr{\`o}}, Antonello and {Chiba}, Rimpei and {Costantin}, Luca and {Drout}, Maria R. and {Fragkoudi}, Francesca and {Guo}, Yuchen and {Holwerda}, B.~W. and {Jogee}, Shardha and {Koekemoer}, Anton M. and {Lucas}, Ray A. and {Pacucci}, Fabio},
        title = "{Galaxy Zoo CEERS: Bar Fractions Up to z {\ensuremath{\sim}} 4.0}",
      journal = {\apj},
         year = 2025,
        month = jul,
       volume = {987},
       number = {1},
          eid = {74},
        pages = {74},
          doi = {10.3847/1538-4357/add7d0},
archivePrefix = {arXiv},
       eprint = {2505.01421},
 primaryClass = {astro-ph.GA},
       adsurl = {https://ui.adsabs.harvard.edu/abs/2025ApJ...987...74G}
}

@ARTICLE{2024LC,
       author = {{Le Conte}, Zoe A. and {Gadotti}, Dimitri A. and {Ferreira}, Leonardo and {Conselice}, Christopher J. and {de S{\'a}-Freitas}, Camila and {Kim}, Taehyun and {Neumann}, Justus and {Fragkoudi}, Francesca and {Athanassoula}, E. and {Adams}, Nathan J.},
        title = "{A JWST investigation into the bar fraction at redshifts 1 {\ensuremath{\leq}} z {\ensuremath{\leq}} 3}",
      journal = {\mnras},
         year = 2024,
        month = may,
       volume = {530},
       number = {2},
        pages = {1984-2000},
          doi = {10.1093/mnras/stae921},
archivePrefix = {arXiv},
       eprint = {2309.10038},
 primaryClass = {astro-ph.GA},
       adsurl = {https://ui.adsabs.harvard.edu/abs/2024MNRAS.530.1984L}
}

@ARTICLE{2025HC,
       author = {{Huertas-Company}, M. and {Shuntov}, M. and {Dong}, Y. and {Walmsley}, M. and {Ilbert}, O. and {McCracken}, H.~J. and {Akins}, H.~B. and {Allen}, N. and {Casey}, C.~M. and {Costantin}, L. and {Daddi}, E. and {Dekel}, A. and {Franco}, M. and {Garland}, I.~L. and {G{\'e}ron}, T. and {Gozaliasl}, G. and {Hirschmann}, M. and {Kartaltepe}, J.~S. and {Koekemoer}, A.~M. and {Lintott}, C. and {Liu}, D. and {Lucas}, R. and {Masters}, K. and {Pacucci}, F. and {Paquereau}, L. and {P{\'e}rez-Gonz{\'a}lez}, P.~G. and {Rhodes}, J.~D. and {Robertson}, B.~E. and {Simmons}, B. and {Smethurst}, R. and {Toft}, S. and {Yang}, L.},
        title = "{COSMOS-Web: The emergence of the Hubble sequence}",
      journal = {\aap},
         year = 2025,
        month = dec,
       volume = {704},
          eid = {A94},
        pages = {A94},
          doi = {10.1051/0004-6361/202553782},
archivePrefix = {arXiv},
       eprint = {2502.03532},
 primaryClass = {astro-ph.GA},
       adsurl = {https://ui.adsabs.harvard.edu/abs/2025A&A...704A..94H}
}

@ARTICLE{2008Sh,
       author = {{Sheth}, Kartik and {Elmegreen}, Debra Meloy and {Elmegreen}, Bruce G. and {Capak}, Peter and {Abraham}, Roberto G. and {Athanassoula}, E. and {Ellis}, Richard S. and {Mobasher}, Bahram and {Salvato}, Mara and {Schinnerer}, Eva and {Scoville}, Nicholas Z. and {Spalsbury}, Lori and {Strubbe}, Linda and {Carollo}, Marcella and {Rich}, Michael and {West}, Andrew A.},
        title = "{Evolution of the Bar Fraction in COSMOS: Quantifying the Assembly of the Hubble Sequence}",
      journal = {\apj},
         year = 2008,
        month = mar,
       volume = {675},
       number = {2},
        pages = {1141-1155},
          doi = {10.1086/524980},
archivePrefix = {arXiv},
       eprint = {0710.4552},
 primaryClass = {astro-ph},
       adsurl = {https://ui.adsabs.harvard.edu/abs/2008ApJ...675.1141S}
}

@ARTICLE{2000Es,
       author = {{Eskridge}, Paul B. and {Frogel}, Jay A. and {Pogge}, Richard W. and {Quillen}, Alice C. and {Davies}, Roger L. and {DePoy}, D.~L. and {Houdashelt}, Mark L. and {Kuchinski}, Leslie E. and {Ram{\'\i}rez}, Solange V. and {Sellgren}, K. and {Terndrup}, Donald M. and {Tiede}, Glenn P.},
        title = "{The Frequency of Barred Spiral Galaxies in the Near-Infrared}",
      journal = {\aj},
         year = 2000,
        month = feb,
       volume = {119},
       number = {2},
        pages = {536-544},
          doi = {10.1086/301203},
archivePrefix = {arXiv},
       eprint = {astro-ph/9910479},
 primaryClass = {astro-ph},
       adsurl = {https://ui.adsabs.harvard.edu/abs/2000AJ....119..536E}
}

@ARTICLE{2018Er,
       author = {{Erwin}, Peter},
        title = "{The dependence of bar frequency on galaxy mass, colour, and gas content - and angular resolution - in the local universe}",
      journal = {\mnras},
         year = 2018,
        month = mar,
       volume = {474},
       number = {4},
        pages = {5372-5392},
          doi = {10.1093/mnras/stx3117},
archivePrefix = {arXiv},
       eprint = {1711.04867},
 primaryClass = {astro-ph.GA},
       adsurl = {https://ui.adsabs.harvard.edu/abs/2018MNRAS.474.5372E}
}

@ARTICLE{2007Ma,
       author = {{Marinova}, Irina and {Jogee}, Shardha},
        title = "{Characterizing Bars at z \raisebox{-0.5ex}\textasciitilde 0 in the Optical and NIR: Implications for the Evolution of Barred Disks with Redshift}",
      journal = {\apj},
         year = 2007,
        month = apr,
       volume = {659},
       number = {2},
        pages = {1176-1197},
          doi = {10.1086/512355},
archivePrefix = {arXiv},
       eprint = {astro-ph/0608039},
 primaryClass = {astro-ph},
       adsurl = {https://ui.adsabs.harvard.edu/abs/2007ApJ...659.1176M}
}

@ARTICLE{2007Me,
       author = {{Men{\'e}ndez-Delmestre}, Kar{\'\i}n and {Sheth}, Kartik and {Schinnerer}, Eva and {Jarrett}, Thomas H. and {Scoville}, Nick Z.},
        title = "{A Near-Infrared Study of 2MASS Bars in Local Galaxies: An Anchor for High-Redshift Studies}",
      journal = {\apj},
         year = 2007,
        month = mar,
       volume = {657},
       number = {2},
        pages = {790-804},
          doi = {10.1086/511025},
archivePrefix = {arXiv},
       eprint = {astro-ph/0611540},
 primaryClass = {astro-ph},
       adsurl = {https://ui.adsabs.harvard.edu/abs/2007ApJ...657..790M}
}

@ARTICLE{1987Sp,
       author = {{Sparke}, Linda S. and {Sellwood}, J.~A.},
        title = "{Dissection of an N-body bar}",
      journal = {\mnras},
         year = 1987,
        month = apr,
       volume = {225},
        pages = {653-675},
          doi = {10.1093/mnras/225.3.653},
       adsurl = {https://ui.adsabs.harvard.edu/abs/1987MNRAS.225..653S}
}

@ARTICLE{2015Bu,
       author = {{Buta}, Ronald J. and {Sheth}, Kartik and {Athanassoula}, E. and {Bosma}, A. and {Knapen}, Johan H. and {Laurikainen}, Eija and {Salo}, Heikki and {Elmegreen}, Debra and {Ho}, Luis C. and {Zaritsky}, Dennis and {Courtois}, Helene and {Hinz}, Joannah L. and {Mu{\~n}oz-Mateos}, Juan-Carlos and {Kim}, Taehyun and {Regan}, Michael W. and {Gadotti}, Dimitri A. and {Gil de Paz}, Armando and {Laine}, Jarkko and {Men{\'e}ndez-Delmestre}, Kar{\'\i}n and {Comer{\'o}n}, S{\'e}bastien and {Erroz Ferrer}, Santiago and {Seibert}, Mark and {Mizusawa}, Trisha and {Holwerda}, Benne and {Madore}, Barry F.},
        title = "{A Classical Morphological Analysis of Galaxies in the Spitzer Survey of Stellar Structure in Galaxies (S4G)}",
      journal = {\apjs},
         year = 2015,
        month = apr,
       volume = {217},
       number = {2},
          eid = {32},
        pages = {32},
          doi = {10.1088/0067-0049/217/2/32},
archivePrefix = {arXiv},
       eprint = {1501.00454},
 primaryClass = {astro-ph.GA},
       adsurl = {https://ui.adsabs.harvard.edu/abs/2015ApJS..217...32B}
}

@ARTICLE{2015Ga,
       author = {{Gadotti}, Dimitri A. and {Seidel}, Marja K. and {S{\'a}nchez-Bl{\'a}zquez}, Patricia and {Falc{\'o}n-Barroso}, Jesus and {Husemann}, Bernd and {Coelho}, Paula and {P{\'e}rez}, Isabel},
        title = "{MUSE tells the story of NGC 4371: The dawning of secular evolution}",
      journal = {\aap},
         year = 2015,
        month = dec,
       volume = {584},
          eid = {A90},
        pages = {A90},
          doi = {10.1051/0004-6361/201526677},
archivePrefix = {arXiv},
       eprint = {1509.00032},
 primaryClass = {astro-ph.GA},
       adsurl = {https://ui.adsabs.harvard.edu/abs/2015A&A...584A..90G}
}

@ARTICLE{2025DSF,
       author = {{de S{\'a}-Freitas}, Camila and {Gadotti}, Dimitri A. and {Fragkoudi}, Francesca and {Coelho}, Paula and {de Lorenzo-C{\'a}ceres}, Adriana and {Falc{\'o}n-Barroso}, Jes{\'u}s and {S{\'a}nchez-Bl{\'a}zquez}, Patricia and {Kim}, Taehyun and {Mendez-Abreu}, Jairo and {Neumann}, Justus and {Querejeta}, Miguel and {van de Ven}, Glenn},
        title = "{Bar ages derived for the first time in nearby galaxies: Insights into secular evolution from the TIMER sample}",
      journal = {\aap},
         year = 2025,
        month = jun,
       volume = {698},
          eid = {A5},
        pages = {A5},
          doi = {10.1051/0004-6361/202453367},
archivePrefix = {arXiv},
       eprint = {2503.20864},
 primaryClass = {astro-ph.GA},
       adsurl = {https://ui.adsabs.harvard.edu/abs/2025A&A...698A...5D}
}

@ARTICLE{2019Bo,
       author = {{Bovy}, Jo and {Leung}, Henry W. and {Hunt}, Jason A.~S. and {Mackereth}, J. Ted and {Garc{\'\i}a-Hern{\'a}ndez}, Domingo A. and {Roman-Lopes}, Alexandre},
        title = "{Life in the fast lane: a direct view of the dynamics, formation, and evolution of the Milky Way's bar}",
      journal = {\mnras},
         year = 2019,
        month = dec,
       volume = {490},
       number = {4},
        pages = {4740-4747},
          doi = {10.1093/mnras/stz2891},
archivePrefix = {arXiv},
       eprint = {1905.11404},
 primaryClass = {astro-ph.GA},
       adsurl = {https://ui.adsabs.harvard.edu/abs/2019MNRAS.490.4740B}
}

@ARTICLE{2020Gr,
       author = {{Grady}, J. and {Belokurov}, V. and {Evans}, N.~W.},
        title = "{Age demographics of the Milky Way disc and bulge}",
      journal = {\mnras},
         year = 2020,
        month = feb,
       volume = {492},
       number = {3},
        pages = {3128-3142},
          doi = {10.1093/mnras/stz3617},
archivePrefix = {arXiv},
       eprint = {1912.02816},
 primaryClass = {astro-ph.GA},
       adsurl = {https://ui.adsabs.harvard.edu/abs/2020MNRAS.492.3128G}
}

@ARTICLE{1976Sa,
       author = {{Sanders}, R.~H. and {Huntley}, J.~M.},
        title = "{Gas response to oval distortions in disk galaxies.}",
      journal = {\apj},
         year = 1976,
        month = oct,
       volume = {209},
        pages = {53-65},
          doi = {10.1086/154692},
       adsurl = {https://ui.adsabs.harvard.edu/abs/1976ApJ...209...53S}
}

@ARTICLE{2003At,
       author = {{Athanassoula}, E.},
        title = "{What determines the strength and the slowdown rate of bars?}",
      journal = {\mnras},
         year = 2003,
        month = jun,
       volume = {341},
       number = {4},
        pages = {1179-1198},
          doi = {10.1046/j.1365-8711.2003.06473.x},
archivePrefix = {arXiv},
       eprint = {astro-ph/0302519},
 primaryClass = {astro-ph},
       adsurl = {https://ui.adsabs.harvard.edu/abs/2003MNRAS.341.1179A}
}

@ARTICLE{1980Se,
       author = {{Sellwood}, J.~A.},
        title = "{Galaxy models with live halos}",
      journal = {\aap},
         year = 1980,
        month = sep,
       volume = {89},
       number = {3},
        pages = {296-307},
       adsurl = {https://ui.adsabs.harvard.edu/abs/1980A&A....89..296S}
}

@ARTICLE{1987Ta,
       author = {{Tagger}, M. and {Sygnet}, J.~F. and {Athanassoula}, E. and {Pellat}, R.},
        title = "{Nonlinear Coupling of Galactic Spiral Modes}",
      journal = {\apjl},
         year = 1987,
        month = jul,
       volume = {318},
        pages = {L43},
          doi = {10.1086/184934},
       adsurl = {https://ui.adsabs.harvard.edu/abs/1987ApJ...318L..43T}
}

@ARTICLE{2010Mi,
       author = {{Minchev}, I. and {Famaey}, B.},
        title = "{A New Mechanism for Radial Migration in Galactic Disks: Spiral-Bar Resonance Overlap}",
      journal = {\apj},
         year = 2010,
        month = oct,
       volume = {722},
       number = {1},
        pages = {112-121},
          doi = {10.1088/0004-637X/722/1/112},
archivePrefix = {arXiv},
       eprint = {0911.1794},
 primaryClass = {astro-ph.GA},
       adsurl = {https://ui.adsabs.harvard.edu/abs/2010ApJ...722..112M}
}

@ARTICLE{2011Br,
       author = {{Brunetti}, M. and {Chiappini}, C. and {Pfenniger}, D.},
        title = "{Stellar diffusion in barred spiral galaxies}",
      journal = {\aap},
         year = 2011,
        month = oct,
       volume = {534},
          eid = {A75},
        pages = {A75},
          doi = {10.1051/0004-6361/201117566},
archivePrefix = {arXiv},
       eprint = {1108.5631},
 primaryClass = {astro-ph.GA},
       adsurl = {https://ui.adsabs.harvard.edu/abs/2011A&A...534A..75B}
}

@ARTICLE{2000De,
       author = {{Dehnen}, Walter},
        title = "{The Effect of the Outer Lindblad Resonance of the Galactic Bar on the Local Stellar Velocity Distribution}",
      journal = {\aj},
         year = 2000,
        month = feb,
       volume = {119},
       number = {2},
        pages = {800-812},
          doi = {10.1086/301226},
archivePrefix = {arXiv},
       eprint = {astro-ph/9911161},
 primaryClass = {astro-ph},
       adsurl = {https://ui.adsabs.harvard.edu/abs/2000AJ....119..800D}
}

@ARTICLE{2003Bi,
       author = {{Bissantz}, Nicolai and {Englmaier}, Peter and {Gerhard}, Ortwin},
        title = "{Gas dynamics in the Milky Way: second pattern speed and large-scale morphology}",
      journal = {\mnras},
         year = 2003,
        month = apr,
       volume = {340},
       number = {3},
        pages = {949-968},
          doi = {10.1046/j.1365-8711.2003.06358.x},
archivePrefix = {arXiv},
       eprint = {astro-ph/0212516},
 primaryClass = {astro-ph},
       adsurl = {https://ui.adsabs.harvard.edu/abs/2003MNRAS.340..949B}
}

@ARTICLE{2002We,
       author = {{Weinberg}, Martin D. and {Katz}, Neal},
        title = "{Bar-driven Dark Halo Evolution: A Resolution of the Cusp-Core Controversy}",
      journal = {\apj},
         year = 2002,
        month = dec,
       volume = {580},
       number = {2},
        pages = {627-633},
          doi = {10.1086/343847},
archivePrefix = {arXiv},
       eprint = {astro-ph/0110632},
 primaryClass = {astro-ph},
       adsurl = {https://ui.adsabs.harvard.edu/abs/2002ApJ...580..627W}
}

@ARTICLE{2013At,
       author = {{Athanassoula}, E. and {Machado}, Rubens E.~G. and {Rodionov}, S.~A.},
        title = "{Bar formation and evolution in disc galaxies with gas and a triaxial halo: morphology, bar strength and halo properties}",
      journal = {\mnras},
         year = 2013,
        month = mar,
       volume = {429},
       number = {3},
        pages = {1949-1969},
          doi = {10.1093/mnras/sts452},
archivePrefix = {arXiv},
       eprint = {1211.6754},
 primaryClass = {astro-ph.CO},
       adsurl = {https://ui.adsabs.harvard.edu/abs/2013MNRAS.429.1949A}
}

@ARTICLE{1999Sa,
       author = {{Sakamoto}, K. and {Okumura}, S.~K. and {Ishizuki}, S. and {Scoville}, N.~Z.},
        title = "{Bar-driven Transport of Molecular Gas to Galactic Centers and Its Consequences}",
      journal = {\apj},
         year = 1999,
        month = nov,
       volume = {525},
       number = {2},
        pages = {691-701},
          doi = {10.1086/307910},
archivePrefix = {arXiv},
       eprint = {astro-ph/9906454},
 primaryClass = {astro-ph},
       adsurl = {https://ui.adsabs.harvard.edu/abs/1999ApJ...525..691S}
}

@ARTICLE{2024Se,
       author = {{Semczuk}, Marcin and {Dehnen}, Walter and {Sch{\"o}nrich}, Ralph and {Athanassoula}, E.},
        title = "{Pattern speed evolution of barred galaxies in TNG50}",
      journal = {\aap},
         year = 2024,
        month = dec,
       volume = {692},
          eid = {A159},
        pages = {A159},
          doi = {10.1051/0004-6361/202451521},
archivePrefix = {arXiv},
       eprint = {2407.11154},
 primaryClass = {astro-ph.GA},
       adsurl = {https://ui.adsabs.harvard.edu/abs/2024A&A...692A.159S}
}

@ARTICLE{2006Se,
       author = {{Sellwood}, J.~A. and {Debattista}, Victor P.},
        title = "{Bar-Halo Friction in Galaxies. II. Metastability}",
      journal = {\apj},
         year = 2006,
        month = mar,
       volume = {639},
       number = {2},
        pages = {868-878},
          doi = {10.1086/499482},
archivePrefix = {arXiv},
       eprint = {astro-ph/0511155},
 primaryClass = {astro-ph},
       adsurl = {https://ui.adsabs.harvard.edu/abs/2006ApJ...639..868S}
}

@ARTICLE{1985We,
       author = {{Weinberg}, M.~D.},
        title = "{Evolution of barred galaxies by dynamical friction.}",
      journal = {\mnras},
         year = 1985,
        month = mar,
       volume = {213},
        pages = {451-471},
          doi = {10.1093/mnras/213.3.451},
       adsurl = {https://ui.adsabs.harvard.edu/abs/1985MNRAS.213..451W}
}

@ARTICLE{2000De2,
       author = {{Debattista}, Victor P. and {Sellwood}, J.~A.},
        title = "{Constraints from Dynamical Friction on the Dark Matter Content of Barred Galaxies}",
      journal = {\apj},
         year = 2000,
        month = nov,
       volume = {543},
       number = {2},
        pages = {704-721},
          doi = {10.1086/317148},
archivePrefix = {arXiv},
       eprint = {astro-ph/0006275},
 primaryClass = {astro-ph},
       adsurl = {https://ui.adsabs.harvard.edu/abs/2000ApJ...543..704D}
}

@ARTICLE{2003ON,
       author = {{O'Neill}, J.~K. and {Dubinski}, John},
        title = "{Detailed comparison of the structures and kinematics of simulated and observed barred galaxies}",
      journal = {\mnras},
         year = 2003,
        month = nov,
       volume = {346},
       number = {1},
        pages = {251-264},
          doi = {10.1046/j.1365-2966.2003.07085.x},
archivePrefix = {arXiv},
       eprint = {astro-ph/0305169},
 primaryClass = {astro-ph},
       adsurl = {https://ui.adsabs.harvard.edu/abs/2003MNRAS.346..251O}
}

@ARTICLE{1992He,
       author = {{Hernquist}, Lars and {Weinberg}, Martin D.},
        title = "{Bar-Spheroid Interaction in Galaxies}",
      journal = {\apj},
         year = 1992,
        month = nov,
       volume = {400},
        pages = {80},
          doi = {10.1086/171975},
       adsurl = {https://ui.adsabs.harvard.edu/abs/1992ApJ...400...80H}
}

@ARTICLE{1991Li,
       author = {{Little}, Blane and {Carlberg}, R.~G.},
        title = "{The long-term evolution of barred galaxies}",
      journal = {\mnras},
         year = 1991,
        month = may,
       volume = {250},
        pages = {161-170},
          doi = {10.1093/mnras/250.1.161},
       adsurl = {https://ui.adsabs.harvard.edu/abs/1991MNRAS.250..161L}
}

@ARTICLE{1980Co,
       author = {{Contopoulos}, G.},
        title = "{How far do bars extend}",
      journal = {\aap},
         year = 1980,
        month = jan,
       volume = {81},
       number = {1-2},
        pages = {198-209},
       adsurl = {https://ui.adsabs.harvard.edu/abs/1980A&A....81..198C}
}

@ARTICLE{2021Cu,
       author = {{Cuomo}, Virginia and {Lee}, Yun Hee and {Buttitta}, Chiara and {Aguerri}, J. Alfonso L. and {Corsini}, Enrico Maria and {Morelli}, Lorenzo},
        title = "{Bar pattern speeds in CALIFA galaxies. III. Solving the puzzle of ultrafast bars}",
      journal = {\aap},
         year = 2021,
        month = may,
       volume = {649},
          eid = {A30},
        pages = {A30},
          doi = {10.1051/0004-6361/202040261},
archivePrefix = {arXiv},
       eprint = {2103.00343},
 primaryClass = {astro-ph.GA},
       adsurl = {https://ui.adsabs.harvard.edu/abs/2021A&A...649A..30C}
}

@ARTICLE{2019Gu,
       author = {{Guo}, Rui and {Mao}, Shude and {Athanassoula}, E. and {Li}, Hongyu and {Ge}, Junqiang and {Long}, R.~J. and {Merrifield}, Michael and {Masters}, Karen},
        title = "{SDSS-IV MaNGA: pattern speeds of barred galaxies}",
      journal = {\mnras},
         year = 2019,
        month = jan,
       volume = {482},
       number = {2},
        pages = {1733-1756},
          doi = {10.1093/mnras/sty2715},
archivePrefix = {arXiv},
       eprint = {1810.03257},
 primaryClass = {astro-ph.GA},
       adsurl = {https://ui.adsabs.harvard.edu/abs/2019MNRAS.482.1733G}
}

@ARTICLE{2015Ag,
       author = {{Aguerri}, J.~A.~L. and {M{\'e}ndez-Abreu}, J. and {Falc{\'o}n-Barroso}, J. and {Amorin}, A. and {Barrera-Ballesteros}, J. and {Cid Fernandes}, R. and {Garc{\'\i}a-Benito}, R. and {Garc{\'\i}a-Lorenzo}, B. and {Gonz{\'a}lez Delgado}, R.~M. and {Husemann}, B. and {Kalinova}, V. and {Lyubenova}, M. and {Marino}, R.~A. and {M{\'a}rquez}, I. and {Mast}, D. and {P{\'e}rez}, E. and {S{\'a}nchez}, S.~F. and {van de Ven}, G. and {Walcher}, C.~J. and {Backsmann}, N. and {Cortijo-Ferrero}, C. and {Bland-Hawthorn}, J. and {del Olmo}, A. and {Iglesias-P{\'a}ramo}, J. and {P{\'e}rez}, I. and {S{\'a}nchez-Bl{\'a}zquez}, P. and {Wisotzki}, L. and {Ziegler}, B.},
        title = "{Bar pattern speeds in CALIFA galaxies. I. Fast bars across the Hubble sequence}",
      journal = {\aap},
         year = 2015,
        month = apr,
       volume = {576},
          eid = {A102},
        pages = {A102},
          doi = {10.1051/0004-6361/201423383},
archivePrefix = {arXiv},
       eprint = {1501.05498},
 primaryClass = {astro-ph.GA},
       adsurl = {https://ui.adsabs.harvard.edu/abs/2015A&A...576A.102A}
}

@ARTICLE{2009Bu,
       author = {{Buta}, Ronald J. and {Zhang}, Xiaolei},
        title = "{Pattern Corotation Radii from Potential-Density Phase-Shifts for 153 OSUBGS Sample Galaxies}",
      journal = {\apjs},
         year = 2009,
        month = jun,
       volume = {182},
       number = {2},
        pages = {559-583},
          doi = {10.1088/0067-0049/182/2/559},
archivePrefix = {arXiv},
       eprint = {0903.4695},
 primaryClass = {astro-ph.CO},
       adsurl = {https://ui.adsabs.harvard.edu/abs/2009ApJS..182..559B}
}

@ARTICLE{2020Hi,
       author = {{Hilmi}, T. and {Minchev}, I. and {Buck}, T. and {Martig}, M. and {Quillen}, A.~C. and {Monari}, G. and {Famaey}, B. and {de Jong}, R.~S. and {Laporte}, C.~F.~P. and {Read}, J. and {Sanders}, J.~L. and {Steinmetz}, M. and {Wegg}, C.},
        title = "{Fluctuations in galactic bar parameters due to bar-spiral interaction}",
      journal = {\mnras},
         year = 2020,
        month = sep,
       volume = {497},
       number = {1},
        pages = {933-955},
          doi = {10.1093/mnras/staa1934},
archivePrefix = {arXiv},
       eprint = {2003.05457},
 primaryClass = {astro-ph.GA},
       adsurl = {https://ui.adsabs.harvard.edu/abs/2020MNRAS.497..933H}
}

@ARTICLE{2008Ra,
       author = {{Rautiainen}, P. and {Salo}, H. and {Laurikainen}, E.},
        title = "{Model-based pattern speed estimates for 38 barred galaxies}",
      journal = {\mnras},
         year = 2008,
        month = aug,
       volume = {388},
       number = {4},
        pages = {1803-1818},
          doi = {10.1111/j.1365-2966.2008.13522.x},
archivePrefix = {arXiv},
       eprint = {0806.0471},
 primaryClass = {astro-ph},
       adsurl = {https://ui.adsabs.harvard.edu/abs/2008MNRAS.388.1803R}
}

@ARTICLE{2011Co,
       author = {{Corsini}, E.~M.},
        title = "{Direct measurements of bar pattern speeds.}",
      journal = {Memorie della Societa Astronomica Italiana Supplementi},
         year = 2011,
        month = jan,
       volume = {18},
        pages = {23},
          doi = {10.48550/arXiv.1002.1245},
archivePrefix = {arXiv},
       eprint = {1002.1245},
 primaryClass = {astro-ph.GA},
       adsurl = {https://ui.adsabs.harvard.edu/abs/2011MSAIS..18...23C}
}

@ARTICLE{2019Pe,
       author = {{Peschken}, Nicolas and {{\L}okas}, Ewa L.},
        title = "{Tidally induced bars in Illustris galaxies}",
      journal = {\mnras},
         year = 2019,
        month = feb,
       volume = {483},
       number = {2},
        pages = {2721-2735},
          doi = {10.1093/mnras/sty3277},
archivePrefix = {arXiv},
       eprint = {1804.06241},
 primaryClass = {astro-ph.GA},
       adsurl = {https://ui.adsabs.harvard.edu/abs/2019MNRAS.483.2721P}
}

@ARTICLE{2017Al,
       author = {{Algorry}, David G. and {Navarro}, Julio F. and {Abadi}, Mario G. and {Sales}, Laura V. and {Bower}, Richard G. and {Crain}, Robert A. and {Dalla Vecchia}, Claudio and {Frenk}, Carlos S. and {Schaller}, Matthieu and {Schaye}, Joop and {Theuns}, Tom},
        title = "{Barred galaxies in the EAGLE cosmological hydrodynamical simulation}",
      journal = {\mnras},
         year = 2017,
        month = jul,
       volume = {469},
       number = {1},
        pages = {1054-1064},
          doi = {10.1093/mnras/stx1008},
archivePrefix = {arXiv},
       eprint = {1609.05909},
 primaryClass = {astro-ph.GA},
       adsurl = {https://ui.adsabs.harvard.edu/abs/2017MNRAS.469.1054A}
}

@ARTICLE{2021Ro,
       author = {{Roshan}, Mahmood and {Ghafourian}, Neda and {Kashfi}, Tahere and {Banik}, Indranil and {Haslbauer}, Moritz and {Cuomo}, Virginia and {Famaey}, Benoit and {Kroupa}, Pavel},
        title = "{Fast galaxy bars continue to challenge standard cosmology}",
      journal = {\mnras},
         year = 2021,
        month = nov,
       volume = {508},
       number = {1},
        pages = {926-939},
          doi = {10.1093/mnras/stab2553},
archivePrefix = {arXiv},
       eprint = {2106.10304},
 primaryClass = {astro-ph.GA},
       adsurl = {https://ui.adsabs.harvard.edu/abs/2021MNRAS.508..926R}
}

@ARTICLE{2008RD,
       author = {{Romano-D{\'\i}az}, Emilio and {Shlosman}, Isaac and {Heller}, Clayton and {Hoffman}, Yehuda},
        title = "{Disk Evolution and Bar Triggering Driven by Interactions with Dark Matter Substructure}",
      journal = {\apjl},
         year = 2008,
        month = nov,
       volume = {687},
       number = {1},
        pages = {L13},
          doi = {10.1086/593168},
archivePrefix = {arXiv},
       eprint = {0809.2785},
 primaryClass = {astro-ph},
       adsurl = {https://ui.adsabs.harvard.edu/abs/2008ApJ...687L..13R}
}

@ARTICLE{2023Be,
       author = {{Beane}, Angus and {Hernquist}, Lars and {D'Onghia}, Elena and {Marinacci}, Federico and {Conroy}, Charlie and {Qi}, Jia and {Sales}, Laura V. and {Torrey}, Paul and {Vogelsberger}, Mark},
        title = "{Stellar Bars in Isolated Gas-rich Spiral Galaxies Do Not Slow Down}",
      journal = {\apj},
         year = 2023,
        month = aug,
       volume = {953},
       number = {2},
          eid = {173},
        pages = {173},
          doi = {10.3847/1538-4357/ace2b9},
archivePrefix = {arXiv},
       eprint = {2209.03364},
 primaryClass = {astro-ph.GA},
       adsurl = {https://ui.adsabs.harvard.edu/abs/2023ApJ...953..173B}
}

@ARTICLE{2007Be,
       author = {{Berentzen}, Ingo and {Shlosman}, Isaac and {Martinez-Valpuesta}, Inma and {Heller}, Clayton H.},
        title = "{Gas Feedback on Stellar Bar Evolution}",
      journal = {\apj},
         year = 2007,
        month = sep,
       volume = {666},
       number = {1},
        pages = {189-200},
          doi = {10.1086/520531},
archivePrefix = {arXiv},
       eprint = {astro-ph/0703028},
 primaryClass = {astro-ph},
       adsurl = {https://ui.adsabs.harvard.edu/abs/2007ApJ...666..189B}
}

@ARTICLE{2010VV,
       author = {{Villa-Vargas}, Jorge and {Shlosman}, Isaac and {Heller}, Clayton},
        title = "{Dark Matter Halos and Evolution of Bars in Disk Galaxies: Varying Gas Fraction and Gas Spatial Resolution}",
      journal = {\apj},
         year = 2010,
        month = aug,
       volume = {719},
       number = {2},
        pages = {1470-1480},
          doi = {10.1088/0004-637X/719/2/1470},
archivePrefix = {arXiv},
       eprint = {1004.4899},
 primaryClass = {astro-ph.CO},
       adsurl = {https://ui.adsabs.harvard.edu/abs/2010ApJ...719.1470V}
}

@ARTICLE{2014At,
       author = {{Athanassoula}, E.},
        title = "{Bar slowdown and the distribution of dark matter in barred galaxies}",
      journal = {\mnras},
         year = 2014,
        month = feb,
       volume = {438},
       number = {1},
        pages = {L81-L85},
          doi = {10.1093/mnrasl/slt163},
archivePrefix = {arXiv},
       eprint = {1312.1690},
 primaryClass = {astro-ph.GA},
       adsurl = {https://ui.adsabs.harvard.edu/abs/2014MNRAS.438L..81A}
}

@ARTICLE{1993Su,
       author = {{Sundin}, M. and {Donner}, K.~J. and {Sundelius}, B.},
        title = "{Change in angular velocity of perturbed galactic bars}",
      journal = {\aap},
         year = 1993,
        month = dec,
       volume = {280},
       number = {1},
        pages = {105-116},
       adsurl = {https://ui.adsabs.harvard.edu/abs/1993A&A...280..105S}
}

@INPROCEEDINGS{1996At,
       author = {{Athanassoula}, E.},
        title = "{Evolution of Bars in Isolated and in Interacting Disk Galaxies}",
    booktitle = {IAU Colloquium 157: Barred Galaxies},
         year = 1996,
       editor = {{Buta}, R. and {Crocker}, D.~A. and {Elmegreen}, B.~G.},
       series = {Astronomical Society of the Pacific Conference Series},
       volume = {91},
        month = jan,
        pages = {309},
       adsurl = {https://ui.adsabs.harvard.edu/abs/1996ASPC...91..309A}
}

@ARTICLE{2014Lo,
       author = {{{\L}okas}, E.~L. and {Athanassoula}, E. and {Debattista}, V.~P. and {Valluri}, M. and {Pino}, A. del and {Semczuk}, M. and {Gajda}, G. and {Kowalczyk}, K.},
        title = "{Adventures of a tidally induced bar}",
      journal = {\mnras},
         year = 2014,
        month = dec,
       volume = {445},
       number = {2},
        pages = {1339-1350},
          doi = {10.1093/mnras/stu1846},
archivePrefix = {arXiv},
       eprint = {1404.1211},
 primaryClass = {astro-ph.GA},
       adsurl = {https://ui.adsabs.harvard.edu/abs/2014MNRAS.445.1339L}
}

@ARTICLE{2018Za,
       author = {{Zana}, Tommaso and {Dotti}, Massimo and {Capelo}, Pedro R. and {Mayer}, Lucio and {Haardt}, Francesco and {Shen}, Sijing and {Bonoli}, Silvia},
        title = "{Bar resilience to flybys in a cosmological framework}",
      journal = {\mnras},
         year = 2018,
        month = oct,
       volume = {479},
       number = {4},
        pages = {5214-5219},
          doi = {10.1093/mnras/sty1850},
archivePrefix = {arXiv},
       eprint = {1805.03658},
 primaryClass = {astro-ph.GA},
       adsurl = {https://ui.adsabs.harvard.edu/abs/2018MNRAS.479.5214Z}
}

@ARTICLE{2025An,
       author = {{Ansar}, Sioree and {Pearson}, Sarah and {Sanderson}, Robyn E. and {Arora}, Arpit and {Hopkins}, Philip F. and {Wetzel}, Andrew and {Cunningham}, Emily C. and {Quinn}, Jamie},
        title = "{Bar Formation and Destruction in the FIRE-2 Simulations}",
      journal = {\apj},
         year = 2025,
        month = jan,
       volume = {978},
       number = {1},
          eid = {37},
        pages = {37},
          doi = {10.3847/1538-4357/ad8b45},
archivePrefix = {arXiv},
       eprint = {2309.16811},
 primaryClass = {astro-ph.GA},
       adsurl = {https://ui.adsabs.harvard.edu/abs/2025ApJ...978...37A}
}

@ARTICLE{2022Ca,
       author = {{Cavanagh}, Mitchell K. and {Bekki}, Kenji and {Groves}, Brent A. and {Pfeffer}, Joel},
        title = "{The evolution of barred galaxies in the EAGLE simulations}",
      journal = {\mnras},
         year = 2022,
        month = mar,
       volume = {510},
       number = {4},
        pages = {5164-5178},
          doi = {10.1093/mnras/stab3786},
archivePrefix = {arXiv},
       eprint = {2112.12935},
 primaryClass = {astro-ph.GA},
       adsurl = {https://ui.adsabs.harvard.edu/abs/2022MNRAS.510.5164C}
}

@ARTICLE{2015Ka,
       author = {{Kannan}, Rahul and {Macci{\`o}}, Andrea V. and {Fontanot}, Fabio and {Moster}, Benjamin P. and {Karman}, Wouter and {Somerville}, Rachel S.},
        title = "{From discs to bulges: effect of mergers on the morphology of galaxies}",
      journal = {\mnras},
         year = 2015,
        month = oct,
       volume = {452},
       number = {4},
        pages = {4347-4360},
          doi = {10.1093/mnras/stv1633},
archivePrefix = {arXiv},
       eprint = {1507.04746},
 primaryClass = {astro-ph.GA},
       adsurl = {https://ui.adsabs.harvard.edu/abs/2015MNRAS.452.4347K}
}

@ARTICLE{2022Bi,
       author = {{Bi}, Da and {Shlosman}, Isaac and {Romano-D{\'\i}az}, Emilio},
        title = "{Modeling Evolution of Galactic Bars at Cosmic Dawn}",
      journal = {\apj},
         year = 2022,
        month = jul,
       volume = {934},
       number = {1},
          eid = {52},
        pages = {52},
          doi = {10.3847/1538-4357/ac779b},
archivePrefix = {arXiv},
       eprint = {2112.09718},
 primaryClass = {astro-ph.GA},
       adsurl = {https://ui.adsabs.harvard.edu/abs/2022ApJ...934...52B}
}

@ARTICLE{2017Mo,
       author = {{Moetazedian}, R. and {Polyachenko}, E.~V. and {Berczik}, P. and {Just}, A.},
        title = "{Effects of galaxy-satellite interactions on bar formation}",
      journal = {\aap},
         year = 2017,
        month = aug,
       volume = {604},
          eid = {A75},
        pages = {A75},
          doi = {10.1051/0004-6361/201630024},
archivePrefix = {arXiv},
       eprint = {1703.06002},
 primaryClass = {astro-ph.GA},
       adsurl = {https://ui.adsabs.harvard.edu/abs/2017A&A...604A..75M}
}

@ARTICLE{1987No,
       author = {{Noguchi}, Masafumi},
        title = "{Close encounter between galaxies - II. Tidal deformation of a disc galaxy stabilized by massive halo.}",
      journal = {\mnras},
         year = 1987,
        month = oct,
       volume = {228},
        pages = {635-651},
          doi = {10.1093/mnras/228.3.635},
       adsurl = {https://ui.adsabs.harvard.edu/abs/1987MNRAS.228..635N}
}

@ARTICLE{1990Ge,
       author = {{Gerin}, M. and {Combes}, F. and {Athanassoula}, E.},
        title = "{The influence of galaxy interactions on stellar bars.}",
      journal = {\aap},
         year = 1990,
        month = apr,
       volume = {230},
        pages = {37-54},
       adsurl = {https://ui.adsabs.harvard.edu/abs/1990A&A...230...37G}
}

@ARTICLE{2024Me,
       author = {{Merrow}, Alex and {Grand}, Robert J.~J. and {Fragkoudi}, Francesca and {Martig}, Marie},
        title = "{Did the Gaia Enceladus/Sausage merger form the Milky Way's bar?}",
      journal = {\mnras},
         year = 2024,
        month = jun,
       volume = {531},
       number = {1},
        pages = {1520-1533},
          doi = {10.1093/mnras/stae1250},
archivePrefix = {arXiv},
       eprint = {2312.02318},
 primaryClass = {astro-ph.GA},
       adsurl = {https://ui.adsabs.harvard.edu/abs/2024MNRAS.531.1520M}
}

@ARTICLE{1998Mi,
       author = {{Miwa}, Toshinobu and {Noguchi}, Masafumi},
        title = "{Dynamical Properties of Tidally Induced Galactic Bars}",
      journal = {\apj},
         year = 1998,
        month = may,
       volume = {499},
       number = {1},
        pages = {149-166},
          doi = {10.1086/305611},
       adsurl = {https://ui.adsabs.harvard.edu/abs/1998ApJ...499..149M}
}

@ARTICLE{2017MV,
       author = {{Martinez-Valpuesta}, Inma and {Aguerri}, J. Alfonso L. and {Gonz{\'a}lez-Garc{\'\i}a}, A. C{\'e}sar and {Dalla Vecchia}, Claudio and {Stringer}, Martin},
        title = "{A numerical study of interactions and stellar bars}",
      journal = {\mnras},
         year = 2017,
        month = jan,
       volume = {464},
       number = {2},
        pages = {1502-1511},
          doi = {10.1093/mnras/stw2500},
archivePrefix = {arXiv},
       eprint = {1610.02326},
 primaryClass = {astro-ph.GA},
       adsurl = {https://ui.adsabs.harvard.edu/abs/2017MNRAS.464.1502M}
}

@ARTICLE{2025Zh,
       author = {{Zheng}, Yirui and {Shen}, Juntai},
        title = "{Comparison of Bar Formation Mechanisms. I. Does a Tidally Induced Bar Rotate Slower than an Internally Induced Bar?}",
      journal = {\apj},
         year = 2025,
        month = jan,
       volume = {979},
       number = {1},
          eid = {60},
        pages = {60},
          doi = {10.3847/1538-4357/ad9bae},
archivePrefix = {arXiv},
       eprint = {2412.04770},
 primaryClass = {astro-ph.GA},
       adsurl = {https://ui.adsabs.harvard.edu/abs/2025ApJ...979...60Z}
}

@ARTICLE{2017Ga,
       author = {{Gajda}, Grzegorz and {{\L}okas}, Ewa L. and {Athanassoula}, E.},
        title = "{Tidally Induced Bars in Dwarf Galaxies on Different Orbits around a Milky Way-like Host}",
      journal = {\apj},
         year = 2017,
        month = jun,
       volume = {842},
       number = {1},
          eid = {56},
        pages = {56},
          doi = {10.3847/1538-4357/aa74b4},
archivePrefix = {arXiv},
       eprint = {1703.02933},
 primaryClass = {astro-ph.GA},
       adsurl = {https://ui.adsabs.harvard.edu/abs/2017ApJ...842...56G}
}

@ARTICLE{2018Pe,
       author = {{Pettitt}, Alex R. and {Wadsley}, J.~W.},
        title = "{Bars and spirals in tidal interactions with an ensemble of galaxy mass models}",
      journal = {\mnras},
         year = 2018,
        month = mar,
       volume = {474},
       number = {4},
        pages = {5645-5671},
          doi = {10.1093/mnras/stx3129},
archivePrefix = {arXiv},
       eprint = {1712.00882},
 primaryClass = {astro-ph.GA},
       adsurl = {https://ui.adsabs.harvard.edu/abs/2018MNRAS.474.5645P}
}

@ARTICLE{2011Qu,
       author = {{Quillen}, Alice C. and {Dougherty}, Jamie and {Bagley}, Micaela B. and {Minchev}, Ivan and {Comparetta}, Justin},
        title = "{Structure in phase space associated with spiral and bar density waves in an N-body hybrid galactic disc}",
      journal = {\mnras},
         year = 2011,
        month = oct,
       volume = {417},
       number = {1},
        pages = {762-784},
          doi = {10.1111/j.1365-2966.2011.19349.x},
archivePrefix = {arXiv},
       eprint = {1010.5745},
 primaryClass = {astro-ph.GA},
       adsurl = {https://ui.adsabs.harvard.edu/abs/2011MNRAS.417..762Q}
}

@ARTICLE{2002Bi,
       author = {{Bissantz}, Nicolai and {Gerhard}, Ortwin},
        title = "{Spiral arms, bar shape and bulge microlensing in the Milky Way}",
      journal = {\mnras},
         year = 2002,
        month = mar,
       volume = {330},
       number = {3},
        pages = {591-608},
          doi = {10.1046/j.1365-8711.2002.05116.x},
archivePrefix = {arXiv},
       eprint = {astro-ph/0110368},
 primaryClass = {astro-ph},
       adsurl = {https://ui.adsabs.harvard.edu/abs/2002MNRAS.330..591B}
}

@ARTICLE{2012Kr,
       author = {{Kraljic}, Katarina and {Bournaud}, Fr{\'e}d{\'e}ric and {Martig}, Marie},
        title = "{The Two-phase Formation History of Spiral Galaxies Traced by the Cosmic Evolution of the Bar Fraction}",
      journal = {\apj},
         year = 2012,
        month = sep,
       volume = {757},
       number = {1},
          eid = {60},
        pages = {60},
          doi = {10.1088/0004-637X/757/1/60},
archivePrefix = {arXiv},
       eprint = {1207.0351},
 primaryClass = {astro-ph.GA},
       adsurl = {https://ui.adsabs.harvard.edu/abs/2012ApJ...757...60K}
}

@ARTICLE{1998De,
       author = {{Debattista}, Victor P. and {Sellwood}, J.~A.},
        title = "{Dynamical Friction and the Distribution of Dark Matter in Barred Galaxies}",
      journal = {\apjl},
         year = 1998,
        month = jan,
       volume = {493},
       number = {1},
        pages = {L5-L8},
          doi = {10.1086/311118},
archivePrefix = {arXiv},
       eprint = {astro-ph/9710039},
 primaryClass = {astro-ph},
       adsurl = {https://ui.adsabs.harvard.edu/abs/1998ApJ...493L...5D}
}

@ARTICLE{1977Co,
       author = {{Contopoulos}, G. and {Mertzanides}, C.},
        title = "{Inner Lindblad resonance in galaxies. Nonlinear theory. II. Bars.}",
      journal = {\aap},
         year = 1977,
        month = nov,
       volume = {61},
       number = {4},
        pages = {477-485},
       adsurl = {https://ui.adsabs.harvard.edu/abs/1977A&A....61..477C}
}

@ARTICLE{1984Sc,
       author = {{Schwarz}, M.~P.},
        title = "{How bar strength and pattern speed affect galactic spiral structure.}",
      journal = {\mnras},
         year = 1984,
        month = jul,
       volume = {209},
        pages = {93-109},
          doi = {10.1093/mnras/209.1.93},
       adsurl = {https://ui.adsabs.harvard.edu/abs/1984MNRAS.209...93S}
}

@ARTICLE{1977Ma,
       author = {{Matsuda}, T. and {Nelson}, A.~H.},
        title = "{A galactic vacuum cleaner?}",
      journal = {\nat},
         year = 1977,
        month = apr,
       volume = {266},
       number = {5603},
        pages = {607-608},
          doi = {10.1038/266607a0},
       adsurl = {https://ui.adsabs.harvard.edu/abs/1977Natur.266..607M}
}

@ARTICLE{1980Si,
       author = {{Simkin}, S.~M. and {Su}, H.~J. and {Schwarz}, M.~P.},
        title = "{Nearby Seyfert galaxies.}",
      journal = {\apj},
         year = 1980,
        month = apr,
       volume = {237},
        pages = {404-413},
          doi = {10.1086/157882},
       adsurl = {https://ui.adsabs.harvard.edu/abs/1980ApJ...237..404S}
}

@ARTICLE{2005Sh,
       author = {{Sheth}, Kartik and {Vogel}, Stuart N. and {Regan}, Michael W. and {Thornley}, Michele D. and {Teuben}, Peter J.},
        title = "{Secular Evolution via Bar-driven Gas Inflow: Results from BIMA SONG}",
      journal = {\apj},
         year = 2005,
        month = oct,
       volume = {632},
       number = {1},
        pages = {217-226},
          doi = {10.1086/432409},
archivePrefix = {arXiv},
       eprint = {astro-ph/0505393},
 primaryClass = {astro-ph},
       adsurl = {https://ui.adsabs.harvard.edu/abs/2005ApJ...632..217S}
}

@ARTICLE{2014Co,
       author = {{Cole}, David R. and {Debattista}, Victor P. and {Erwin}, Peter and {Earp}, Samuel W.~F. and {Ro{\v{s}}kar}, Rok},
        title = "{The formation of stellar nuclear discs in bar-induced gas inflows}",
      journal = {\mnras},
         year = 2014,
        month = dec,
       volume = {445},
       number = {4},
        pages = {3352-3369},
          doi = {10.1093/mnras/stu1985},
archivePrefix = {arXiv},
       eprint = {1410.4339},
 primaryClass = {astro-ph.GA},
       adsurl = {https://ui.adsabs.harvard.edu/abs/2014MNRAS.445.3352C}
}

@ARTICLE{2007El,
       author = {{Elmegreen}, Bruce G. and {Elmegreen}, Debra Meloy and {Knapen}, Johan H. and {Buta}, Ronald J. and {Block}, David L. and {Puerari}, Iv{\^a}nio},
        title = "{Variation of Galactic Bar Length with Amplitude and Density as Evidence for Bar Growth over a Hubble Time}",
      journal = {\apjl},
         year = 2007,
        month = dec,
       volume = {670},
       number = {2},
        pages = {L97-L100},
          doi = {10.1086/524359},
archivePrefix = {arXiv},
       eprint = {0711.3055},
 primaryClass = {astro-ph},
       adsurl = {https://ui.adsabs.harvard.edu/abs/2007ApJ...670L..97E}
}

@ARTICLE{1984Tr2,
       author = {{Tremaine}, S. and {Weinberg}, M.~D.},
        title = "{Dynamical friction in spherical systems.}",
      journal = {\mnras},
         year = 1984,
        month = aug,
       volume = {209},
        pages = {729-757},
          doi = {10.1093/mnras/209.4.729},
       adsurl = {https://ui.adsabs.harvard.edu/abs/1984MNRAS.209..729T}
}

@ARTICLE{2006MV,
       author = {{Martinez-Valpuesta}, Inma and {Shlosman}, Isaac and {Heller}, Clayton},
        title = "{Evolution of Stellar Bars in Live Axisymmetric Halos: Recurrent Buckling and Secular Growth}",
      journal = {\apj},
         year = 2006,
        month = jan,
       volume = {637},
       number = {1},
        pages = {214-226},
          doi = {10.1086/498338},
archivePrefix = {arXiv},
       eprint = {astro-ph/0507219},
 primaryClass = {astro-ph},
       adsurl = {https://ui.adsabs.harvard.edu/abs/2006ApJ...637..214M}
}

@ARTICLE{2005Bo,
       author = {{Bournaud}, F. and {Combes}, F. and {Semelin}, B.},
        title = "{The lifetime of galactic bars: central mass concentrations and gravity torques}",
      journal = {\mnras},
         year = 2005,
        month = nov,
       volume = {364},
       number = {1},
        pages = {L18-L22},
          doi = {10.1111/j.1745-3933.2005.00096.x},
archivePrefix = {arXiv},
       eprint = {astro-ph/0509126},
 primaryClass = {astro-ph},
       adsurl = {https://ui.adsabs.harvard.edu/abs/2005MNRAS.364L..18B}
}

@ARTICLE{2023Ge,
       author = {{G{\'e}ron}, Tobias and {Smethurst}, Rebecca J. and {Lintott}, Chris and {Kruk}, Sandor and {Masters}, Karen L. and {Simmons}, Brooke and {Mantha}, Kameswara Bharadwaj and {Walmsley}, Mike and {Garma-Oehmichen}, L. and {Drory}, Niv and {Lane}, Richard R.},
        title = "{Galaxy Zoo: kinematics of strongly and weakly barred galaxies}",
      journal = {\mnras},
         year = 2023,
        month = may,
       volume = {521},
       number = {2},
        pages = {1775-1793},
          doi = {10.1093/mnras/stad501},
archivePrefix = {arXiv},
       eprint = {2302.05464},
 primaryClass = {astro-ph.GA},
       adsurl = {https://ui.adsabs.harvard.edu/abs/2023MNRAS.521.1775G}
}

@ARTICLE{2022GO,
       author = {{Garma-Oehmichen}, L. and {Hern{\'a}ndez-Toledo}, H. and {Aquino-Ort{\'\i}z}, E. and {Martinez-Medina}, L. and {Puerari}, I. and {Cano-D{\'\i}az}, M. and {Valenzuela}, O. and {V{\'a}zquez-Mata}, J.~A. and {G{\'e}ron}, T. and {Mart{\'\i}nez-V{\'a}zquez}, L.~A. and {Lane}, R.},
        title = "{SDSS IV MaNGA: bar pattern speed in Milky Way analogue galaxies}",
      journal = {\mnras},
         year = 2022,
        month = dec,
       volume = {517},
       number = {4},
        pages = {5660-5677},
          doi = {10.1093/mnras/stac3069},
archivePrefix = {arXiv},
       eprint = {2210.11424},
 primaryClass = {astro-ph.GA},
       adsurl = {https://ui.adsabs.harvard.edu/abs/2022MNRAS.517.5660G}
}

@ARTICLE{2015Bu2,
       author = {{Bundy}, Kevin and {Bershady}, Matthew A. and {Law}, David R. and {Yan}, Renbin and {Drory}, Niv and {MacDonald}, Nicholas and {Wake}, David A. and {Cherinka}, Brian and {S{\'a}nchez-Gallego}, Jos{\'e} R. and {Weijmans}, Anne-Marie and {Thomas}, Daniel and {Tremonti}, Christy and {Masters}, Karen and {Coccato}, Lodovico and {Diamond-Stanic}, Aleksandar M. and {Arag{\'o}n-Salamanca}, Alfonso and {Avila-Reese}, Vladimir and {Badenes}, Carles and {Falc{\'o}n-Barroso}, J{\'e}sus and {Belfiore}, Francesco and {Bizyaev}, Dmitry and {Blanc}, Guillermo A. and {Bland-Hawthorn}, Joss and {Blanton}, Michael R. and {Brownstein}, Joel R. and {Byler}, Nell and {Cappellari}, Michele and {Conroy}, Charlie and {Dutton}, Aaron A. and {Emsellem}, Eric and {Etherington}, James and {Frinchaboy}, Peter M. and {Fu}, Hai and {Gunn}, James E. and {Harding}, Paul and {Johnston}, Evelyn J. and {Kauffmann}, Guinevere and {Kinemuchi}, Karen and {Klaene}, Mark A. and {Knapen}, Johan H. and {Leauthaud}, Alexie and {Li}, Cheng and {Lin}, Lihwai and {Maiolino}, Roberto and {Malanushenko}, Viktor and {Malanushenko}, Elena and {Mao}, Shude and {Maraston}, Claudia and {McDermid}, Richard M. and {Merrifield}, Michael R. and {Nichol}, Robert C. and {Oravetz}, Daniel and {Pan}, Kaike and {Parejko}, John K. and {Sanchez}, Sebastian F. and {Schlegel}, David and {Simmons}, Audrey and {Steele}, Oliver and {Steinmetz}, Matthias and {Thanjavur}, Karun and {Thompson}, Benjamin A. and {Tinker}, Jeremy L. and {van den Bosch}, Remco C.~E. and {Westfall}, Kyle B. and {Wilkinson}, David and {Wright}, Shelley and {Xiao}, Ting and {Zhang}, Kai},
        title = "{Overview of the SDSS-IV MaNGA Survey: Mapping nearby Galaxies at Apache Point Observatory}",
      journal = {\apj},
         year = 2015,
        month = jan,
       volume = {798},
       number = {1},
          eid = {7},
        pages = {7},
          doi = {10.1088/0004-637X/798/1/7},
archivePrefix = {arXiv},
       eprint = {1412.1482},
 primaryClass = {astro-ph.GA},
       adsurl = {https://ui.adsabs.harvard.edu/abs/2015ApJ...798....7B}
}

@ARTICLE{2023Ha,
       author = {{Hamilton}, Chris and {Tolman}, Elizabeth A. and {Arzamasskiy}, Lev and {Duarte}, Vin{\'\i}cius N.},
        title = "{Galactic Bar Resonances with Diffusion: An Analytic Model with Implications for Bar-Dark Matter Halo Dynamical Friction}",
      journal = {\apj},
         year = 2023,
        month = sep,
       volume = {954},
       number = {1},
          eid = {12},
        pages = {12},
          doi = {10.3847/1538-4357/acd69b},
archivePrefix = {arXiv},
       eprint = {2208.03855},
 primaryClass = {astro-ph.GA},
       adsurl = {https://ui.adsabs.harvard.edu/abs/2023ApJ...954...12H}
}

@ARTICLE{2018Ho,
       author = {{Hopkins}, Philip F. and {Wetzel}, Andrew and {Kere{\v{s}}}, Du{\v{s}}an and {Faucher-Gigu{\`e}re}, Claude-Andr{\'e} and {Quataert}, Eliot and {Boylan-Kolchin}, Michael and {Murray}, Norman and {Hayward}, Christopher C. and {Garrison-Kimmel}, Shea and {Hummels}, Cameron and {Feldmann}, Robert and {Torrey}, Paul and {Ma}, Xiangcheng and {Angl{\'e}s-Alc{\'a}zar}, Daniel and {Su}, Kung-Yi and {Orr}, Matthew and {Schmitz}, Denise and {Escala}, Ivanna and {Sanderson}, Robyn and {Grudi{\'c}}, Michael Y. and {Hafen}, Zachary and {Kim}, Ji-Hoon and {Fitts}, Alex and {Bullock}, James S. and {Wheeler}, Coral and {Chan}, T.~K. and {Elbert}, Oliver D. and {Narayanan}, Desika},
        title = "{FIRE-2 simulations: physics versus numerics in galaxy formation}",
      journal = {\mnras},
         year = 2018,
        month = oct,
       volume = {480},
       number = {1},
        pages = {800-863},
          doi = {10.1093/mnras/sty1690},
archivePrefix = {arXiv},
       eprint = {1702.06148},
 primaryClass = {astro-ph.GA},
       adsurl = {https://ui.adsabs.harvard.edu/abs/2018MNRAS.480..800H}
}

@ARTICLE{2014Vo,
       author = {{Vogelsberger}, M. and {Genel}, S. and {Springel}, V. and {Torrey}, P. and {Sijacki}, D. and {Xu}, D. and {Snyder}, G. and {Bird}, S. and {Nelson}, D. and {Hernquist}, L.},
        title = "{Properties of galaxies reproduced by a hydrodynamic simulation}",
      journal = {\nat},
         year = 2014,
        month = may,
       volume = {509},
       number = {7499},
        pages = {177-182},
          doi = {10.1038/nature13316},
archivePrefix = {arXiv},
       eprint = {1405.1418},
 primaryClass = {astro-ph.CO},
       adsurl = {https://ui.adsabs.harvard.edu/abs/2014Natur.509..177V}
}

@ARTICLE{2018Ma,
       author = {{Marinacci}, Federico and {Vogelsberger}, Mark and {Pakmor}, R{\"u}diger and {Torrey}, Paul and {Springel}, Volker and {Hernquist}, Lars and {Nelson}, Dylan and {Weinberger}, Rainer and {Pillepich}, Annalisa and {Naiman}, Jill and {Genel}, Shy},
        title = "{First results from the IllustrisTNG simulations: radio haloes and magnetic fields}",
      journal = {\mnras},
         year = 2018,
        month = nov,
       volume = {480},
       number = {4},
        pages = {5113-5139},
          doi = {10.1093/mnras/sty2206},
archivePrefix = {arXiv},
       eprint = {1707.03396},
 primaryClass = {astro-ph.CO},
       adsurl = {https://ui.adsabs.harvard.edu/abs/2018MNRAS.480.5113M}
}

@ARTICLE{2020Cu,
       author = {{Cuomo}, V. and {Aguerri}, J.~A.~L. and {Corsini}, E.~M. and {Debattista}, V.~P.},
        title = "{Relations among structural parameters in barred galaxies with a direct measurement of bar pattern speed}",
      journal = {\aap},
         year = 2020,
        month = sep,
       volume = {641},
          eid = {A111},
        pages = {A111},
          doi = {10.1051/0004-6361/202037945},
archivePrefix = {arXiv},
       eprint = {2003.07455},
 primaryClass = {astro-ph.GA},
       adsurl = {https://ui.adsabs.harvard.edu/abs/2020A&A...641A.111C}
}

@ARTICLE{2020GO,
       author = {{Garma-Oehmichen}, L. and {Cano-D{\'\i}az}, M. and {Hern{\'a}ndez-Toledo}, H. and {Aquino-Ort{\'\i}z}, E. and {Valenzuela}, O. and {Aguerri}, J.~A.~L. and {S{\'a}nchez}, S.~F. and {Merrifield}, M.},
        title = "{SDSS-IV MaNGA: bar pattern speed estimates with the Tremaine-Weinberg method and their error sources}",
      journal = {\mnras},
         year = 2020,
        month = jan,
       volume = {491},
       number = {3},
        pages = {3655-3671},
          doi = {10.1093/mnras/stz3101},
archivePrefix = {arXiv},
       eprint = {1911.00090},
 primaryClass = {astro-ph.GA},
       adsurl = {https://ui.adsabs.harvard.edu/abs/2020MNRAS.491.3655G}
}

@ARTICLE{2022Fr,
       author = {{Frankel}, Neige and {Pillepich}, Annalisa and {Rix}, Hans-Walter and {Rodriguez-Gomez}, Vicente and {Sanders}, Jason and {Bovy}, Jo and {Kollmeier}, Juna and {Murray}, Norm and {Mackereth}, Ted},
        title = "{Simulated Bars May Be Shorter but Are Not Slower Than Those Observed: TNG50 versus MaNGA}",
      journal = {\apj},
         year = 2022,
        month = nov,
       volume = {940},
       number = {1},
          eid = {61},
        pages = {61},
          doi = {10.3847/1538-4357/ac9972},
archivePrefix = {arXiv},
       eprint = {2201.08406},
 primaryClass = {astro-ph.GA},
       adsurl = {https://ui.adsabs.harvard.edu/abs/2022ApJ...940...61F}
}

@ARTICLE{2020Ma,
       author = {{Marasco}, A. and {Posti}, L. and {Oman}, K. and {Famaey}, B. and {Cresci}, G. and {Fraternali}, F.},
        title = "{Massive disc galaxies too dominated by dark matter in cosmological hydrodynamical simulations}",
      journal = {\aap},
         year = 2020,
        month = aug,
       volume = {640},
          eid = {A70},
        pages = {A70},
          doi = {10.1051/0004-6361/202038326},
archivePrefix = {arXiv},
       eprint = {2005.01724},
 primaryClass = {astro-ph.GA},
       adsurl = {https://ui.adsabs.harvard.edu/abs/2020A&A...640A..70M}
}

@ARTICLE{2015GB,
       author = {{Garc{\'\i}a-Benito}, R. and {Zibetti}, S. and {S{\'a}nchez}, S.~F. and {Husemann}, B. and {de Amorim}, A.~L. and {Castillo-Morales}, A. and {Cid Fernandes}, R. and {Ellis}, S.~C. and {Falc{\'o}n-Barroso}, J. and {Galbany}, L. and {Gil de Paz}, A. and {Gonz{\'a}lez Delgado}, R.~M. and {Lacerda}, E.~A.~D. and {L{\'o}pez-Fernandez}, R. and {de Lorenzo-C{\'a}ceres}, A. and {Lyubenova}, M. and {Marino}, R.~A. and {Mast}, D. and {Mendoza}, M.~A. and {P{\'e}rez}, E. and {Vale Asari}, N. and {Aguerri}, J.~A.~L. and {Ascasibar}, Y. and {Bekerait{\.{e}}}, S. and {Bland-Hawthorn}, J. and {Barrera-Ballesteros}, J.~K. and {Bomans}, D.~J. and {Cano-D{\'\i}az}, M. and {Catal{\'a}n-Torrecilla}, C. and {Cortijo}, C. and {Delgado-Inglada}, G. and {Demleitner}, M. and {Dettmar}, R. -J. and {D{\'\i}az}, A.~I. and {Florido}, E. and {Gallazzi}, A. and {Garc{\'\i}a-Lorenzo}, B. and {Gomes}, J.~M. and {Holmes}, L. and {Iglesias-P{\'a}ramo}, J. and {Jahnke}, K. and {Kalinova}, V. and {Kehrig}, C. and {Kennicutt}, R.~C. and {L{\'o}pez-S{\'a}nchez}, {\'A}. R. and {M{\'a}rquez}, I. and {Masegosa}, J. and {Meidt}, S.~E. and {Mendez-Abreu}, J. and {Moll{\'a}}, M. and {Monreal-Ibero}, A. and {Morisset}, C. and {del Olmo}, A. and {Papaderos}, P. and {P{\'e}rez}, I. and {Quirrenbach}, A. and {Rosales-Ortega}, F.~F. and {Roth}, M.~M. and {Ruiz-Lara}, T. and {S{\'a}nchez-Bl{\'a}zquez}, P. and {S{\'a}nchez-Menguiano}, L. and {Singh}, R. and {Spekkens}, K. and {Stanishev}, V. and {Torres-Papaqui}, J.~P. and {van de Ven}, G. and {Vilchez}, J.~M. and {Walcher}, C.~J. and {Wild}, V. and {Wisotzki}, L. and {Ziegler}, B. and {Alves}, J. and {Barrado}, D. and {Quintana}, J.~M. and {Aceituno}, J.},
        title = "{CALIFA, the Calar Alto Legacy Integral Field Area survey. III. Second public data release}",
      journal = {\aap},
         year = 2015,
        month = apr,
       volume = {576},
          eid = {A135},
        pages = {A135},
          doi = {10.1051/0004-6361/201425080},
archivePrefix = {arXiv},
       eprint = {1409.8302},
 primaryClass = {astro-ph.GA},
       adsurl = {https://ui.adsabs.harvard.edu/abs/2015A&A...576A.135G}
}

@ARTICLE{2002Pa,
       author = {{Paturel}, G. and {Dubois}, P. and {Petit}, C. and {Woelfel}, F.},
        title = "{Comparison LEDA/SIMBAD octobre 2002. Catalogue to be published in 2003.}",
      journal = {LEDA},
         year = 2002,
        month = jan,
        pages = {0},
       adsurl = {https://ui.adsabs.harvard.edu/abs/2002LEDA.........0P}
}

@ARTICLE{2024Ha,
       author = {{Habibi}, Asiyeh and {Roshan}, Mahmood and {Hosseinirad}, Mohammad and {Khosroshahi}, Habib and {Aguerri}, J.~A.~L. and {Cuomo}, Virginia and {Abbassi}, Shahram},
        title = "{The redshift evolution of galactic bar pattern speed in TNG50}",
      journal = {\aap},
         year = 2024,
        month = nov,
       volume = {691},
          eid = {A122},
        pages = {A122},
          doi = {10.1051/0004-6361/202451028},
archivePrefix = {arXiv},
       eprint = {2409.02456},
 primaryClass = {astro-ph.GA},
       adsurl = {https://ui.adsabs.harvard.edu/abs/2024A&A...691A.122H}
}

@ARTICLE{2019Ka,
       author = {{Kataria}, Sandeep Kumar and {Das}, Mousumi},
        title = "{The Effect of Bulge Mass on Bar Pattern Speed in Disk Galaxies}",
      journal = {\apj},
         year = 2019,
        month = nov,
       volume = {886},
       number = {1},
          eid = {43},
        pages = {43},
          doi = {10.3847/1538-4357/ab48f7},
archivePrefix = {arXiv},
       eprint = {1910.03967},
 primaryClass = {astro-ph.GA},
       adsurl = {https://ui.adsabs.harvard.edu/abs/2019ApJ...886...43K}
}

@ARTICLE{2023Li,
       author = {{Li}, Xingchen and {Shlosman}, Isaac and {Heller}, Clayton and {Pfenniger}, Daniel},
        title = "{Stellar bars in spinning haloes: delayed buckling and absence of slowdown}",
      journal = {\mnras},
         year = 2023,
        month = dec,
       volume = {526},
       number = {2},
        pages = {1972-1986},
          doi = {10.1093/mnras/stad2799},
archivePrefix = {arXiv},
       eprint = {2211.04484},
 primaryClass = {astro-ph.GA},
       adsurl = {https://ui.adsabs.harvard.edu/abs/2023MNRAS.526.1972L}
}

@ARTICLE{2022RG,
       author = {{Rosas-Guevara}, Yetli and {Bonoli}, Silvia and {Dotti}, Massimo and {Izquierdo-Villalba}, David and {Lupi}, Alessandro and {Zana}, Tommaso and {Bonetti}, Matteo and {Nelson}, Dylan and {Springel}, Volker and {Hernquist}, Lars and {Vogelsberger}, Mark},
        title = "{The evolution of the barred galaxy population in the TNG50 simulation}",
      journal = {\mnras},
         year = 2022,
        month = jun,
       volume = {512},
       number = {4},
        pages = {5339-5357},
          doi = {10.1093/mnras/stac816},
archivePrefix = {arXiv},
       eprint = {2110.04537},
 primaryClass = {astro-ph.GA},
       adsurl = {https://ui.adsabs.harvard.edu/abs/2022MNRAS.512.5339R}
}

@ARTICLE{2019Zo,
       author = {{Zou}, Yanfei and {Shen}, Juntai and {Bureau}, Martin and {Li}, Zhao-Yu},
        title = "{Testing the Tremaine-Weinberg Method Applied to Integral-field Spectroscopic Data Using a Simulated Barred Galaxy}",
      journal = {\apj},
         year = 2019,
        month = oct,
       volume = {884},
       number = {1},
          eid = {23},
        pages = {23},
          doi = {10.3847/1538-4357/ab3f34},
archivePrefix = {arXiv},
       eprint = {1908.10524},
 primaryClass = {astro-ph.GA},
       adsurl = {https://ui.adsabs.harvard.edu/abs/2019ApJ...884...23Z}
}

@ARTICLE{2024Gh,
       author = {{Ghosh}, Soumavo and {Di Matteo}, Paola},
        title = "{Looking for a needle in a haystack: Measuring the length of a stellar bar}",
      journal = {\aap},
         year = 2024,
        month = mar,
       volume = {683},
          eid = {A100},
        pages = {A100},
          doi = {10.1051/0004-6361/202347763},
archivePrefix = {arXiv},
       eprint = {2308.10948},
 primaryClass = {astro-ph.GA},
       adsurl = {https://ui.adsabs.harvard.edu/abs/2024A&A...683A.100G}
}

@ARTICLE{2003De,
       author = {{Debattista}, Victor P.},
        title = "{On position angle errors in the Tremaine-Weinberg method}",
      journal = {\mnras},
         year = 2003,
        month = jul,
       volume = {342},
       number = {4},
        pages = {1194-1204},
          doi = {10.1046/j.1365-8711.2003.06620.x},
archivePrefix = {arXiv},
       eprint = {astro-ph/0401137},
 primaryClass = {astro-ph},
       adsurl = {https://ui.adsabs.harvard.edu/abs/2003MNRAS.342.1194D}
}

@ARTICLE{2026LC,
       author = {{Le Conte}, Zoe A. and {Gadotti}, Dimitri A. and {Ferreira}, Leonardo and {Conselice}, Christopher J. and {de S{\'a}-Freitas}, Camila and {Kim}, Taehyun and {Neumann}, Justus and {Fragkoudi}, Francesca and {Athanassoula}, E. and {Adams}, Nathan J.},
        title = "{The evolution of the bar fraction and bar lengths in the last 12 billion years}",
      journal = {\mnras},
         year = 2026,
        month = jan,
       volume = {545},
       number = {1},
          eid = {staf2010},
        pages = {staf2010},
          doi = {10.1093/mnras/staf2010},
archivePrefix = {arXiv},
       eprint = {2510.07407},
 primaryClass = {astro-ph.GA},
       adsurl = {https://ui.adsabs.harvard.edu/abs/2026MNRAS.545f2010L}
}

@ARTICLE{2019Cu,
       author = {{Cuomo}, Virginia and {Lopez Aguerri}, J. Alfonso and {Corsini}, Enrico Maria and {Debattista}, Victor P. and {M{\'e}ndez-Abreu}, Jairo and {Pizzella}, Alessandro},
        title = "{Bar pattern speeds in CALIFA galaxies. II. The case of weakly barred galaxies}",
      journal = {\aap},
         year = 2019,
        month = dec,
       volume = {632},
          eid = {A51},
        pages = {A51},
          doi = {10.1051/0004-6361/201936415},
archivePrefix = {arXiv},
       eprint = {1909.01023},
 primaryClass = {astro-ph.GA},
       adsurl = {https://ui.adsabs.harvard.edu/abs/2019A&A...632A..51C}
}

@ARTICLE{2023Bu,
       author = {{Buttitta}, C. and {Corsini}, E.~M. and {Aguerri}, J.~A.~L. and {Coccato}, L. and {Costantin}, L. and {Cuomo}, V. and {Debattista}, V.~P. and {Morelli}, L. and {Pizzella}, A.},
        title = "{The bar rotation rate as a diagnostic of dark matter content in the centre of disc galaxies}",
      journal = {\mnras},
         year = 2023,
        month = may,
       volume = {521},
       number = {2},
        pages = {2227-2238},
          doi = {10.1093/mnras/stad646},
archivePrefix = {arXiv},
       eprint = {2303.11441},
 primaryClass = {astro-ph.GA},
       adsurl = {https://ui.adsabs.harvard.edu/abs/2023MNRAS.521.2227B}
}

@ARTICLE{2009Kl,
       author = {{Klypin}, Anatoly and {Valenzuela}, Octavio and {Col{\'\i}n}, Pedro and {Quinn}, Thomas},
        title = "{Dynamics of barred galaxies: effects of disc height}",
      journal = {\mnras},
         year = 2009,
        month = sep,
       volume = {398},
       number = {2},
        pages = {1027-1040},
          doi = {10.1111/j.1365-2966.2009.15187.x},
archivePrefix = {arXiv},
       eprint = {0808.3422},
 primaryClass = {astro-ph},
       adsurl = {https://ui.adsabs.harvard.edu/abs/2009MNRAS.398.1027K}
}

@ARTICLE{2011Ga,
       author = {{Gadotti}, Dimitri A.},
        title = "{Secular evolution and structural properties of stellar bars in galaxies}",
      journal = {\mnras},
         year = 2011,
        month = aug,
       volume = {415},
       number = {4},
        pages = {3308-3318},
          doi = {10.1111/j.1365-2966.2011.18945.x},
archivePrefix = {arXiv},
       eprint = {1003.1719},
 primaryClass = {astro-ph.CO},
       adsurl = {https://ui.adsabs.harvard.edu/abs/2011MNRAS.415.3308G}
}

@ARTICLE{2018Ne,
       author = {{Nelson}, Dylan and {Pillepich}, Annalisa and {Springel}, Volker and {Weinberger}, Rainer and {Hernquist}, Lars and {Pakmor}, R{\"u}diger and {Genel}, Shy and {Torrey}, Paul and {Vogelsberger}, Mark and {Kauffmann}, Guinevere and {Marinacci}, Federico and {Naiman}, Jill},
        title = "{First results from the IllustrisTNG simulations: the galaxy colour bimodality}",
      journal = {\mnras},
         year = 2018,
        month = mar,
       volume = {475},
       number = {1},
        pages = {624-647},
          doi = {10.1093/mnras/stx3040},
archivePrefix = {arXiv},
       eprint = {1707.03395},
 primaryClass = {astro-ph.GA},
       adsurl = {https://ui.adsabs.harvard.edu/abs/2018MNRAS.475..624N}
}

@ARTICLE{2018Pi,
       author = {{Pillepich}, Annalisa and {Nelson}, Dylan and {Hernquist}, Lars and {Springel}, Volker and {Pakmor}, R{\"u}diger and {Torrey}, Paul and {Weinberger}, Rainer and {Genel}, Shy and {Naiman}, Jill P. and {Marinacci}, Federico and {Vogelsberger}, Mark},
        title = "{First results from the IllustrisTNG simulations: the stellar mass content of groups and clusters of galaxies}",
      journal = {\mnras},
         year = 2018,
        month = mar,
       volume = {475},
       number = {1},
        pages = {648-675},
          doi = {10.1093/mnras/stx3112},
archivePrefix = {arXiv},
       eprint = {1707.03406},
 primaryClass = {astro-ph.GA},
       adsurl = {https://ui.adsabs.harvard.edu/abs/2018MNRAS.475..648P}
}

@ARTICLE{2018Sp,
       author = {{Springel}, Volker and {Pakmor}, R{\"u}diger and {Pillepich}, Annalisa and {Weinberger}, Rainer and {Nelson}, Dylan and {Hernquist}, Lars and {Vogelsberger}, Mark and {Genel}, Shy and {Torrey}, Paul and {Marinacci}, Federico and {Naiman}, Jill},
        title = "{First results from the IllustrisTNG simulations: matter and galaxy clustering}",
      journal = {\mnras},
         year = 2018,
        month = mar,
       volume = {475},
       number = {1},
        pages = {676-698},
          doi = {10.1093/mnras/stx3304},
archivePrefix = {arXiv},
       eprint = {1707.03397},
 primaryClass = {astro-ph.GA},
       adsurl = {https://ui.adsabs.harvard.edu/abs/2018MNRAS.475..676S}
}

@ARTICLE{2018Na,
       author = {{Naiman}, Jill P. and {Pillepich}, Annalisa and {Springel}, Volker and {Ramirez-Ruiz}, Enrico and {Torrey}, Paul and {Vogelsberger}, Mark and {Pakmor}, R{\"u}diger and {Nelson}, Dylan and {Marinacci}, Federico and {Hernquist}, Lars and {Weinberger}, Rainer and {Genel}, Shy},
        title = "{First results from the IllustrisTNG simulations: a tale of two elements - chemical evolution of magnesium and europium}",
      journal = {\mnras},
         year = 2018,
        month = jun,
       volume = {477},
       number = {1},
        pages = {1206-1224},
          doi = {10.1093/mnras/sty618},
archivePrefix = {arXiv},
       eprint = {1707.03401},
 primaryClass = {astro-ph.GA},
       adsurl = {https://ui.adsabs.harvard.edu/abs/2018MNRAS.477.1206N}
}

@ARTICLE{2019Ne,
       author = {{Nelson}, Dylan and {Pillepich}, Annalisa and {Springel}, Volker and {Pakmor}, R{\"u}diger and {Weinberger}, Rainer and {Genel}, Shy and {Torrey}, Paul and {Vogelsberger}, Mark and {Marinacci}, Federico and {Hernquist}, Lars},
        title = "{First results from the TNG50 simulation: galactic outflows driven by supernovae and black hole feedback}",
      journal = {\mnras},
         year = 2019,
        month = dec,
       volume = {490},
       number = {3},
        pages = {3234-3261},
          doi = {10.1093/mnras/stz2306},
archivePrefix = {arXiv},
       eprint = {1902.05554},
 primaryClass = {astro-ph.GA},
       adsurl = {https://ui.adsabs.harvard.edu/abs/2019MNRAS.490.3234N}
}

@ARTICLE{2019Pi,
       author = {{Pillepich}, Annalisa and {Nelson}, Dylan and {Springel}, Volker and {Pakmor}, R{\"u}diger and {Torrey}, Paul and {Weinberger}, Rainer and {Vogelsberger}, Mark and {Marinacci}, Federico and {Genel}, Shy and {van der Wel}, Arjen and {Hernquist}, Lars},
        title = "{First results from the TNG50 simulation: the evolution of stellar and gaseous discs across cosmic time}",
      journal = {\mnras},
         year = 2019,
        month = dec,
       volume = {490},
       number = {3},
        pages = {3196-3233},
          doi = {10.1093/mnras/stz2338},
archivePrefix = {arXiv},
       eprint = {1902.05553},
 primaryClass = {astro-ph.GA},
       adsurl = {https://ui.adsabs.harvard.edu/abs/2019MNRAS.490.3196P}
}

@ARTICLE{2015Cr,
       author = {{Crain}, Robert A. and {Schaye}, Joop and {Bower}, Richard G. and {Furlong}, Michelle and {Schaller}, Matthieu and {Theuns}, Tom and {Dalla Vecchia}, Claudio and {Frenk}, Carlos S. and {McCarthy}, Ian G. and {Helly}, John C. and {Jenkins}, Adrian and {Rosas-Guevara}, Yetli M. and {White}, Simon D.~M. and {Trayford}, James W.},
        title = "{The EAGLE simulations of galaxy formation: calibration of subgrid physics and model variations}",
      journal = {\mnras},
         year = 2015,
        month = jun,
       volume = {450},
       number = {2},
        pages = {1937-1961},
          doi = {10.1093/mnras/stv725},
archivePrefix = {arXiv},
       eprint = {1501.01311},
 primaryClass = {astro-ph.GA},
       adsurl = {https://ui.adsabs.harvard.edu/abs/2015MNRAS.450.1937C}
}

@ARTICLE{2023We,
       author = {{Wetzel}, Andrew and {Hayward}, Christopher C. and {Sanderson}, Robyn E. and {Ma}, Xiangcheng and {Angl{\'e}s-Alc{\'a}zar}, Daniel and {Feldmann}, Robert and {Chan}, T.~K. and {El-Badry}, Kareem and {Wheeler}, Coral and {Garrison-Kimmel}, Shea and {Nikakhtar}, Farnik and {Panithanpaisal}, Nondh and {Arora}, Arpit and {Gurvich}, Alexander B. and {Samuel}, Jenna and {Sameie}, Omid and {Pandya}, Viraj and {Hafen}, Zachary and {Hummels}, Cameron and {Loebman}, Sarah and {Boylan-Kolchin}, Michael and {Bullock}, James S. and {Faucher-Gigu{\`e}re}, Claude-Andr{\'e} and {Kere{\v{s}}}, Du{\v{s}}an and {Quataert}, Eliot and {Hopkins}, Philip F.},
        title = "{Public Data Release of the FIRE-2 Cosmological Zoom-in Simulations of Galaxy Formation}",
      journal = {\apjs},
         year = 2023,
        month = apr,
       volume = {265},
       number = {2},
          eid = {44},
        pages = {44},
          doi = {10.3847/1538-4365/acb99a},
archivePrefix = {arXiv},
       eprint = {2202.06969},
 primaryClass = {astro-ph.GA},
       adsurl = {https://ui.adsabs.harvard.edu/abs/2023ApJS..265...44W}
}

@ARTICLE{2008Li,
       author = {{Lintott}, Chris J. and {Schawinski}, Kevin and {Slosar}, An{\v{z}}e and {Land}, Kate and {Bamford}, Steven and {Thomas}, Daniel and {Raddick}, M. Jordan and {Nichol}, Robert C. and {Szalay}, Alex and {Andreescu}, Dan and {Murray}, Phil and {Vandenberg}, Jan},
        title = "{Galaxy Zoo: morphologies derived from visual inspection of galaxies from the Sloan Digital Sky Survey}",
      journal = {\mnras},
         year = 2008,
        month = sep,
       volume = {389},
       number = {3},
        pages = {1179-1189},
          doi = {10.1111/j.1365-2966.2008.13689.x},
archivePrefix = {arXiv},
       eprint = {0804.4483},
 primaryClass = {astro-ph},
       adsurl = {https://ui.adsabs.harvard.edu/abs/2008MNRAS.389.1179L}
}



\appendix

\section{Bar length measurement}
\label{lengthapp}

\begin{figure*}
	\includegraphics[width=2\columnwidth]{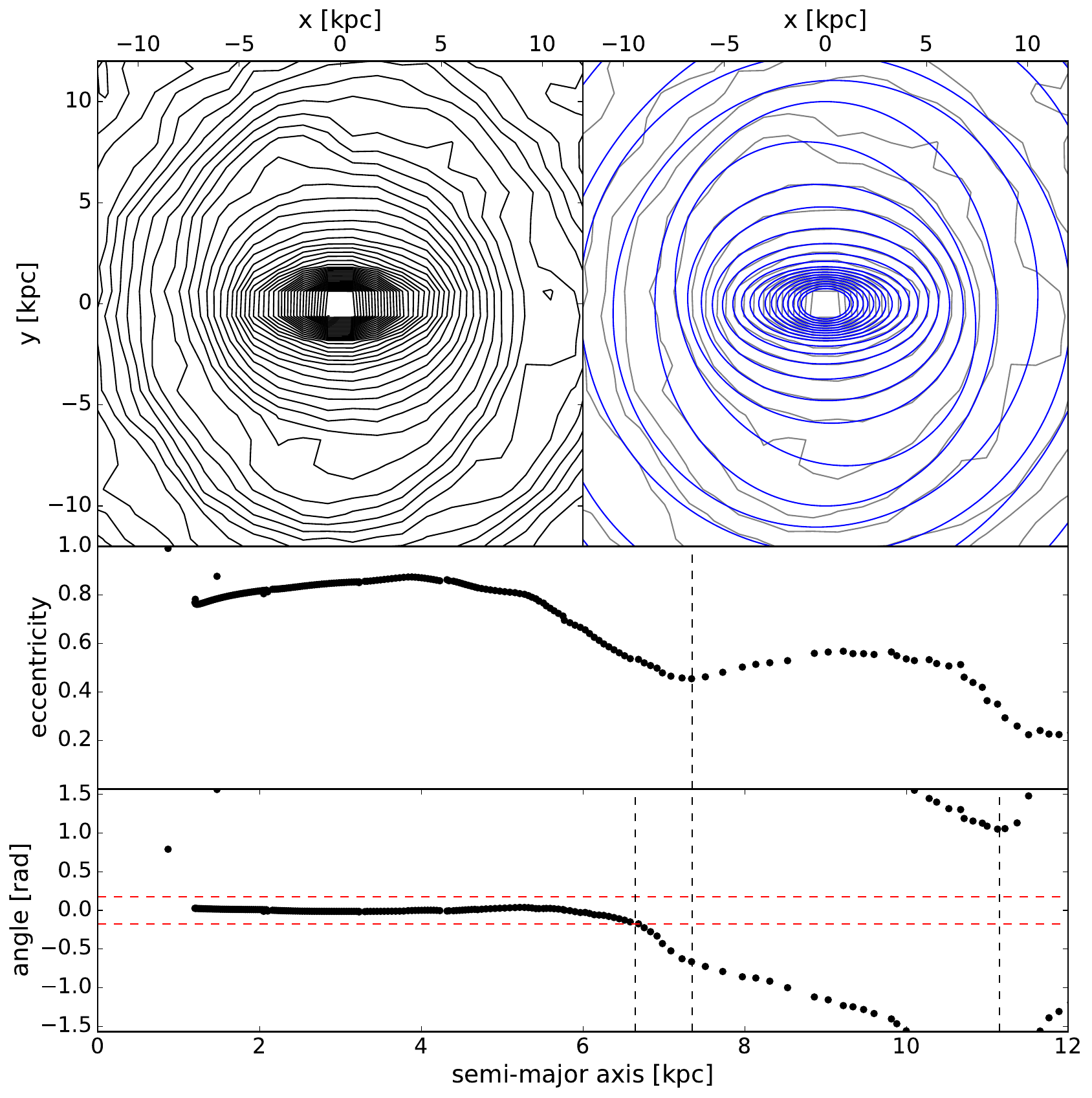}
    \caption{An illustration of of our method for finding the bar length for Au24 at $z=0$. Top left: the contours of the face-on stellar surface density profile within $12\,\mathrm{kpc}$. Top right: the ellipse fitting (blue) to the largest closed loop of each contour level from the previous panel (grey). Middle: the eccentricity plotted against semi-major axis of each ellipse fitted. Bottom: the angle plotted against semi-major axis of each ellipse fitted, with $\pm\frac{\pi}{12}\,\mathrm{rad}$ shown by the red dashed lines. In both the lower two panels, the vertical grey dashed lines indicate the radius of the 3 features used to determine the bar length.}
    \label{fig:barexample}
\end{figure*}

This section gives a description of the method we use to calculate the bar lengths in our sample. Fig.~\ref{fig:barexample} illustrates an example of this procedure for Au24 at $z=0$, in the final snapshot of its simulation. 

After aligning the disc to the x-y plane and the bar to the x-axis as described in Sec.~\ref{method:internal}, we find the contours to the face-on stellar surface density. We show every $10^\mathrm{th}$ calculated contour in the upper left panel of Fig.~\ref{fig:barexample}. For each contour level, we fit an ellipse to the longest closed loop, centred on the origin, as shown in the upper right panel in blue.

The lower two panels show the eccentricity and angle from the aligned bar axis respectively, against the semi-major axis of each ellipse up to $12\,\mathrm{kpc}$. The bar length is then the minimum of three radii: the first minimum in eccentricity outside the visually strong bar region, the first turning point in angle outside the visually strong bar region, and the first deviation in angle of more than $15^\circ$/$\frac{\pi}{12}\,\mathrm{rad}$ from the bar axis (excluding the inner $\mathrm{1\,kpc}$ which shows little bar signal). These three radii are shown in Fig.~\ref{fig:barexample} as the middle, right, and left vertical dashed lines respectively. In this case, we find a bar length of $6.5\,\mathrm{kpc}$, corresponding to the left-most dashed line.

To match observations where only one time is available, we take our bar length measurements from the individual snipshot for each initial and final time. However, there are effects which can introduce inaccuracies, most notably artificial lengthening of the bar from spiral arms temporarily aligning with the ends of the bar \citep[see][]{2002Bi,2011Qu,2020Hi}. To acknowledge this in our results, we compute a measure of uncertainty as the range of bar lengths found for a few snipshots within $100\,\mathrm{Myr}$ of the quoted time. For temporary/oscillating inaccuracies in the bar length measurement, these uncertainties should reliably encompass the true bar length.

\section{Comprehensive correlations}
\label{corrapp}

\begin{figure*}
	\includegraphics[width=2\columnwidth]{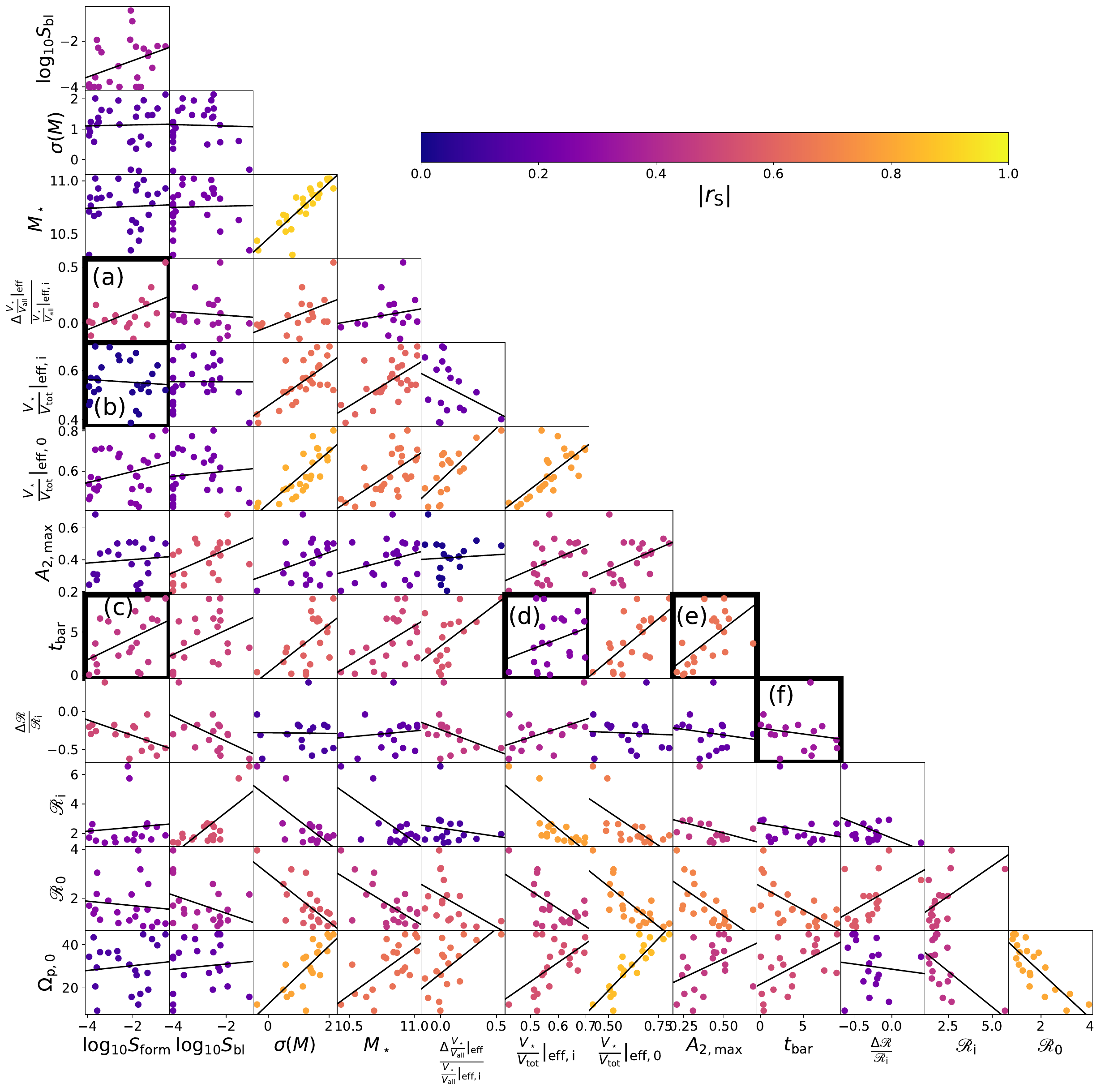}
    \caption{Plots comparing, in pairs, the parameters shown in figures within Sec.~\ref{allresults}: $\Omega_\mathrm{p,0}$ the pattern speed at $z=0$ in units of $\mathrm{rad}\,\mathrm{Gyr}^{-1}$; $\mathscr{R}_0$ the rotation rate at $z=0$; $\mathscr{R}_\mathrm{i}$ the rotation rate at time $t_\mathrm{i}$; $\frac{\Delta\mathscr{R}}{\mathscr{R}_\mathrm{i}}$ the fractional change in rotation rate from $t_\mathrm{i}$ until $z=0$; $t_\mathrm{bar}$ the lookback time to bar formation in units of $\mathrm{Gyr}$; $A_{2\mathrm{,max}}$ the bar strength at $z=0$; $\frac{V_\star}{V_\mathrm{all}}|_{\mathrm{eff,}0}$ the baryon dominance within the effective radius at $z=0$; $\frac{V_\star}{V_\mathrm{all}}|_\mathrm{eff,i}$ the baryon dominance within the effective radius at $t_\mathrm{i}$; $\frac{\Delta\frac{V_\star}{V_\mathrm{all}}|_\mathrm{eff}}{\frac{V_\star}{V_\mathrm{all}}|_\mathrm{eff,i}}$ the relative change in baryon dominance within the effective radius from $t_\mathrm{i}$ until $z=0$; $M_\star$ the stellar mass in units of $\mathrm{M}_\odot$; $\sigma(M)$ the vertical offset from the abundance matching relation in units of the $1\sigma$ dispersion in the relation (see Sec.~\ref{globalresults}); $\log_{10}S_\mathrm{bl}$ the logarithm of the maximum interaction strength from $t_\mathrm{i}$ until $z=0$; $\log_{10}S_\mathrm{form}$ the logarithm of the maximum interaction strength from $1\,\mathrm{Gyr}$ before to $0.5\,\mathrm{Gyr}$ after $t_\mathrm{bar}$. In each panel, we include all halos in our sample for which the parameters involved have a physical meaning (e.g. unbarred galaxies are not included in panels involving pattern speed). Each panel contains a linear line of best fit. The colours of points indicates the magnitude of the Spearman's rank correlation coefficient $r_\mathrm{S}$ between the two parameters in the panel, with high correlation in yellow and no correlation in blue. The 6 labelled panels with thick borders indicate the panels referred to from the main body of this paper.}
    \label{fig:app_wide}
\end{figure*}

We present in Fig.~\ref{fig:app_wide} an overview of the correlations existing between each of the parameters presented above. This includes interdependencies between each pair of parameters, in addition to the relations with pattern speed and rotation rate shown in the lower 4 rows. Each panel contains points for each relevant galaxy, a line of best fit, and a colour to indicate the strength of the correlation between the pair of parameters, with yellow as the most correlated and blue as the least. For example, the panel in the third column from the left and third row from the top shows the relation between the most strongly correlated pair of parameters, $M_\star$ and $\sigma(M)$ -- as seen in Fig.~\ref{fig:abundance} and explored further in \citet{2025Fr}, galaxies in the Auriga simulations with higher stellar masses tend to lie further above the abundance matching relation from \citet{2013Mo}. In contrast, the panel in the 5th column from the left and 6th row from the bottom shows the relation between the least correlated pair of parameters, $A_{2\mathrm{,max}}$ and $\frac{\Delta\frac{V_\star}{V_\mathrm{all}}|_\mathrm{eff}}{\frac{V_\star}{V_\mathrm{all}}|_\mathrm{eff,i}}$ -- this indicates that the relative change in baryon dominance in the inner regions since the bar formed has no relation to the final strength of the bar.

In the above, we directly mention 6 of the relations shown here, outside of the focal relations with measures of bar speed. In Fig.~\ref{fig:app_wide}, these 6 relations are labelled (a)-(f) and each have a thick border. In Sec.~\ref{barresults} we refer to the weak correlation between $t_\mathrm{bar}$ and $\frac{\Delta\mathscr{R}}{\mathscr{R}_\mathrm{i}}$, shown in panel (f), as well as the positive correlation between $A_{2\mathrm{,max}}$ and $t_\mathrm{bar}$, shown in panel (e). In Sec.~\ref{externalresults} we refer to 4 relations. The positive correlation between $\log_{10}S_\mathrm{form}$ and $t_\mathrm{bar}$ is shown in panel (c). The weak positive correlation between $\frac{V_\star}{V_\mathrm{all}}|_\mathrm{eff,i}$ and $t_\mathrm{bar}$ is shown in panel (d). The lack of correlation between $\log_{10}S_\mathrm{form}$ and $\frac{V_\star}{V_\mathrm{all}}|_\mathrm{eff,i}$ is shown in panel (b). Lastly, the positive relation between $\log_{10}S_\mathrm{form}$ and $\frac{\Delta\frac{V_\star}{V_\mathrm{all}}|_\mathrm{eff}}{\frac{V_\star}{V_\mathrm{all}}|_\mathrm{eff,i}}$, which is also referred to in Sec.~\ref{ratediscuss}, is shown in panel (a).


\bsp	
\label{lastpage}
\end{document}